\documentclass[apj]{emulateapj}
\slugcomment{{\sc Accepted to ApJS:} August 27, 2012}

\shorttitle{Near-infrared properties of Sgr~A*}
\shortauthors{Witzel et al.}
\usepackage{natbib}
\usepackage{lscape}

\begin{document}

\title{Source-intrinsic near-infrared properties of Sgr~A*: \\ Total intensity measurements}

\author{G. Witzel\altaffilmark{1}, A. Eckart\altaffilmark{1,2}, M. Bremer\altaffilmark{1}, M. Zamaninasab\altaffilmark{2}, B. Shahzamanian\altaffilmark{1,2}, M. Valencia-S.\altaffilmark{1,2}, \\
R. Sch\"odel\altaffilmark{3}, V. Karas\altaffilmark{4}, R. Lenzen\altaffilmark{5}, N. Marchili\altaffilmark{2}, N. Sabha\altaffilmark{1}, M. Garcia-Marin\altaffilmark{1}, \\
R. M. Buchholz\altaffilmark{1}, D. Kunneriath\altaffilmark{4} and C. Straubmeier\altaffilmark{1}}

\altaffiltext{1}{I. Physikalisches Institut der Universit\"at zu K\"oln (PH1), Z\"ulpicher Stra\ss e 77, 50937 K\"oln, Germany}
\altaffiltext{2}{Max-Planck-Institut f\"ur Radioastronomie (MPIfR), Auf dem H\"ugel 69, 53121 Bonn, Germany}
\altaffiltext{3}{Instituto de Astrof\'{i}sica de Andaluc\'{i}a (CSIC), Glorieta de la Astronom\'{i}a S/N, 18008 Granada, Spain}
\altaffiltext{4}{Astronomical Institute of the Academy of Sciences, Bocni II 1401/1a, CZ-141 31 Praha 4, Czech Republic}
\altaffiltext{5}{Max-Planck-Institut f\"ur Astronomie (MPIA), K\"onigstuhl 17, 69117 Heidelberg, Germany}
\email{witzel@astro.ucla.edu}

\begin{abstract}
%\begin{scriptsize}
We present a comprehensive data description for Ks-band measurements of Sgr~A*. We characterize the statistical properties of the variability of Sgr~A* in the near-infrared, which we find to be consistent with a single-state process forming a power-law distribution of the flux density. We discover a linear rms-flux relation for the flux-density range up to 12 mJy on a timescale of 24 minutes. This and the power-law flux density distribution implies a phenomenological, formally non-linear statistical variability model with which we can simulate the observed variability and extrapolate its behavior to higher flux levels and longer timescales. 
%We can show that a bright outburst within the last 400 years that has been discussed as the possible reason for the X-ray emission from massive molecular clouds surrounding the Galactic center can be explained as an extreme value of our statistics without the need for an extraordinary event.
We present reasons why data with our cadence cannot be used to decide on the question whether the power spectral density of the underlying random process shows more structure at timescales between 25 min and 100 min compared to what is expected from a red noise random process.  
%\end{scriptsize}
\end{abstract}


\section{Introduction} \label{intro}

Sagittarius (Sgr~A*) at the center of the our galaxy is a highly variable near-infrared (NIR) and X-ray source which is associated with a $ 4 \times 10^{6} M_{\sun}$ supermassive central black hole (\citealt{1996Natur.383..415E,1997MNRAS.284..576E}; \citealt{2002MNRAS.331..917E}; \citealt{2002Natur.419..694S}; \citealt{2003ApJ...597L.121E}; \citealt{1998ApJ...509..678G,2000Natur.407..349G,2005ApJ...620..744G,2008ApJ...689.1044G}; \citealt{2009ApJ...692.1075G}). While first detected as a bright, ultra compact, and comparatively steady radio source, the strong variability at shorter wave-lengths, the variable polarization of the NIR emission, and the correlation between fluctuations in the the sub-mm, NIR and X-ray regimes provide evidence that this variable emission originates in the direct surrounding of the black hole. Therefore, properties of the black hole and of the emission and accretion mechanisms in its close surrounding can be studied at these wavelengths (\citealt{2001Natur.413...45B}; \citealt{2003A&A...407L..17P,2008A&A...488..549P}; \citealt{2003Natur.425..934G}; \citealt{2004ApJ...601L.159G,2005ApJ...635.1087G} ;\citealt{2004A&A...427....1E, 2006A&A...450..535E, 2006A&A...455....1E, 2006Msngr.125....2E, 2008A&A...479..625E, 2008JPhCS.131a2002E, 2008A&A...492..337E}; \citealt{2006ApJ...644..198Y, 2006ApJ...650..189Y, 2007ApJ...668L..47Y, 2008ApJ...682..361Y}; \citealt{2007ApJ...667..900H} \citealt{2009ApJ...698..676D}; \citealt{2010A&A...512A...2S}). 

To explain the observed variability and its correlation between the NIR and X-ray regimes several authors suggest Synchrotron Self-Compton (SSC) or inverse Compton emission as the responsible radiation mechanisms (\citealt{2004A&A...427....1E, 2006A&A...450..535E, 2006A&A...455....1E}; \citealt{2004ApJ...606..894Y}; \citealt{2006ApJ...648.1020L}; \citealt{2012A&A...537A..52E}). Relativistic models that assume the variability to be linked to emission from single or multiple hot spots in the accretion disk near the last stable orbit of the black hole have been applied successfully to individual datasets (\citealt{2006A&A...460...15M, 2006A&A...458L..25M, 2007A&A...473..707M}; \citealt{2008JPhCS.131a2008Z}). These models interpret the shorter timescales of the variability (between 10 and 30 minutes) to be dominated by the timescales of the orbital motion\footnote{$ \sim 20 $~minutes at the innermost stable orbit for Sgr~A*.}. Orbital motion close to the black hole (and an associated quasi periodic signal in the light curves) is of special interest: it could used as a timing experiment in a strong gravitational field that might allow for determining black hole parameters like spin or inclination. 

Potential quasi-periodic oscillations (QPOs) have been found in some light curves (see, e.g., \citealt{2003Natur.425..934G}). However, these QPOs are not statistically significant within the overall variability. Based on NIR light curves of 7 nights observed with Keck telescope \cite{2009ApJ...691.1021D} analyzed the flux density distribution of Sgr~A* and the significance of quasi periodic oscillations. They showed that QPOs in total intensity light curves cannot be established with sufficient significance against random fluctuations, finding a pure red noise power spectral density sufficient to account for the time correlation of the fluctuations. 

On the other hand, based on relativistic models, \cite{2010A&A...510A...3Z} predicted a correlation between the modulations of the observed flux density light curves and changes in polarimetric data (also see \citealt{2006A&A...450..535E}, \citealt{2006A&A...460...15M, 2006A&A...458L..25M, 2007A&A...473..707M}; \citealt{2008A&A...479..625E}, \citealt{1973ApJ...183..237C}; \citealt{1977Natur.266..429S}; \citealt{1991A&A...245..454A}; \citealt{1992A&A...257..531K}; \citealt{1995ApJ...448L..21H}; \citealt{2004ApJS..153..205D,2008MNRAS.384..361D}). A comparison of predicted and observed light curve features (obtained from 6 nights of polarimetric observations with VLT and Subaru telescope) through a pattern recognition algorithm resulted in the detection of a signature possibly associated with orbiting matter under the influence of strong gravity. 

Since the first discovery of variable emission of Sgr~A* in the NIR in 2003 (\citealt{2003Natur.425..934G}) a number of publications concentrated on the statistical properties of the flaring activity rather than on interpreting individual observations. These papers investigated the timing properties of the light curves as well as the radiation mechanisms involved (\citealt{2009ApJ...706..348Y}, \citealt{2010A&A...510A...3Z}, \citealt{2011A&A...532A..26B}, \citealt{2011A&A...532A..83S}, \citealt{2012A&A...537A..52E}). On the base of 14 light curves observed between 2004 and 2008, including alternating observations with VLT and Keck, \cite{2009ApJ...694L..87M} discovered first a dominant timescale at about 150 min, supporting linear scaling relations of break timescales in the power spectral density with the black hole mass. They determined the power-law slope of the high-frequency part of the power spectral density (PSD) to be $ 2.1 \pm 0.5 $. 

A comprehensive statistical approach in the analysis of the Ks-band total intensity variability observed with NACO at the VLT has been conducted by \cite{2011ApJ...728...37D}. The authors analyzed VLT K-band data between 2004 and 2009. They presented a detailed investigation of the flux density statistics and described the time-variable stellar confusion at the position of Sgr~A* that makes an investigation of the faint emission difficult. They also emphasized the importance of these faint states for the overall statistical evaluation of the variability of Sgr~A*.  Based on the flux density histogram the authors claim the evidence for two states of variability, a log-normal distributed quiescent state for flux densities $ <5 $ mJy and a power-law distributed flaring state for flux densities $ >5 $ mJy. They argue that it is very unlikely that the same variability process is responsible for both high and low flux density emission from Sgr~A* (the statistical model the authors promote is summarized in Appendix~\ref{domo}). With reference to this model \cite{2010RvMP...82.3121G} state that ``it is a key issue whether the brightest variable emission from Sgr A* are statistical fluctuations from the probability distribution at low flux or flare events with distinct properties" and that ``the transition from the log-normal distribution at low-flux levels to the tail of high fluxes may also explain the apparent mismatch between the detection vs. non- detection of quasi-periodic substructures in different near-infrared light curve studies". 

Our statistical analysis presented in this paper serves the following goals: 

\begin{itemize}
\item to provide a more comprehensive, uniformly reduced data set of Ks-band observations from 2003 to early 2010;

\item to conduct a rigorous analysis of the observed flux density distribution;

\item to explain why a proper statistical analysis of the Ks-band light curves cannot reproduce the results found by \cite{2011ApJ...728...37D};

\item to conduct a rigorous time series analysis on the base of a representative dataset;

\item  to propose a comprehensive statistical model that, using standard methods for generating Fourier transform based surrogate data, describes all aspects of the observed (total intensity) data and lets us simulate light curves with the observed time behavior and flux density distribution;

\item and to investigate extreme variability events in the context of our statistics.
\end{itemize}

\section{Data reduction} 
\label{obsre}

In the following, we describe the data and the reduction methods we applied in order to obtain time-resolved photometric information on Sgr~A*. Whereas large portions of the dataset are the same as used in the analysis by \cite{2011ApJ...728...37D}, we have chosen different reduction methods: Lucy-Richardson deconvolution in order to guarantee the best possible isolation of the target sources from nearby point-like sources, a well controlled flux density calibration with 13 stars, and an objective quality cut based upon seeing and Strehl ratio values.

\subsection{The dataset} \label{base}

Our analysis is based on ESO archive data. All observations have been conducted with the NIR adaptive optics (AO) instrument NAOS/CONICA at the Very Large Telescope (VLT) in Chile (\citealt{2003SPIE.4841..944L,2003SPIE.4839..140R}). We included all available Ks-band frames of the central cluster of the Galactic Center (GC) from early 2003 to mid-2010. For all observations, the NIR wavefront sensor of the NAOS adaptive optics system was used to lock on the NIR bright supergiant IRS 7 (variable, $ m_{\rm{Ks}} $ = 6.5 in the 1990s, $ m_{\rm{Ks}} $ = 7.4 in 2006 and $ m_{\rm{Ks}} $ = 7.7 in 2011, $5.6^{\prime\prime}$ north of Sgr~A*\footnote{see \cite{2010A&A...511A..18S}, \cite{1999ApJ...523..248O}}). Two different cameras, S13 and S27, with $13^{\prime\prime}$ and $27^{\prime\prime}$ field of view\footnote{and accordingly with different pixel sampling of 13 mas/pix and 27 mas/pix} respectively, and a polarimetric mode with inserted Wollaston prism and mask have been used. The small field of view of the polarimetric mode restricted the set of calibrators to the innermost arcsecond around Sgr~A*.

We concentrated on datasets with a length of more than 40 minutes (shorter datasets are often severely affected by bad weather conditions in which case the observer decided to change to another wavelength or target). Problematic frames with obviously bad AO-correction (most of the stars not visible at all) or frames, which do not show Sgr~A* or a sufficient number of calibrators, have not been included for the photometric analysis. Ultimately we investigated 12855 frames photometrically. Table \ref{datasets} shows a list of all datasets that are part of this analysis.

\subsection{Data reduction and flux density calibration} 
\label{red}

We performed every reduction step for every frame uniformly: The reduction included basic steps like sky subtraction, flat fielding, and correction for bad pixels. For total intensity data, we used sky flat fields (where available), for polarimetric data a lamp flat field (\citealt{2011A&A...525A.130W}). Most of the data was observed using a jitter routine with random offsets to prevent systematic influences on the measurements by detector artifacts. These offsets need to be detected and corrected for to guarantee stable aperture photometry at a constant position. This was achieved with a cross correlation algorithm for sub-pixel accuracy alignment (ESO Eclipse Jitter, \citealt{1999ASPC..172..333D}). For each aligned frame, we determined an estimate of the point spread function (PSF) with the IDL routine Starfinder (\citealt{2000SPIE.4007..879D}) using separated stars like S30 or S65 in the $ 2^{\prime\prime} $ surrounding of Sgr~A*. The PSF-fitting algorithm of Starfinder provided an estimate of the extended background in each single frame and a list of detected stars with position and relative flux density. We decided not to use the values resulting from Starfinder photometry (for a detailed reasoning see below). Instead we used the Lucy-Richardson (LR) deconvolution algorithm to separate neighboring point-like sources. In the case of polarization data, we aligned all (four to six) polarization channels of the individual observation night with the cross correlation algorithm. Data observed with the different pixel scale of the S27 camera ($0.02715^{\prime\prime}$) were resampled to $0.01327^{\prime\prime}$ (S13). Finally a beam restoration was carried out with a Gaussian beam of a FWHM corresponding to the diffraction limited resolution at $ 2.2 \mu $m ($ \sim 60 $ mas).

\begin{figure}[ht]
   \centering
   \includegraphics[width=7.5cm]{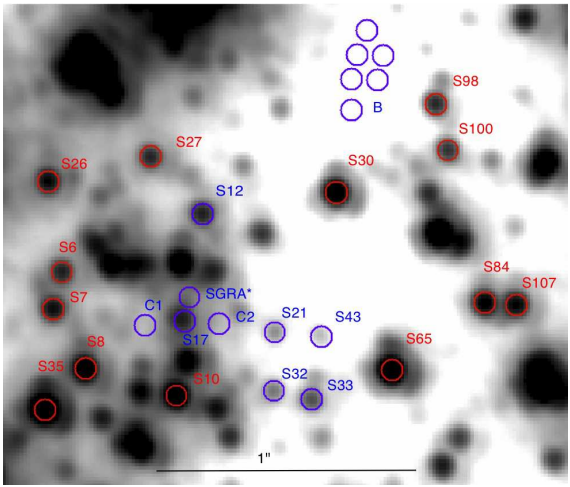}
      \caption[Map of calibrators.]{Ks-band image from 2004 September 30. The red circles mark the constant stars (see variability study in \citealt{2007ApJ...659.1241R}) which have been used as calibrators, the blue the position of photometric measurements of Sgr~A*, comparison stars and comparison apertures for background estimation. Source identifications from \cite{2009ApJ...692.1075G}.}
         \label{calib}
\end{figure}

After the described preparative steps we conducted aperture photometry at the position of  13 constant calibrators (\citealt{2007ApJ...659.1241R}), of 6 comparison stars, at the position of Sgr~A*, and at 8 positions where no star was detected (B- and C-apertures), to measure the background flux (see Figure~\ref{calib}). The background was estimated at the locations of lowest background (6 apertures) and close to Sgr~A* (2 apertures) where no obvious point-like source is visible. We applied a circular masking of radius $ 0.04^{\prime\prime} $ at all the measurement positions. For a small number of observations, according to the available field of view, we accepted a smaller set of calibrators (at least seven). 

The positions of the apertures in each night have been defined as consistently as possible: For Sgr~A* with the help of its brighter states, for the stars with the help of mosaics (averages over the single frames of one night in order to increase the signal to noise and to also estimate the centroid of the fainter comparison stars), carefully following their proper motions. For the background apertures and the aperture of Sgr~A* when it was faint we conducted triangulation relative to nearby stars. One set of positions then was used for all frames of the corresponding night. For some polarimetric  observations NACO was rotated. In these cases, we determined a rotation matrix for the position coordinates, making them comparable to the closest unrotated observations. 

\begin{table}[h!]
\begin{center}
\caption{List of calibrators\label{caliblist}}
\begin{tabular}{ccc}
\tableline
\tableline
&&\\
star & $ m_{\rm{Ks}} $ & flux density \\
&& [mJy] \\
&&\\
\tableline
&&\\
S26 & 14.94 & 6.79 \\
S27 & 15.41 & 4.41 \\
S6 & 15.35 & 4.66 \\
S7 & 14.92 & 6.92 \\
S8 & 14.21 & 13.31 \\
S35 & 13.20 & 33.74 \\
S10 & 13.95 & 16.91 \\
S65 & 13.58 & 23.78 \\
S30 & 14.12 & 14.46 \\
S98 & 15.27 & 5.01 \\
S100 & 15.29 & 4.92 \\
S84 & 14.66 & 8.79 \\
S107 & 14.82 & 7.59 \\
&&\\
\tableline
\end{tabular}
\tablecomments{The flux density for each star was calculated correcting for extinction with $ m_{\rm{ext}} = 2.46 $.}
\end{center}
\end{table}

For each aperture, we summed up its total content in analog-to-digital units (ADU). For the polarimetric data, the obtained ADU values of orthogonal channels were added. We subtracted the average background value (B-apertures) in ADU from the calibrator values and conducted a flux density calibration using the photometric values for the calibrators in Tab. \ref{caliblist} (\citealt{2010A&A...511A..18S}). Because of the high proper motions of the stars within this field, the state of confusion of the calibrators changes from epoch to epoch. We applied the following algorithm to reduce the epoch-dependent systematic error of the calibration: 

First we calculated for each calibrator $ k $ the quantity:

\begin{equation}\label{fk}
\frac{f_{k}}{\rm{ADU}} = \frac{c_{k}}{\rm{ADU}} \cdot 10^{0.4 \cdot m_{\rm{ref}}}~~,
\end{equation}    
with $ c_{k} $ the background subtracted ADU values for the k-th calibrator, and $ m_{\rm{ref}} $ its reference magnitude.
We sorted these values, rejected the three largest and the three smallest values (for the data sets with less calibrators we accordingly reject a smaller number), and took the arithmetic average $ f_{\rm{0}} $ over the remaining values. According to \cite{2000asqu.book..143T} we then obtained the magnitude $ m_{\rm{A}} $  and flux density $ F_{\rm{A}} $ for each aperture~A  by

\begin{eqnarray}\label{f0}
 m_{\rm{A}} & = & -2.5 \cdot \log\left(\frac{c_{k}}{f_{\rm{0}}}\right) \nonumber ~~,\\
 F_{\rm{A}}  & = & 667 \cdot 10^{3} \cdot 10^{-0.4 \cdot ( m_{\rm{A}}-m_{\rm{ext}})}~~,
\end{eqnarray}    
with $ m_{\rm{ext}}=2.46 $ the K-band extinction as determined in \cite{2010A&A...511A..18S}. This procedure ensures a best possible constance of the flux density calibration under the changing conditions of each dataset.

As a last step, we collected parameters for each frame that provide information on the data and calibration quality, which allowed us to reject data points based on objective criteria. These parameters are: Julian date, integration time (NDITxDIT), rotator position angle (orientation of NACO), airmass, FWHM of the active optics guide star PSF, coherence time of the atmosphere, camera, all obtained from the header of the fits data, and the number of stars detected by Starfinder, the Strehl ratio calculated from the extracted PSF using the ESO Eclipse routine STREHL, the RMS of the values $ f_{k} $, and, as the most important quality check, the average normalized flux of the calibrators. This last quantity is obtained for each frame by dividing the measured flux density of the individual by its reference value and averaging over all available (i.e. not rejected) calibrators.  

We emphasize that both methods - PSF-fitting and Lucy-Richardson aperture photometry - are in general equivalent for estimating the flux density of a point source (\citealt{2008ApJ...688L..17M}). However, in the case of the presence of extended flux underlying a dim and confused point source observed under varying correction performance of the AO-system, it is more difficult to control
\begin{itemize} 
\item how Starfinder divides the given flux at a position into background and point source flux, 
\item how the possible over-estimation of the flux due to noise-peaks influences the statistics of low flux densities (if the center of the fit is not forced to a given position), and 
\item what non-detections due to the quality thresholds set in Starfinder mean for the overall statistics.
\end{itemize}

Also LR-deconvolution has drawbacks, especially in handling extended flux that is added to existing point sources or gathered into artificial sources by the algorithm. We can account for this effect with a suitable big aperture and by monitoring apertures at positions without obvious point sources. Thus, the statistics of the interplay between the background (coming from unresolved sources, truly extended emission and PSF contributions from the surrounding point sources), the AO and point-like flux at a given position is directly propagated to the statistics of the measured values, which is crucial for understanding the instrumental influence on our flux density statistics and our statistics do not suffer from non-detections that might introduce a selection effect. We will come back to the measurement statistics in section \ref{stat}.

\subsection{Light curves of Sgr~A*} \label{lc}

\begin{figure*}
   \centering
   \includegraphics[width=15.5cm]{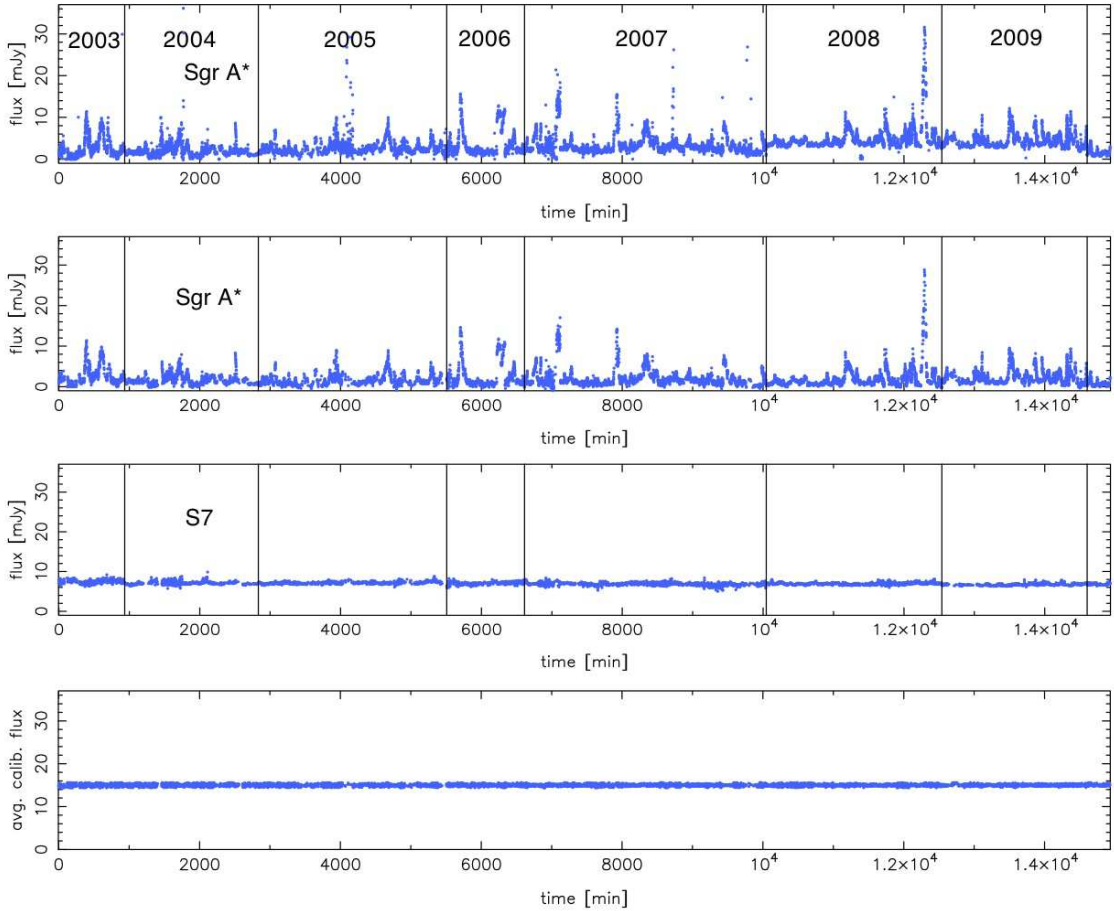}
      \caption[The concatenated light curve of Sgr~A*.]{The concatenated light curve of Sgr~A* and S7 with time gaps longer than 30 minutes reduced to 1.2 min. The top panel shows the result of aperture photometry before the quality cut, the next panel the same data after quality cut and removal of offsets. The third shows a light curve of S 7, a nearby star that has been used as a calibrator. The time gaps are the result of the rejection procedure described in section~\ref{red}. The lower panel shows the average ratio between the measured flux of each calibrator and its reference value (Tab.~\ref{caliblist}), scaled  with a factor of 15. Its noise corresponds to the error a hypothetical noise-free aperture with a flux density of 15 mJy would have only due to the uncertainties of the calibrators.}
         \label{conclc}
\end{figure*}

As a result of the reduction procedure described in section \ref{red} we obtained the data shown in the upper panel of Fig.~\ref{conclc}. For convenience and following the visualization used in \cite{2011ApJ...728...37D}, we show a concatenated light curve with all time gaps longer than 30 minutes reduced to the average sampling of the individual datasets (1.2 min). This visualization shows the data of all nights as a pseudo-continuous light curve allowing for a comparison of the variability and the confusion in each epoch. A visualization of the true cadence is presented in Fig.~\ref{realtime}. The timing analysis in section \ref{tsa} is based on the true cadence, and not on the concatenated light curve.

These 12855 data points still include points of bad observation conditions and insufficient calibration reliability. We used an objective frame rejection algorithm by incorporating information on seeing, Strehl ratio , fraction of stars detected by Starfinder\footnote{This is a method to reject frames of lower quality with respect to the majority of the individual observation night. For each sub-frame of the night (i.e. the part of the jittered frames that is common to all) we count the stars detected by Starfinder. We then calculate the average number of detections over the night and the standard deviation and reject all frames with a negative deviation from the average of more than $ 1.7 $ times the standard deviation.}, the standard deviation of the $ f_{\rm{0}} $-values obtained from the individual calibrators in each frame , and the normalized average calibrator flux density as described in section~\ref{red}. Only frames with seeing $ < 2^{\prime\prime} $ , a Strehl-ratio $ > 6 \% $, a $ f_{\rm{0}} $-standard deviation $ < 16 \% $ of the  average $ f_{\rm{0}} $-value, and a normalized  average calibrator flux density between $ > 0.96$ and $< 1.04 $ have been accepted.

%As a result of the reduction procedure described in section \ref{red} we obtained the data shown in the upper panel of Fig~\ref{conclc}. For convenience and following the visualization used in \cite{2011ApJ...728...37D} we show a concatenated light curve for which we removed all time gaps longer than 30 minutes. These 12855 data points still include points of bad observation conditions and insufficient calibration reliability. We used an objective frame rejection algorithm by incorporating information on seeing (only points with seeing $ < 2^{\prime\prime} $ accepted), Strehl ratio (only points $ > 6 \% $), fraction of stars detected by Starfinder\footnote{This is a method to reject frames of worse quality than the majority of the observation night. For each sub-frame of the night (exhibiting the part of the jittered frames that is common to all) we count the stars detected by Starfinder. We calculate then the average number of detections over these sub-frames and the standard deviation and reject all points with a negative deviation from the average of more than $ 1.7 $ times the standard deviation.}, the standard deviation of the $ f_{\rm{0}} $-values for each frame normalized by the average $ f_{\rm{0}} $-values (only points $ < 16 \% $), and the normalized average calibrator flux density as described in section~\ref{red} (only points $ > 0.96$ and $< 1.04 $).  

The top panel of Figure~\ref{conclc} additionally shows a long-time trend of the data from epoch to epoch. As far as these ``offsets" are concerned we agree with the conclusions of \cite{2011ApJ...728...37D} that confusion with stellar sources is responsible for this long-time trend in a first approximation. In order to make the different years comparable we subtracted the 2.5 percentile value of the flux in each epoch, resulting in 0 mJy for 2003, 0.3 mJy for 2004, 1.0 mJy for 2005, 1.0 mJy for 2006, 1.3 mJy for 2007, 2.8 mJy for 2008, 2.1 mJy for 2009, and 0.7 mJy for 2010. In 2003 and 2004 Sgr~A*  could be  sufficiently separated from the near star S2\footnote{which is the reason why \cite{2011ApJ...728...37D} did not include the 2003 data and which is clearly visible in the 2004 epoch of their uncorrected light curve} by the deconvolution reduction step.

The second panel of Figure~\ref{conclc} shows the data after quality cut and subtraction of the faint stellar contribution. For comparison and as an indicator of the calibration stability we show additionally the light curve of one of the calibrators, S7, in the third panel, and, in the lowest panel,  the average calibrator flux density scaled to 15 mJy. On average, the calibration is very stable and the data points after quality cut exhibit a relative standard deviations of the average calibrator flux of $ 1.4 \% $.

For a more detailed inspection we present 112 data blocks (defined by continuity without gaps of more than 30 minutes) in Appendix~\ref{detlc}.

\section{Statistical analysis of the flux density distribution} 
\label{stat}
In this section we investigate the properties of the flux density statistics of the variability of Sgr~A*. This analysis gives information on both the intrinsic flux density distribution of the variability as well as the instrumental effects within our measurements. The differentiation between instrumental features within the distribution and the intrinsic component turns out to be crucial in the context of the question whether the intrinsic flux density distribution provides evidence for two physical mechanisms at work. Additionally, this analysis allows us to develop a full statistical model of the variability of Sgr~A* in the next section. 

\subsection{Optimal data visualization} \label{rep}

\begin{figure} [ht]
   \centering
   \includegraphics[width=7.5cm]{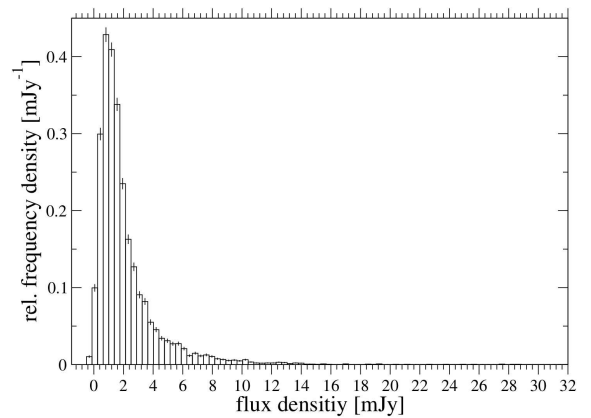}
      \caption[Flux density histogram of Sgr~A*.]{Flux density histogram of Sgr~A*, based on the data shown in Figure~\ref{conclc}}
         \label{simphis}
\end{figure}

The first step in the analysis of the flux density distribution of Sgr~A* is a proper graphical representation of the data in form of a histogram. A representation of our data in a simple flux density histogram, normalized by total number of points and bin size, is shown in Figure~\ref{simphis}.

To investigate the high flux density tail of this distribution a logarithmic histogram with an equally spaced logarithmic binning is best suited. The number of bins for a given data range is a crucial parameter for the evaluation of the structure of the sample distribution. Following the study of \cite{2006physics...5197K} we first determine the best bin size. As the author points out, the idea is to choose a number of bins sufficiently large to capture the major features in the data while ignoring fine details due to random sampling fluctuations. By considering the histogram as a piecewise-constant model of the probability density function from which $ n $ data points $ x_{i} $ were sampled the author derives an expression for the relative logarithmic posterior probability (RLP) for each bin number:

\begin{eqnarray}
\rm{RLP} & = & n \log N + N \log \Gamma\left( \frac{N}{2}\right) \nonumber \\
& + & \log \Gamma\left( \frac{1}{2}\right) - \log \Gamma\left( n + \frac{N}{2}\right) \nonumber \\
& + & \sum_{\lambda=1}^{N} \log \Gamma\left( n_{\lambda} + \frac{1}{2}\right)~~.
\end{eqnarray}
with $ N $ the number of bins and $ n_{\lambda} $ the value of the $ \lambda $th bin. To find the best number of bins $ M $ the posterior probability has to be maximized:

\begin{equation}
M = \arg\max_{N} \lbrace \rm{RLP} \rbrace ~~.
\end{equation}

The best estimator for the bin value $ \mu_{\lambda} $ and its variance $ \sigma^{2}_{\lambda} $ given the bin values $ n_{\lambda} $ is deduced to be:

\begin{equation}
\label{bh}
\mu_{\lambda} = \left( \frac{M}{v}\right) \left(  \frac{n_{\lambda}+\frac{1}{2}}{n + \frac{M}{2}}\right)
\end{equation}
 and 
\begin{equation}
\label{bherr}
\sigma^{2}_{\lambda} = \left( \frac{M}{v}\right)^{2} \left[\frac{  \left(n_{\lambda}+\frac{1}{2}\right) \left(n - n_{\lambda} + \frac{M-1}{2} \right)  }{\left(n + \frac{M}{2} + 1 \right)  \left( n + \frac{M}{2}\right)^{2}} \right]~~.
\end{equation}
with $ v $ the interval between the maximum and the minimum measurement value.  

\cite{2006physics...5197K} demonstrates in his study that these results outperform several other rules for choosing bin sizes, e.g. ``Scott's rule" or ``Stone's rule".

%By considering the histogram as a piecewise-constant model of the probability density function from which $ n $ data points were sampled the author derives an expression for the relative logarithmic posterior probability (RLP) for each bin number:
%
%\begin{eqnarray}
%\rm{RLP} & = & n \log M + M \log \Gamma\left( \frac{M}{2}\right) \nonumber \\
%& + & \log \Gamma\left( \frac{1}{2}\right) - \log \Gamma\left( n + \frac{M}{2}\right) \nonumber \\
%& + & \sum_{k=1}^{M} \log \Gamma\left( x_{k} + \frac{1}{2}\right)~~.
%\end{eqnarray}
%
%To find the best number of bins $ M $ the posterior probability has to be maximized:
%
%\begin{equation}
%M = \arg\max_{M} \lbrace \rm{RLP} \rbrace ~~.
%\end{equation}
%
%The best estimator for the bin height and its error given the data points $ x_{k} $ is deduced to be:
%
%\begin{equation}
%\label{bh}
%\mu_{k} = \left( \frac{M}{v}\right) \left(  \frac{x_{k}+\frac{1}{2}}{n + \frac{M}{2}}\right)
%\end{equation}
% and 
%\begin{equation}
%\label{bherr}
%\sigma^{2}_{k} = \left( \frac{M}{v}\right)^{2} \left[\frac{  \left(x_{k}+\frac{1}{2}\right) \left(n - x_{k} + \frac{M-1}{2} \right)  }{\left(n + \frac{M}{2} + 1 \right)  \left( n + \frac{M}{2}\right)^{2}} \right]~~,
%\end{equation}
%with $ v $ the interval between the maximum and the minimum measurement value.  
%
%\cite{2006physics...5197K} demonstrates in his study that these results outperform several other rules for choosing bin sizes, e.g. ``Scott's rule" or ``Stone's rule".

\begin{figure} [h]
   \centering
   \includegraphics[width=7.5cm]{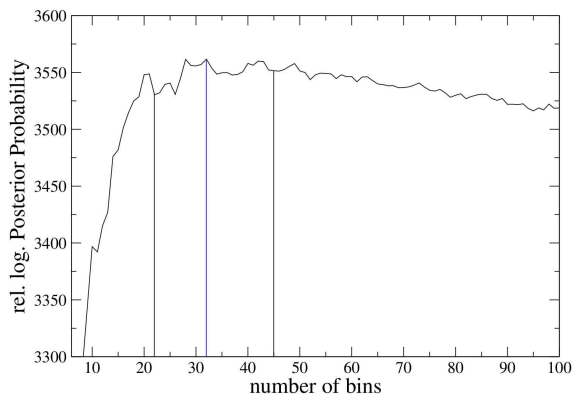}
      \caption[Log posterior probability as a function of the number of bins.]{Optimal data based binning. We show the log posterior probability as a function of the number of bins. The maximum for the logarithmic flux density of Sgr~A* is reached for 32 bins (blue line). The black lines represent bin numbers used for the histograms in Appendix~\ref{apstat} which clearly show that our results do not strongly depend on the binning.}
         \label{bestbin}
\end{figure}

\begin{figure} [h]
   \centering
   \includegraphics[width=7.5cm]{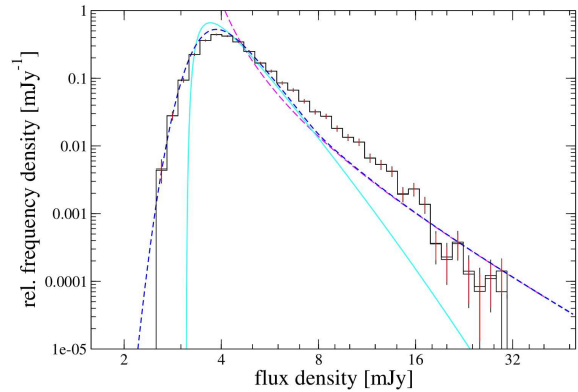}
       \caption[Best piecewise-constant probability density model for the flux densities of Sgr~A*.]{Best piecewise-constant probability density model for the flux densities of Sgr~A*. The red error bars are the uncertainty of the bin height for the full amount of 10\,639 data points. The second bin height visible in some cases and the black error bars belong to the average histograms of 1000 datasets with 6774 data points (\citealt{2011ApJ...728...37D}), generated by randomly removing points from our full dataset. The over-plotted cyan and magenta dashed lines show the log-normal distribution and the power-law distribution found in the analysis by \cite{2011ApJ...728...37D}, the blue the combined distribution of those components, convolved with a Gaussian with a flux density-dependent $\sigma$ (compare Eqn.~\ref{dodds1} to \ref{dodds2}).}      
%      \caption[Best piecewise-constant probability density model for the flux densities of Sgr~A*.]{Best piecewise-constant probability density model for the flux densities of Sgr~A*. The red error bars are the uncertainty of the bin height for the full amount of 10\,639 data points. The second bin height visible in some cases and the black error bars belong to the average histograms of 1000 datasets with 6774 data points, generated by randomly removing points from our full dataset. The over-plotted orange and magenta dashed lines show the log-normal distribution and the power-law distribution found in the analysis by \cite{2011ApJ...728...37D}, the blue the combined distribution of those components, convolved with a Gaussian with a flux density-dependent $\sigma$ (compare Equations~(\ref{dodds1}) to (\ref{dodds2})).}
         \label{bestbinmod}
\end{figure}

We applied the described binning method to our data of Sgr~A*. To make the flux density distribution comparable to the results by \cite{2011ApJ...728...37D}, which include the flux density of the star S17 due to a double aperture method, we added 3 mJy to the flux density of Sgr~A*. We find a best bin number of $ M = 32 $. The dependence of the log posterior on the bin number is shown in Figure~\ref{bestbin}. The best piece-wise constant model (i.e., an estimator of the best histogram representation) describing our sample is shown in Figure~\ref{bestbinmod}. The histograms in this work have been created using this method.

Additionally to the best histogram model obtained from Equation~(\ref{bh}) and Equation~(\ref{bherr}) we over-plotted a graph of the model\footnote{slightly shifted on the $x$-axis to account for the proper offset due to the double aperture method used by \cite{2011ApJ...728...37D}, which before we included only roughly by adding 3 mJy, and to make it fit best the extremes of our histogram} proposed by \cite{2011ApJ...728...37D} . It is obvious that our sample is more populated in the middle flux density range between 7 mJy and 15 mJy and shows a linear behavior between 4 mJy and 17 mJy, not showing any break or change of slope. That the shape of our histogram is not sensitive to the binning is shown in Appendix~\ref{apstat}, Figure~\ref{nonoptbin}, where we present histograms with 22 bins for the range of the observed flux density values (resulting in a comparable bin width as used by \citealt{2011ApJ...728...37D}) and 45 bins\footnote{The latter we show in an unweighted and in an integration time-weighted version.}, both reproducing the linear trend. Rather than being a matter of representation, this difference is related to the different sample selection (10639 data points in this work in comparison to 6774 points in the case of \citealt{2011ApJ...728...37D}). To better understand the character of the selected subsample in \cite{2011ApJ...728...37D} (their quality cut is based on the visual impression of the individual frame), we selected randomly 6774 points from our sample (by nonparametric bootstrapping, i.e. sampling with replacement) and generated 1000 surrogate datasets in this way, binned each dataset in a histogram with the same bin size as for our total sample, averaged the bin values over all surrogate sets, obtained an error from the standard deviation, and plotted the result as the second bin height (now with black error bars) in Figure~\ref{bestbinmod}. One can clearly see that for most of the bins there is barely any difference, showing that a random influence cannot be responsible for the difference of both distributions. This test shows the robustness of the linear behavior of the histogram also in the case of smaller datasets under random selection. Futhermore, in our dataset we do not see the observation conditions to be correlated with the flux density states of Sgr~A* (compare Figure~\ref{conclc} and Figure~\ref{quality} in Appendix~\ref{apqual}). In general also data worse than average should be represented in the uncertainties, and not simply eliminated, because otherwise errors might be underestimated due to an introduced bias, and we have to conclude here that probably the subsample used by \cite{2011ApJ...728...37D} shows a severe selection effects. 

The linear behavior in the log-log diagram points towards a power-law distribution $ p \propto x^{-\alpha} $ as a possible description for all flux densities higher than $ \sim4 $mJy. The statistical significance of this visual impression is analyzed in the next section. We mention here that a power-law distribution is only showing a linear behavior in a log-log diagram if it is of the form:

\begin{equation}
p \propto (x-x_{\rm{0}})^{-\alpha}, ~~x_{\rm{0}} = 0~~.
\end{equation}
Otherwise the logarithm $ \log(p) $ of the probability density is only linear as a function of $ \log(x) $ for large values of $ x $:

\begin{eqnarray}
\log(p) & = & -\alpha \cdot \log(x-x_{\rm{0}}) + \rm{const} \\
& = & -\alpha \cdot \log(x) - \alpha \cdot \log\left(1-\frac{x_{\rm{0}}}{x}\right) +  \rm{const}~. \nonumber
\end{eqnarray}

This is the main reason why the distribution claimed in \cite{2011ApJ...728...37D} shows a break: In this case the high flux density tail is described by a power-law with $ x_{0} \approx 0.8 $~mJy, and this power-law does not show a linear behavior in the log-log diagram if plotted versus the sum of intrinsic flux density, background and flux density of S2. It starts to deviate from a linear behavior (it, so to say, ``breaks") close to the transition value $ F_t $ of the total distribution (see Fig.~\ref{bestbinmod}, blue and magenta dashed line). Thus, even if we accept the selection of data points by \cite{2011ApJ...728...37D}, the visual impression of the necessity of introducing a break in the distribution is a feature of the data visualization, and the power-law actually is suited also in their case to describe the data down to a flux density value of about 4 mJy quite well. This means, even under the assumption that the data selection of \cite{2011ApJ...728...37D} is valid, the discussion of a double state model for Sgr~A* and its significance is very much depending on the evidence of a log-normal model describing the low flux density part of the histogram. We come back to this point in the next section.

The reason why we see the distribution behaving linearly inour visualization is a coincidental equality of $ -x_{0} = 3 $~mJy in our power-law model (see next section) and the flux density of S17 ($ \approx 3 $~mJy) which we added to the flux density of Sgr~A* to make it comparable with the distribution proposed by \cite{2011ApJ...728...37D}.

%This is the main reason why the distribution claimed by \cite{2011ApJ...728...37D} shows a break: In this case the high flux density tail is described by a power-law with $ x_{\rm{0}} \approx 0.8 $~mJy, and this power-law ``breaks" at the same flux density value as the total distribution (see Figure~\ref{bestbinmod}, magenta and blue dashed line). Thus, even if we accept the selection of data points by \cite{2011ApJ...728...37D}, the visual impression of the necessity of introducing a distribution break is a feature of the data visualization and the power-law actually is suited also in their case to describe quite well the data down to a flux density value of about 4 mJy. This means, even under the assumption that the data selection of \cite{2011ApJ...728...37D} is valid, the discussion of a double state model for Sgr~A* and its significance is very much depending on the plausibility of the assumption of a log-normal distribution for low intrinsic flux densities. We come back to this point in the next section.
%
%The reason for the fact that we see the distribution falling on a straight line is a coincidental equality of $ -x_{\rm{0}} = 3 $ for our data (see next section) and the flux density of S17 ($ \approx 3 $) which we added to the flux density of Sgr~A* to make it comparable with the distribution proposed by \cite{2011ApJ...728...37D}.

\subsection{Power-law representation of the intrinsic flux density distribution} 
\label{pl}

\begin{figure}[h]
   \centering
   \includegraphics[width=7.5cm]{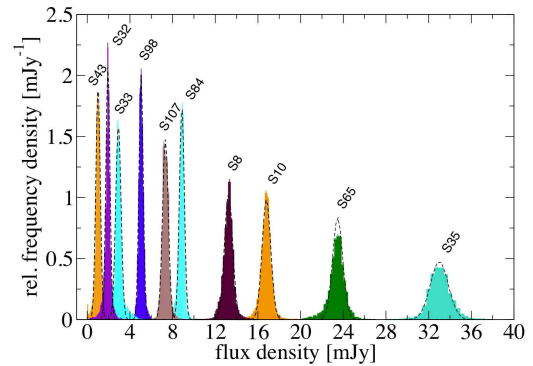}
      \caption[The normalized flux density histograms of ten calibration stars.]{The normalized flux density histograms of ten calibration stars. The dashed curves are Gaussian fits to the flanks of the distribution, suppressing the broader tails of the distributions. This guarantees a proper measurement of the FWHM of the distributions.}
         \label{starhis}
\end{figure}

\begin{figure}[h]
   \centering
   \includegraphics[width=7.5cm]{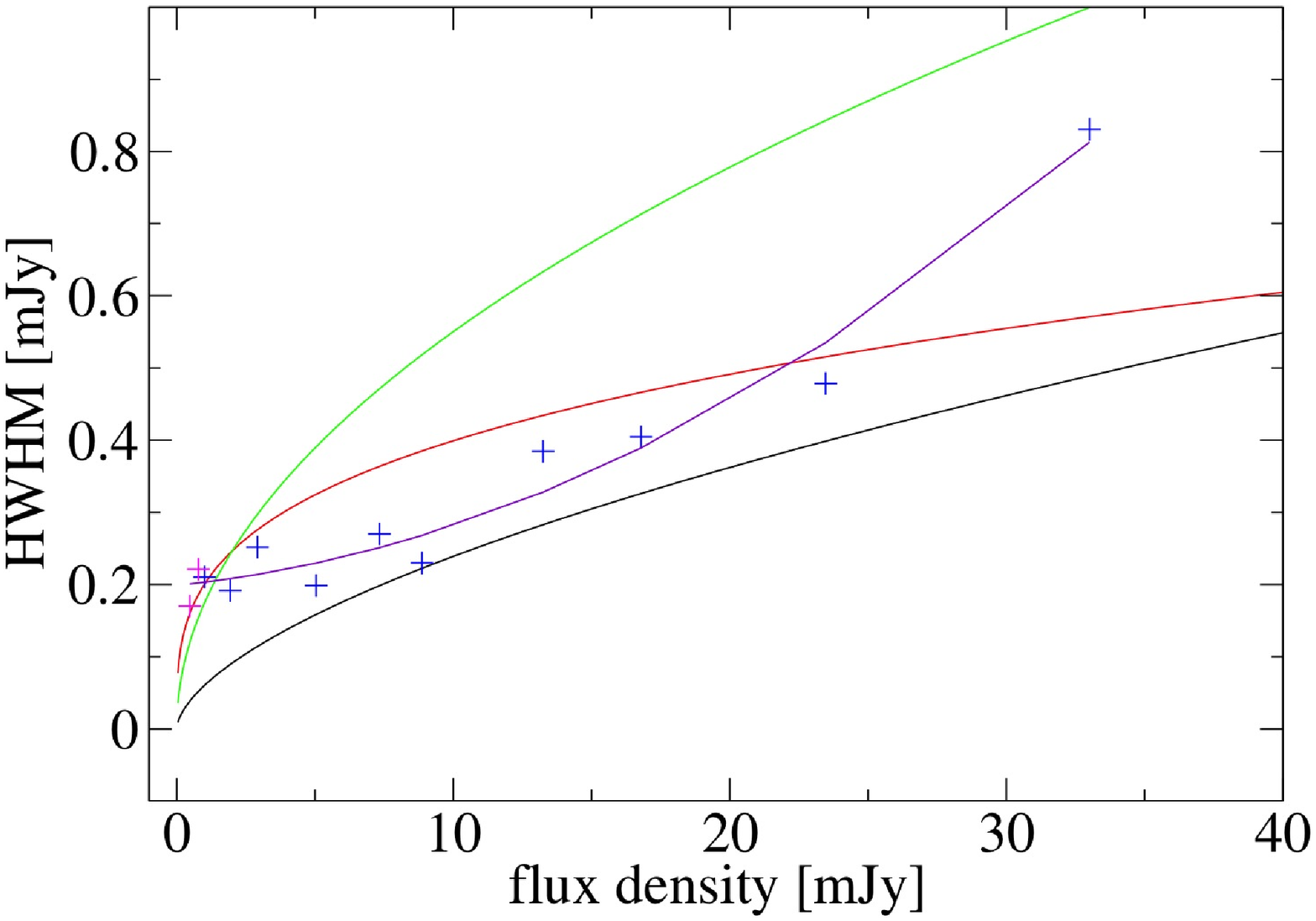}
      \caption[The uncertainty of the photometry as a function of flux density.]{The measurement error as a function of flux density. The purple line is a quadratic fit to the measured $ \sigma$-values of the calibration stars shown in Figure~\ref{starhis} (blue crosses). The red line is the power-law dependency found by \cite{2009ApJ...691.1021D} for their data, the green line and the black line the dependency found by \cite{2011ApJ...728...37D} for their aperture photometry and their PSF-fitting photometry, respectively. The two magenta crosses represent the measured $ \sigma$-values at the position of the C-apertures, compare Figure~\ref{calib}.}
         \label{errorflux}
\end{figure}

We begin with a description of the instrumental effects and uncertainties of our photometry. Figure~\ref{starhis} shows the flux density histograms of 10 stars (calibrators and comparison stars). As expected, the width of the distribution is decreasing with decreasing flux density of the source. To estimate the FWHM of the scatter of our photometry at a given flux density, we fit Gaussian distributions to the histograms. The $ \sigma $ values of these fits as a function of the mean flux density is shown in Figure~\ref{errorflux}. For our photometry we clearly find smaller uncertainties than \cite{2011ApJ...728...37D} for their aperture photometry, very similar to their Starfinder photometry, or the photometry of \cite{2009ApJ...691.1021D}. In contrast to \cite{2011ApJ...728...37D}, the functional dependency shows a clear flattening towards small flux densities\footnote{This of course corresponds to a hyperbolic increase of the relative error toward small flux densities.} and we can not confirm the dominance of photon noise (\citealt{2011ApJ...728...37D}). We find a parabola to be a suited (phenomenological) description up to flux densities of 32 mJy. Actually the uncertainties in the flux density range between 0 and 10 mJy are more or less constant. This corresponds well to the visual impression that the variable AO-correction and its influence on the background contribution due to PSF halos of bright sources (halo noise, \citealt{2010MNRAS.401.1177F}) as well as differential tip/tilt jitter and their interplay with the deconvolution are the dominant reasons for the uncertainties.

\begin{figure}[h]
   \centering
   \includegraphics[width=7.5cm]{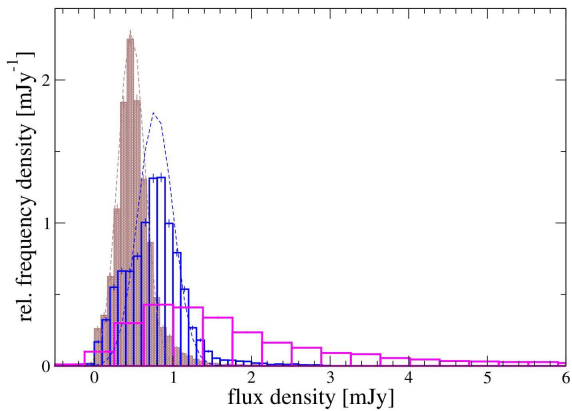}
      \caption[Flux density histograms of two background (C-) apertures.]{The flux density histograms of the two background (C-) apertures (blue, brown) close to Sgr~A* in comparison to the flux distribution of Sgr~A* (magenta). The dashed curves are Gaussian fits to the flankes of the distributions.}
         \label{backhis}
\end{figure}

The role of halo noise becomes evident when looking at the control apertures close to Sgr~A* (Figure~\ref{calib}, C-apertures). Their average flux density is clearly not zero, and the flux density values are scattered around the mean with a FWHM comparable to the width of all flux density distributions of the stars fainter than 10 mJy. The aperture west of Sgr~A* shows a varying contribution of faint confusion on the level of a few tenth of a mJy (Figure~\ref{backhis}, blue histogram). The average flux density of these ``empty" apertures is about 0.6 mJy.

In the following we investigate if a power-law distribution  indeed is suited to describe our sample. We follow the strategy described in \cite{2007arXiv0706.1062C} for identifying power-law distributions and determining their parameters. The authors of this study point out that a least square regression to a histogram in log-log representation can generate significant systematic errors, mainly due to the non-Gaussianity of the variation of the logarithmic bin height. Furthermore, binning the data in a histogram introduces further parameters corrupting standard goodness-of-fit estimators. The procedure described in the following overcomes this problems.

The probability density of a power-law distribution is defined as:

\begin{equation}
\label{powlawdef}
p(x) = \left \{ \begin{array}{r@{\quad:\quad}l}
0 & x \leq x_{\rm{min}} + x_{0} \\
\frac{\alpha-1}{x_{\rm{min}}} \cdot \left(\frac{x-x_{\rm{0}}}{x_{\rm{min}}} \right)^{-\alpha} & x > x_{\rm{min}} + x_{0}~~,
\end{array} \right.
\end{equation}
with $ x_{\rm{min}} = x_{\rm{min,intr}} - x_{0}$ and  $ x_{\rm{min,intr}} $ the lowest value to which the data is power-law distributed, making the power-law normalizable with normalization factor $ (\alpha-1) \cdot x_{\rm{min}}^{\alpha-1} $. The corresponding cumulative distribution is given by:

\begin{equation}
P(x) = \left \{ \begin{array}{r@{\quad:\quad}l}
1 & x \leq x_{\rm{min}}  + x_{0}\\
\left(\frac{x-x_{\rm{0}}}{x_{\rm{min}}} \right)^{1-\alpha} & x > x_{\rm{min}} + x_{0} ~~.
\end{array} \right.
\end{equation}

The steps for analyzing power-law distributed data are as follows (\citealt{2007arXiv0706.1062C}): 

\begin{itemize}
\item Find estimators for $ x_{\rm{min}} $ and $ x_{0} $ and the maximum likelihood estimator for $ \alpha $. The maximum likelihood estimator for $ \alpha $ given any value for $ x_{\rm{min}} $ and $ x_{0} $ is calculated using the equations

\begin{equation}
\alpha_{\rm{est}} = 1 + n_{\rm{tail}}\left[ \sum^{n_{\rm{tail}}}_{i=1}\ln\frac{x_{i}-x_{0}}{x_{\rm{min}}}\right]^{-1},
\label{alphafit}
\end{equation}

\begin{equation}
\sigma_{\rm{est}} = \frac{\alpha_{\rm{est}}-1}{\sqrt{n_{\rm{tail}}}}~~,
\end{equation}

with $ x > x_{\rm{min}}+x_{0} $ and $ n_{\rm{tail}} $ the number of data points higher than $ x_{\rm{min}}+ x_{0}$.
Estimators for $ x_{\rm{min}} $ and $ x_{0} $ are obtained by choosing $ x_{\rm{min}} $ and $ x_{0} $ in a way that makes the probability density and the best-fit power-law model (i.e. the power-law with $ \alpha $ the maximum likelihood estimator) as similar as possible. The similarity is estimated by Kolmogorov-Smirnov (KS) statistics:

\begin{equation}
D = \max_{x\geq x_{\rm{min}}+x_{0}}\vert C(x)-P(x) \vert~~,
\end{equation}
with $ P(x) $ the cumulative distribution of the best-fit power-law model and $ C(x) $ the relative cumulative frequency of the empirical data sample.
The parameter D has to be minimized. The error on $ x_{\rm{min}} $ and $ x_{0} $ can be found by a nonparametric ``bootstrap" method, i.e. (given $ n $ empirical data points)  by drawing a new set of $ n $ data points uniformly at random from the empirical data and determining the standard deviation of $ x_{\rm{min}} $ and $ x_{0} $ for these surrogate samples. To replicate the act of drawing an independent, identically distributed sample from the population the surrogate data has to be drawn with replacement.

\item Test for plausibility by calculating the goodness-of-fit between the empirical data and the power-law. The goodness-of-fit parameter $ q $ is defined as the fraction of synthetic data drawn from the best-fit probability model that has a worse KS statistics than the empirical data sample. Here it is important to obtain the KS value $ D $ for each synthetic sample in the same way as for the empirical data, i.e. the estimators for $ x_{\rm{min}} $, $ x_{0} $ and $ \alpha $ have to be found for each synthetic sample individually and the KS values has to be calculated relative to the individual best-fit power-law. To ensure that the $ x_{\rm{min}} $ values are determined under the same conditions like for the empirical data sample we have to make sure that the synthetic sample follows the empirical data sample distribution for the values smaller than $ x_{\rm{min}}+ x_{0} $. This is realized by choosing with probability $ 1/n_{\rm{tail}} $ a random number from the best-fit empirical power-law and with probability $ 1-1/n_{\rm{tail}} $ a value from the empirical data below the $ x_{\rm{min}}+ x_{0} $ estimate (with $ n_{\rm{tail}} $ the number of data points higher than $ x_{\rm{min}}+ x_{0} $). A $ q $-value $ > 0.05 $, or more conservative $ > 0.1 $, makes the power-law model a plausible assumption.
\end{itemize}

With a simulation of 1000 surrogate samples (again obtained by nonparametric bootstrapping) for determining the errors of $ x_{\rm{min}} $ and $ x_{\rm{0}} $ and 6000 synthetic samples for testing for plausibility we obtained the values:

\begin{eqnarray}
x_{\rm{0}} & = & (-2.94 \pm 0.1) \; \rm{mJy} \nonumber \\ 
x_{\rm{min}} & = & (4.22 \pm 0.1) \; \rm{mJy} \nonumber \\ 
\alpha & = & (4.215 \pm 0.05) \nonumber \\ 
q & = & 0.2~~~~~~~~~~~~~~~~~~~~~.
\label{powerre}
\end{eqnarray}

\begin{figure} [h]
   \centering
   \includegraphics[width=7.5cm]{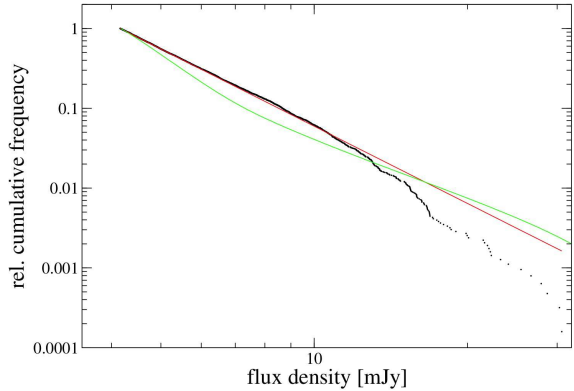}
      \caption[Estimation of the goodness parameter $p$ by a Kolmogrov statistic.]{Estimation of the goodness parameter $p$ by a Kolmogrov statistic. The value $p$ is defined by the maximum difference of the measured CDF (black) with respect to the CDF of the best fitting power-law (red). In green we show the CDF of the model proposed by \cite{2011ApJ...728...37D} restricted to flux density values higher than $ x_{\rm{min}} $.}
         \label{cdflog}
\end{figure}

\begin{figure}[ht]
   \centering
   \includegraphics[width=7.5cm]{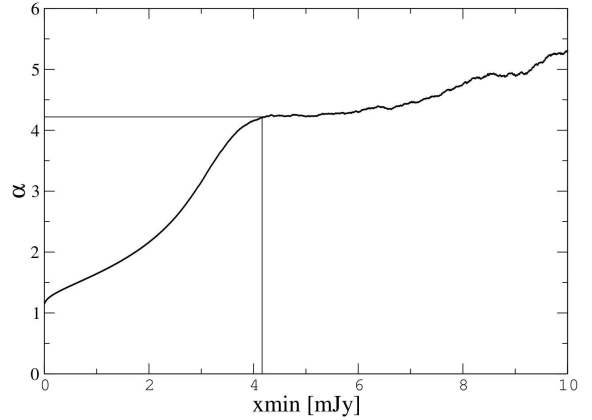}
      \caption[The scaling parameter $\alpha$ as a function of $x_{\rm{min}}$.]{The scaling parameter $\alpha$ as a function of the value for $x_{\rm{min}}$ (compare Figure~3.3 in \citealt{2007arXiv0706.1062C}).}
         \label{xmina}
\end{figure}

A goodness-of-fit parameter $ q = 0.2 $ means that in a fifth of the cases a sample drawn from a power-law with parameters as in Equation~(\ref{powerre}) will show deviations worse than our empirical sample, making the power-law description plausible for all flux densities higher than 4.2 mJy (note that the exact value of $x$-axis offset is $ x_{\rm{0}} = -2.9 $ mJy, close to the value found due to the linear appearance of the histogram in the log-log diagram). A diagram of the cumulative distribution function (CDF) of the best-fits power-law and the empirical relative cumulative frequency is shown in Figure~\ref{cdflog} . For a comparison we also over-plotted the CDF of the model proposed by \cite{2011ApJ...728...37D} (restricted to flux density values higher than $ x_{\rm{min}} $). Figure~\ref{xmina} shows the value of the power-law slope $ \alpha $ as a function of $ x_{\rm{min}} $. As expected, the best  $ x_{\rm{min}} $ is close to the point where $ \alpha $ starts to be constant for a range of $ x_{\rm{min}} $ values (compare Figure 3.3 in \citealt{2007arXiv0706.1062C}).

Equation~(\ref{alphafit}) only represents the maximum likelihood estimator for the power-law slope if the $ x_{i} $ are independent or at least uncorrelated, otherwise the estimator is biased. However, we know that for our sample the $ x_{i} $ are not uncorrelated, since the flux densities are occurring in ``flares", it means in a time-continuous development. This describes simply the fact that finding the flux density to be at a level of 15 mJy implies very low probability for a level of 5 mJy to be reached within e.g. the next 3 minutes. This predictable behavior disappears on longer timescales, and it is not possible anymore, knowing the flux density at a time point $ t_{\rm{0}} $, to predict the flux density level e.g. 100 minutes later very reliably. This has been investigated by \cite{2009ApJ...694L..87M} in their analysis of the power spectral density discovering a timescale at about 150 minutes. Here we are analyzing data covering about 15\,000 minutes, a hundred times longer than the timescale on which the \mbox{correlation} of the data vanishes. This means to first order and because the estimator in equation Equation~(\ref{alphafit}) is most sensitive to data points close to $ x_{\rm{min}} $ (where the histogram is most populated and the value of each bin can be considered to be fairly independent from the neighboring bins), the bias is negligible.  For the high flux density tail the bias due to the time correlation is significant, because here the histogram bars are populated only with data points of the rare strong outbursts. In the case of  our dataset all histogram bars higher than about 17 mJy are populated only due to one very bright outburst. This is the reason why for higher flux densities the empirical cumulative distribution deviates more and more from the ideal CDF of the model and the statistics becomes incomplete. This effect is also visible in Figure~\ref{CDFindepend} of Appendix~\ref{apstat}, where we show the ideal CDF, the empirical cumulative distribution and the cumulative distribution of some of the uncorrelated synthetic samples we generated for the estimation of the goodness parameter $ q $. For high flux densities many synthetic datasets show a closer development with respect to their best-fit CDF than our empirical sample, even if their KS value $ D $ is worse. In this context it is important to notice that the synthetic samples with a worse KS statistics may differ significantly from the ideal CDF at lower flux density ranges which is not well visible in a log-log diagram (see Figure~\ref{CDFindepend}, Appendix~\ref{apstat}). These arguments also show that a $ \chi^{2} $-minimization fitting to the histogram is questionable, especially if the $ \chi^{2} $-values are used to establish the significance of a distribution break based on the highest, most correlated bins.

Up to this point we found a description for flux densities higher than $x_{\rm{min}}$. We can show that our data sample is consistent with a pure power-law describing the intrinsic flux density distribution under the influence of an instrument with limited resolution and sensitivity and that $x_{\rm{min}}$ can be interpreted as the detection limit of NACO for Sgr~A* due to being embedded in extended flux and its confusion by faint unresolved stars. The argument is simple: If we weight the power-law distribution for fluxes higher than $ x_{\rm{min}} $, with a factor $ n_{\rm{tail}}/n $ with $ n_{\rm{tail}} $ the number of data points with values higher than $ x_{\rm{min}} $, we can extend the power-law to smaller flux densities until its integral becomes unity:

\begin{equation}
x_{\rm{min}}^{\ast} = \left( \frac{n_{\rm{tail}}}{n}\right) ^{\frac{1}{\alpha-1}} \cdot x_{\rm{min}}~~.
\label{xminast}
\end{equation}

In our case with the values of Equation~(\ref{powerre}) we find $ x_{\rm{min}}^{\ast} = 3.57 \pm 0.1$ mJy. Correcting this value for $ x_{\rm{0}} = -2.94 \pm 0.1$ mJy, this power-law distribution shows a cut-off at $ x_{\rm{min}}^{\ast} + x_{\rm{0}} = 0.63 \pm 0.15$, which is identical to the average flux density of the two background apertures close to Sgr~A*. The measured distribution now can indeed be obtained by convolving the power-law distribution with a Gaussian distribution to account for the uncertainty of the photometry: 

\begin{equation}
p_{\rm{obs}}(x) = \int p_{\rm{in}}(z)\frac{1}{\sqrt{2\pi}\sigma^{\ast}} \exp\left[ \frac{-\left( x-z\right)^{2} }{2{\sigma^{\ast}}^{2}}\right] dz ~~,
\end{equation}
with $ \sigma^{\ast} $ the HWHM of the error distribution, which for our data sample can be considered constant up to $ \approx 15 $ mJy (see Figure~\ref{errorflux}). For higher values the histogram starts to be incomplete, and the bias due to the time correlation is dominating the statistical errors. The slope of the power-law is slightly changed by the convolution, but for values of $ \sigma^{\ast} $ of the magnitude of the observational errors this effect is within the errors of $ \alpha $. 

By using again KS-statistics we find a constant value of $ \sigma^{\ast} \approx 0.32$ mJy as a best fit to the histogram (see Figure~\ref{bestfithis}). This value is larger than the observed error $ \sigma^{\ast} \approx 0.2$ mJy of better separated sources (see Figure~\ref{errorflux}). The reason is that the photometry on Sgr~A* for a large fraction of the data is influenced by the nearby star S17, and that we subtracted a constant offset from epoch to epoch, for which it is difficult to find a realistic error. Both influences effectively broaden the expected distribution (which as a result is not Gaussian anymore in general, but somewhat less peaked). That this is not a disadvantage of the deconvolution method with respect to the double-aperture method preferred by \cite{2011ApJ...728...37D} can be seen in Figure~\ref{errorflux}. The error of the double aperture photometry for the low flux density range of Sgr~A* (starting at flux densities $ >3 $ mJy) is also in the range of $ >0.3 $ mJy. 

\begin{figure}
   \centering
   \includegraphics[width=7.5cm]{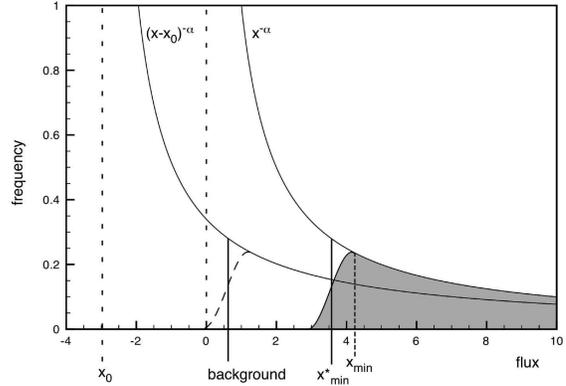}
      \caption[Schematic view of the power-law flux density distribution and the parameters $ x_{\rm{min}} $, $ x^{\ast}_{\rm{min}} $, $ x_{\rm{0}} $ and the background flux density.]{Schematic view of the power-law flux density distribution and the parameters $ x_{\rm{min}} $, $ x^{\ast}_{\rm{min}} $, $ x_{\rm{0}} $ and the background flux density ($y$-axis in arbitrary units, $x$-axis in mJy). We show the measured flux density distribution (grey area) after adding 2.94 mJy to account for $ x_{\rm{0}} $, the pole of the measured non-shifted power-law (which belongs to the non-shifted distribution indicated by the long-dashed line), and the intrinsic distribution with and without correction for $ x_{\rm{0}} $ (described by the continuous lines $ \propto x^{-\alpha} $ and $ \propto (x-x_{\rm{0}})^{-\alpha} $, respectively). As in Eq.~(\ref{powlawdef}), $ x_{\rm{min}} $ is the minimum flux density down to which the shifted measured distribution is a power-law, and $ x^{\ast}_{\rm{min}} $ is the minimum flux density obtained by an extrapolation of the power-law toward lower values, assuming the distribution below $ x_{\rm{min}} $ to be dominated by instrumental effects. Therefore, $ x^{\ast}_{\rm{min}} $ represents the \textit{intrinsic} minimum flux density in the case of the $ x_{\rm{0}} $-shifted distribution. For the case of the non-shifted distribution the intrinsic minimum is represented by $ x^{\ast}_{\rm{min}}+x_{\rm{0}} $, which equals the background flux density within our uncertainties. Thus, the intrinsic minimum $x^{\ast}_{\rm{min,intr.}} = x^{\ast}_{\rm{min}}+x_{\rm{0}} - \rm{backgr.}$ (now additionally corrected for the background) equals zero.}
         \label{sketch}
\end{figure}

\begin{figure}[ht]
   \centering
   \includegraphics[width=7.5cm]{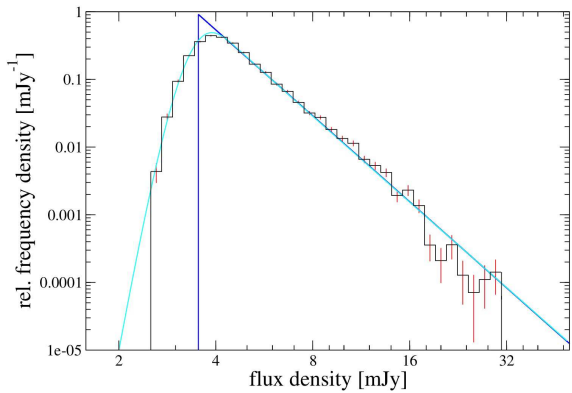}
      \caption[Extrapolation of the best power-law fit.]{Flux density histogram like in Figure~\ref{bestbinmod}. The blue line shows the extrapolation of the best power-law fit, the cyan line the power-law convolved with a Gaussian distribution with $\sigma = 0.32$ mJy.}
         \label{bestfithis}
\end{figure}

We conclude that it is not possible to verify the evidence of an \textit{intrinsic} turnover (that could indicate the peak shape of a log-normal distribution) based on the larger scatter of the low states of Sgr~A* with respect to the typical error of a better separated source of this flux density. The difference of 0.1 mJy between the typical error of a faint source of 0.2 mJy and the error value for our best fit of 0.32 mJy is well within the uncertainties of our knowledge about the true error distribution at the position of Sgr~A* as Fig.~\ref{backhis} demonstrates. Having no evidence for a log-normal distribution for low flux density values\footnote{because the \textit{shape} of the intrinsic distribution close to the detection limit cannot be determined accurately enough to make a difference between a log-normal behavior and a truncated power-law}, the necessity of a break in the distribution to account for the highest flux densities vanishes, even if we accept the data selection of \cite{2011ApJ...728...37D} and ignore the fact that all flux density bins higher than 17 mJy are populated due to one bright event only. Rather than a double state description we prefer a simple power-law with a slope of $ \alpha = 4.2 \pm 0.1$ and an intrinsic pole at $ x_{\rm{0,intr.}} = x_{\rm{0}}-\rm{backgr.} = - x^{\ast}_{\rm{min}} = -3.57 \pm 0.1$~mJy. Since flux density is a positive quantity, this intrinsic power-law naturally breaks at $ x^{\ast}_{\rm{min,intr.}}=x^{\ast}_{\rm{min}} + x_{\rm{0,intr.}} =0 $~mJy. The instrumental effects on the photometry are sufficiently described by a Gaussian distribution centered around the background value of $ 0.6 \pm 0.1 $~mJy with a constant $ \sigma = 0.32 $~mJy. This instrumental effect leads to a detection limit, which here is defined as the limit up to which a reliable photometry is not possible, of $ \sim 0.7 \pm 0.16$~mJy intrinsically, and  $ \sim 1.3 \pm 0.15$~mJy for the actual measurements which include the background flux density of $ 0.63 \pm 0.15 $~mJy. With the power-law model we find a median value for the flux density of $ med(x)_{\rm{intr.}} = 0.9 \pm 0.15$ mJy (corrected for the background), or $ med(x)_{\rm{obs.}} = 1.5 \pm 0.1$ mJy (including the background flux density). This shows that, assuming we can extrapolate the power-law to smaller flux densities below the detection limit, we find the average flux density to be very close to the detection limit, indicating a severe limitation of the knowledge about the variability of Sgr~A* we are able to infer from our data. The relation of $ x_{0} $, $ x_{\rm{min}} $, $ x^{\ast}_{\rm{min}} $ and the background flux density is schematically shown in Fig.~\ref{sketch}.

Of course we cannot \textit{prove} that the flux density distribution is a strict power-law distribution. We only can show that the observable intrinsic flux densities can be well described by this model. This assumption is simpler and needs less parameters than than the assumption of a broken distribution.  Nevertheless, it might well be that the real distribution shows some structure at flux densities below the detection limit. In particular it might even follow a log-normal distribution (with a high multiplicative standard deviation to account for the linear appearance in the log-log plot). The log-normal distribution used in the model of \cite{2011ApJ...728...37D} and an evidence for a break in the distribution at an observable flux density level, however, can be ruled out. 

For the sake of comparability with other objects that show log-normal distributions (e.g. \citealt{2004ApJ...612L..21G}) we want to include here the best fit parameters for a simple log-normal model. It is not easy to estimate the parameters (and their uncertainties) of a log-normal distribution describing the intrinsic flux. The reason is that due to the described instrumental effects we do not have precise knowledge about the shape and position of the intrinsic peak. This is a more severe restriction in the case of a log-normal model that is characterized by its intrinsic turnover than for the power-law model.\footnote{The latter allows for a separation of the part of the histogram that is assumed to represent the intrinsic distribution of flux densities ($ > x_{\rm{min}} $) from the part dominated by instrumental effects.} Fitting the linear histogram with a log-normal model as defined in Equation~(\ref{dodds2}) convolved with a Gaussian (with the width of the Gaussian as a free parameter) we obtain a best fit for $ \sigma_{\ast}=1.00\pm0.05 $, $ \mu_{\ast}=0.12\pm0.07$, $ x_{b}=3.38\pm0.15 $, a width of the Gaussian distribution of $ \sigma^{\ast}=3.38\pm0.06$ and a $ \chi^2/\rm{dof} = 1.6 $. The given uncertainties a larger then the formal uncertainties. In particular, they allow for larger deviations from the histogram at low flux densities than implied by the statistical errors of the bin. This accounts for the fact that the true instrumental effects are only approximately Gaussian (due to the non-uniform white noise contribution in each night and the epoch-wise correction for stellar confusion). The intrinsic median flux density of the log-normal model is $4.51\pm0.2$~mJy and equal to the the corresponding value for the power-law model of $ 4.47\pm0.2 $~mJy (both for the shifted distribution).

%On the other hand it might well be that the real distribution shows some structure at flux densities below the detection limit. In particular it might even follow a log-normal distribution (with a high multiplicative standard deviation to account for the straight appearance in the log-log plot). The log-normal distribution used in the model of \cite{2011ApJ...728...37D} and an evidence for a distribution break at an observable flux density level on the other hand can be ruled out. 

\section{Time series analysis} \label{tsa}

In this section we investigate the nature of the already mentioned time correlation of the flux density measurements. In a more formal description: Assuming that all measurements are samples of the very same random process, and starting with the idea that this random process is (weakly) stationary (for which the impression of a stable mean and variance is indicative), we consider the flux density distribution of section~\ref{stat} as the marginal probability distribution of the random process. Now we want to find a characterization of the joint probability distribution. Whereas in general 'red noise' light curves˝\footnote{I.e., light curves with a time-correlation characterized by a power-law like power spectral density} only can be considered as weakly non-stationary,  the (weak) stationarity for the underlying random process is a consequence of the PSD break found by \cite{2009ApJ...694L..87M} that is far shorter than the covered time period of 15\,000 minutes.

\begin{figure*}[ht!]
   \centering
   \includegraphics[width=14.5cm]{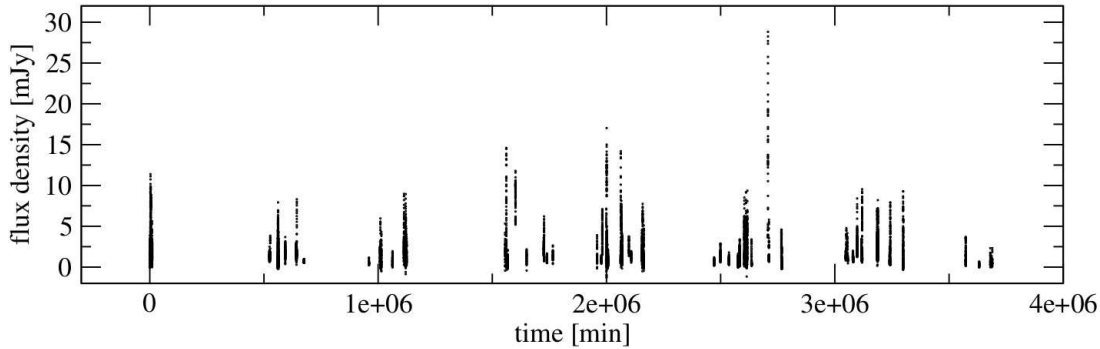}
      \caption[Light curve of Sgr A* with true cadence.]{Light curve of Sgr A* like in Figure~\ref{conclc}. In this case no time gaps have been removed, the data is shown in its true time coverage. A comparison of both plots shows: Only about $0.4 \%$ of the 7 years have been covered by observations.}
         \label{realtime}
\end{figure*}

A first, very simple approach for characterizing the time behavior of the variability is the following: Let us associate the average sampling (of the concatenated light curve) of $ \sim 1.2 $ minutes to every data point. In this way we can relate the total time the source spent in the range of a given flux density bin to the total time covered by observations ($ \sim 15\,000 $ minutes), and we get a rough estimator for the fraction of time the source spends at that flux density. For a more detailed analysis we have to use standard time series analysis tools, like periodograms as estimators for the power spectral density (PSD) of the process, Lomb-Scarle periodograms, the autocorrelation function or the structure function (\citealt{1982ApJ...263..835S, priestley82, 1985ApJ...296...46S}). As \cite{2009ApJ...694L..87M} point out the given window function (covering $ \sim 3.6 \cdot 10^{6} $ min with a coverage fraction of only $ \sim 0.4 \% $) makes standard Fourier transform techniques unsuitable (see Figure~\ref{realtime}). Similarly the Lomb-Scargle periodogram, generally suited as a PSD estimator for non-equally sampled data, is based on the average sampling, which in this case is $ > 3000 $ min. \cite{2009ApJ...691.1021D} used the approach of comparing and averaging the Lomb-Scargle periodograms of data subsets with similar length and sampling to access the PSD of the higher frequencies. In our case this approach again would probably introduce selection effects, and we decided to generally follow the method presented by \cite{2009ApJ...694L..87M}, a Monte Carlo (MC) approach, similar to the PSRESP method by \cite{2002MNRAS.332..231U}. \cite{2011ApJ...728...37D} mention that the MC simulation approach used in \cite{2009ApJ...691.1021D} and \cite{2009ApJ...694L..87M} are based on a comparison sample with a flux density distribution that is Gaussian, and in particular allows negative values, questioning the validity of the method.  In the following we overcome these concerns by developing an algorithm that allows us to simulate time series with the flux density distribution we observe.
 
\subsection{RMS-flux density relation} 
\label{rms}

\begin{figure}[h!]
   \centering
   \includegraphics[width=7.5cm]{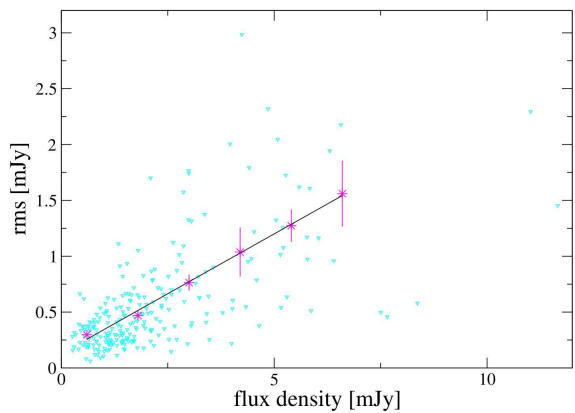}
      \caption[Relation between the rms and average flux density]{Relation between the rms (on timescales of 2 to 16 min) and average flux density for data segments of 24 min (blues points). In red the re-binned data, in black the best linear fit.}
         \label{flurm}
\end{figure}

For the study of X-ray binary variability (linear) rms-flux relations represent an important piece of information. A relation between rms and the radiation flux was initially discovered in observations of black-hole and neutron star X-ray binaries (\citealt{2001MNRAS.323L..26U}). Since then the rms-–flux relation has been studied in a several other observations of black-hole binaries, active galactic nuclei, neutron-star X-ray binaries, and an ultra-luminous X-ray sources (ULX) (see e.g. \citealt{2011MNRAS.411L..66H}, and references cited therein). The rms (root mean square) - as it is used here - is a measure for the magnitude of the variability of the light curve. Following \cite{2005MNRAS.359..345U} the absolute rms amplitude of variability $ \sigma_{\rm{rms}} $ of a time series of $ n $ data points, $ x_{i} $, is defined as as:

\begin{equation}
\sigma_{\rm{rms}} = \sqrt{\frac{1}{n-1}\sum_{i=1}^{n}(x_{i}-\langle x\rangle)^{2}}\;.
\end{equation}
In the case of weakly non-stationary segments of a stationary light curve, $ \sigma_{\rm{rms}} $ varies randomly about a mean value. Under certain circumstances this mean scales with the average flux density of the segment $ \langle x\rangle $.

In particular, \cite{2005MNRAS.359..345U} related the rms-flux relation (which could be observed on all timescales for some of their sources) to a formally non-linear, in their case exponential statistical, model. With this model the authors can convincingly reproduce the behavior of the observed X-ray light curves and rule out additive (shot-noise) models or self-organized criticality as the responsible processes. They conclude that the variability processes must be multiplicative. Because the rms-flux relation is stable for all spectral states of the black hole X-ray binary Cyg X-1, independent of its PSD shape, it is believed to be a more fundamental property of the variability than the PSD shape.

We report here the discovery of the rms-flux relation for Sgr~A* in the NIR. Following the description of the rms analysis in \cite{2005MNRAS.359..345U} and, using time series IDL-codes  written by S.Vaughan, we estimated the flux density dependency of the rms on a frequency range from the average Nyquist-frequency of $ \sim 0.5 \, \rm{min}^{-1}$  to $ 0.06 \, \rm{min}^{-1} $ (corresponding to timescales of 2 to 16 minutes) for data sections with a length of 24 min. The algorithm works as follows: We divided the light curves in continuous data segments of 24 minutes, took the average as a flux density estimate, and determined the PSD of this section. We then obtained the rms for timescales between 2 and 16 minutes by taking the square root of the integral of the PSD over the corresponding frequency range. 

The dominant timescale is of the order of a few hundred minutes (150 min, \citealt{2009ApJ...694L..87M}). Realizations of the random process that are not significantly longer than this timescale are weakly non-stationary. So for a given mean flux density the rms-values of our data sections of 24 minutes are significantly scattered around their average value, and we re-binned the obtained rms-values into flux density-bins with a width  $ \Delta F=1.2 $~mJy. The result is shown in Figure~\ref{flurm}. To a first approximation, we find a linear rms-flux dependence.

\subsection{Simulating light curves} 
\label{simu}

A simple argument for the plausibility of a rms-flux relation is given by \cite{2005MNRAS.359..345U}. The authors point out that a product of two sinusoidal variations with two well separated frequencies, where the lower 'modulates' the amplitude of the higher frequency, would show a linear rms-flux relation. In contrast to this, a linear and Gaussian random process would not show any correlation between flux density and rms. Because the rms-flux relation in their case holds for all observed timescales, the authors choose the ansatz:

\begin{equation}
x(t) = \prod_{i=1}^{\infty}\left[1+A_{i}\sin\left( 2\pi\nu_{i}t + \phi_{i} \right)  \right]~~.
\label{mult}
\end{equation}
For this multiplicative sine model the authors then can show that $ x_{t} $ under general conditions is log-normal distributed, that it can be obtained from a Gaussian linear random process $ l(t) $ by the transform $ x_{t}\approx \exp[l(t)]$, and that for this kind of transform one indeed can derive the rms-flux relation to be a linear function.

Whereas the rms-flux relation often is considered as indicative for a multiplicative process being at work, it actually can be shown (\citealt{2005MNRAS.359..345U} and references therein) that, for every non-Gaussian, skewed distribution, the sample mean and variance are correlated (the distribution is heteroskedastic). Thus, another, less favored possibility to explain the rms-flux relation and the skewness of the flux density distribution is simply a non-Gaussian linear process. It is because of this reason that one has to speak of a \textit{formally} non-linear description, and, without the modeling in the framework of a concrete physical model, it must remain unclear whether the non-linearity has a physical meaning rather than being the property of the mathematical description.

In this case we do not know, if the observed rms-flux relation also is valid for a bigger range of time and flux density scales. As mentioned in section~\ref{stat}, the median flux density of the intrinsic distribution is close to the detections limit, i.e. we only detect the variability of Sgr~A* about half of the time with reliable photometric accuracy. Furthermore, we have to use data segments of a length comparable to or shorter than the dominant timescale. Additionally, the segments have to be comparably short with respect to the typical data length (about 130 minutes on average) to provide a sufficiently big number of rms-flux data pairs. On the other hand, the segments should be long enough to contain enough data points for a reliable rms estimation. In the case of our data, these constraints only allow an investigation of the timescales presented in section~\ref{rms}.
%
%The flux density distribution does not seem to be log-normal distributed (although this cannot be excluded with absolute certainty). Some conditions for the derivation of the log-normal behavior in \cite{2005MNRAS.359..345U} might be violated in our case. One of the conditions in the derivation (under which the central limit theorem is applicable) is the broad, over a wide frequency range almost constant PSD. Since we clearly see a dominant timescale in our data, this condition is not well fulfilled, and from this perspective we would not necessarily expect the flux densities to be log-normal distributed, even if the multiplicative approach of Equation~(\ref{mult}) can be applied to this case. It is also possible that for this case Equation~(\ref{mult}) is not a good representation of the random process, either because we only have a finite product, or the factors are not sinusoidal. 

In \citealt{2005MNRAS.359..345U} the authors use Equation~(\ref{mult}) to show the plausibility of the multiplicative approach in the context of an rms-flux relation and a log-normal distribution, deducing an exponential transform of a linear, Gaussian random process as a good approximation of their random process. Thus, in our case we ask which is the transform that has to be applied to a Gaussian distributed random variable to obtain a random variable that is power-law distributed with the parameters found in section~\ref{pl}. We then assume that we can apply this transform to a linear, Gaussian process to find a description of the observed process. This actually is a standard method for generating Fourier transform based surrogate data with a non-linear appearance (see \citealt{1992PhyD...58...77T}). 

The statistical flux density model we described in section~\ref{pl} has a simple analytic form. This allows us to deduce an analytic transform

\begin{equation}
x_{t} = T(y_{t})~~,
\label{transap}
\end{equation}
with $ y_{t} $ a Gaussian, linear process with unity variance, $ T  $ the transform, and $ x_{t} $ a power-law distributed process. In the following we describe how we find this transform.

Let $ y $ be a random variable with Gaussian probability density $ p_{y} $ of zero mean and unit variance:

\begin{equation}
p_{y}(y) =  \frac{1}{\sqrt{2\pi}} \exp\left( \frac{-y^{2} }{2}\right)
\end{equation}
and $ x $ a random variable with a power-law probability density $ p_{x} $:

\begin{equation}
p_{x}(x) = \left \{ \begin{array}{r@{\quad:\quad}l}
0 & x \leq c \\
\frac{\alpha-1}{c} \cdot \left(\frac{x}{c} \right)^{-\alpha} & x > c~~.
\end{array} \right.
\end{equation}

Let assume that $ \alpha > 1 $. We are looking for a transformation $ x = T(y) $ such that $ p_{y} $ transforms to $ p_{x} $. For such a transform the probability to find a value in the immediate surrounding of $ y $ has to equal the probability to find a value in the immediate surrounding of the corresponding $ x=T(y) $:

\begin{equation}
p_{y}(y) dy =  p_{y}\left[T^{-1}(x)\right] \frac{d\left[T^{-1}(x)\right]}{dx} dx = p_{x}(x) dx~~,
\end{equation}
or
\begin{equation}
p_{x}(x) = p_{y}\left[T^{-1}(x)\right] \frac{d\left[T^{-1}(x)\right]}{dx}~~.
\end{equation}
To solve this equation we use a normalization argument:
\begin{equation}
\int^{y}_{-\infty} p_{y}(y') dy' = \int^{x}_{c} p_{x}(x') dx'~~.
\label{norm}
\end{equation}
With
\begin{equation}
\int^{\infty}_{c} x^{-\alpha} dx = \frac{1}{(\alpha-1)} \cdot c^{1-\alpha}
\end{equation}
Equation~(\ref{norm}) can be reduced to
\begin{equation}
\frac{1}{2} \left[{\rm{erf}}\left(\frac{y}{\sqrt{2}}\right) +1 \right] = 1 - \left(\frac{x}{c}\right)^{(1-\alpha)}~~,
\end{equation}
with $ {\rm{erf}}(y) $ the Gaussian error function. This can be solved for $ x $:
\begin{equation}
x = c \cdot \left\lbrace \frac{1}{2}\left[ 1 + {\rm{erf}}\left(\frac{y}{\sqrt{2}}\right)\right]\right\rbrace^{\frac{1}{\left(1-\alpha \right)}} = T(y)~~.
\end{equation}

To reproduce the variability with and intrinsic power-law distribution on top of a constant background flux as discussed in section~\ref{pl} we have to replace $ c $ with $ x_{\rm{min}}^{\ast} $ and subtract $ x_{\rm{0}} $:

\begin{equation}
T(y_{t}) = x_{\rm{min}}^{\ast} \cdot \left\lbrace \frac{1}{2}\left[ 1 + {\rm{erf}}\left(\frac{y_{t}}{\sqrt{2}}\right)\right]\right\rbrace^{\frac{1}{\left(1-\alpha \right)}} - x_{\rm{0}}~~,
\label{transf}
\end{equation}
with $ x_{\rm{min}}^{\ast} $ defined in Equation~(\ref{xminast}), $ x_{\rm{0}} $ as in Equation~(\ref{powerre}), and $ \alpha $ the slope of the power-law.

With this transform, we are able to generate surrogate light curves (i.e. single realizations of the underlying process) for any input power spectral density (PSD) with the following algorithm:

\begin{itemize}

\item Generate a Gaussian, linear light curve following the method by \cite{1995A&A...300..707T}. This includes drawing Fourier coefficients for each frequency from a Gaussian distribution with a variance proportional to the value of the PSD at that frequency, and Fourier transforming to time domain. 

\item Normalize the obtained Gaussian process to a variance of unity. Optionally re-sample the equally spaced data to the cadence of the observed data.

\item Transform the light curve according to Equation~(\ref{transap}) and Equation~(\ref{transf}).

\item Add an independently drawn quantity (e.g. Gaussian) for each time point to account for the white noise contribution of the measurement. 

\end{itemize}

As we show in the next section, the obtained surrogate data represent the observed flux density distribution (by construction) and the time behavior of the light curves of Sgr~A*. Applied to a linear Gaussian light curve with the PSDs discussed in section~\ref{sim} we obtained light curves of an typical appearance as shown in Figures~\ref{simlc94} and \ref{simlc68}. The non-linearity introduced by a transformation as described here is static. This means that for each time point we apply the same transform, and the non-linearity is only in the amplitude distribution of the observed quantity and not in its dynamics (see \citealt{1992PhyD...58...77T}). This can be illustrated with light curves generated with a double broken power-law PSD of the form: 

\begin{equation}
S(f_{j}) = \left \{ \begin{array}{r@{\quad:\quad}l} 
f_{a}^{-\alpha_{2} + \alpha_{1}}f_{j}^{-\alpha_{1}} & f_{j} < f_{a} \\ 
f_{j}^{-\alpha_{2}} & f_{a} \leq f_{j} < f_{b} \\
f_{b}^{-\alpha_{2} + \alpha_{3}}f_{j}^{-\alpha_{3}} & f_{b} \leq f_{j}~~~~~~~,
\end{array} \right.
\label{brokenpsd}
\end{equation}
with $ f_{a}<f_{b} $ the break frequencies, and $ \alpha_{1}~<~\alpha_{2}~<~\alpha_{3} $ the power-law slopes.

With the algorithm described above we generated 100 equally sampled light curves with a sampling of 0.1 min and a length of 50\,000 minutes. For each surrogate light curve we calculated the periodogram as an estimator of the PSD after transformation according to:
  
\begin{eqnarray}
\rm{Per}(f_{j}) & = & \vert \rm{DFT}(f_{j})\vert^{2} \nonumber \\
 & \propto & \left[ \sum^{n}_{i=1} x_{i} \cos(2\pi f_{j} t_{i}) \right]^{2} \nonumber \\
& + & \left[ \sum^{n}_{i=1} x_{i} \sin(2\pi f_{j} t_{i}) \right]^{2}~~,
\end{eqnarray}
with DFT the discrete Fourier transform. For a review of common conventions of normalization see \cite{2003MNRAS.345.1271V}. Here we want to compare the shape of the PSD, and due to the normalization step applied to the Gaussian linear light curve only relative power is of importance. Because the Fourier based periodogram is not a consistent estimator of the PSD (i.e. for a single realization its standard variation is equal to the mean values at each frequency, irrespective of number of data points) some kind of averaging has to be applied (\citealt{1995A&A...300..707T}, \citealt{2003MNRAS.345.1271V}), so we averaged over the 100 surrogate sets we generated.

\begin{figure}[h]
   \centering
   \includegraphics[width=7.5cm]{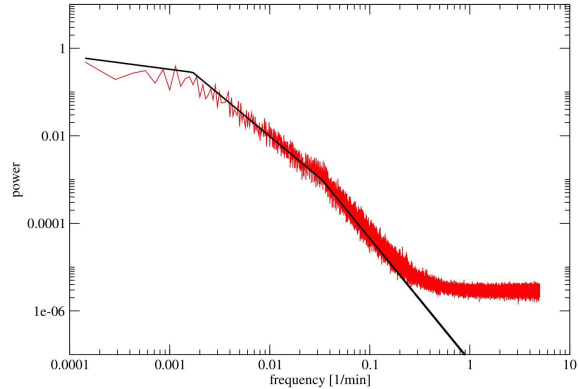}
      \caption[The PSD under transformation.]{The PSD under transformation according to Equation~(\ref{transf}). In black a double broken power-law input PSD, in red the output PSD after applying the transform. The power is given in arbitrary units.}
         \label{PSD}
\end{figure}

A comparison of the input and the output PSD is shown in Figure~\ref{PSD}. In the first approximation and with the exception of the high-frequency part (that is dominated by the white noise contribution) and a calibration factor the PSD is invariant under the transformation.
%With the exception of the high-frequency part that is dominated by the white noise contribution, the PSD is to first order invariant under the transformation.

\subsection{The structure function and the PSD} 
\label{sim}

\begin{figure}[h]
   \centering
   \includegraphics[width=7.5cm]{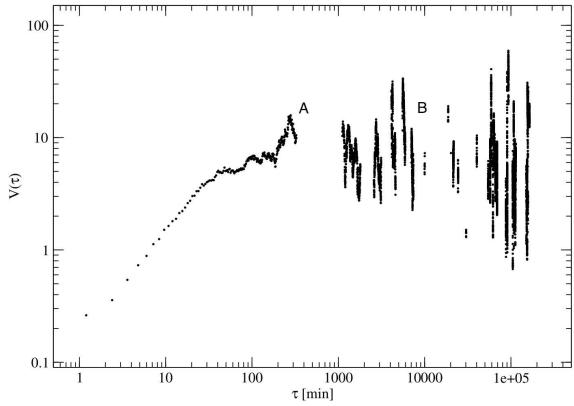}
      \caption[The structure function of the observed data sample.]{The structure function of the observed data sample (time binning 1.2 min).}
         \label{struct}
\end{figure}

Now we can investigate the time correlation within the random process. One way to do this is an investigation of the structure function, a running variance method that measures the mean value of the flux density variance for a given time separation~$ \tau $ (\citealt{1985ApJ...296...46S, 2009ApJ...691.1021D}):

\begin{equation}
V(\tau) = \langle\left[x\left(t + \tau \right) - x(t) \right]^{2} \rangle~~.
\end{equation}
The structure function of our data sample (on its true cadence) is shown in Figure~\ref{struct}. We only considered time separations with more than 300 flux density pairs. Clearly the night-day gap between $ \sim 360 $ min and $ \sim 1200 $ min (A) and the section with low density of data points beginning at $ \sim 7500 $ min due to the typical length of the observation runs (B) are visible. The structure function shows the expected tendency for a flat behavior at small $ \tau $-values (white noise of the measurement), a steeply increasing, power-law like middle section, and a flat behavior at longer timescales (\citealt{2009ApJ...691.1021D}; \citealt{2009ApJ...694L..87M}). Interestingly the structure function also shows a number of features starting at $ \sim25 $ min, which give the impression of a second break at this timescale. This timescale is of particular interest in the discussion of the role of physical processes close to the innermost stable orbit.

The analysis of the structure function is problematic. A comprehensive study about the use and the caveats of structure function methods can be found in \cite{2010MNRAS.404..931E}. The authors point out that spurious breaks may occur for many realizations of random processes with even featureless PSDs, only reflecting the interplay of the PSD of the underlying random process and the data length. This can be easily understood, because timescales much longer than the data length (weakly stationary case) can define the \textit{sample} average (then different from the average of the process) around which the shorter timescales (smaller than the data length) might vary with similar, repeated fluctuations (see Figure~4 in \citealt{2010MNRAS.404..931E}). This makes the true average of the underlying random process an essential piece of information, which in our case can be easily inferred. Although we do not know the true distribution below the detection limit, the average of the flux densities above this limit ($ 2.3 \pm 0.1 $ mJy including the background) is an upper limit to the true average value. Because the true average has to be greater than zero, its uncertainty is small in comparison to the values reached by many outbursts. Thus, the main feature of the structure function, the flattening towards long timescales, is indeed an intrinsic feature. This also is supported by the fact that \cite{2009ApJ...694L..87M} find zero percentage of acceptance for single slope power-law PSDs.

Other concerns of \cite{2010MNRAS.404..931E} are more important for our case: The structure function values for different timescales are not independent and not Gaussian, and for broken intrinsic PSFs the break timescale can occur at systematically lower values, making usual fitting algorithms and their error estimation unsuited. Furthermore, the authors show that one may expect plenty of artificial features in the case of non-equally spaced data. While the latter concern indeed makes it necessary to carefully investigate the discussed features of our structure function at shorter timescales (compare the structure function for dense and sparse sampling in Figure~12 of \citealt{2010MNRAS.404..931E}), the former can be overcome by the procedure introduced by \cite{2009ApJ...691.1021D} (also see\citealt{2009ApJ...694L..87M}).

\begin{figure*}{ht}
   \centering
   \includegraphics[width=12.5cm]{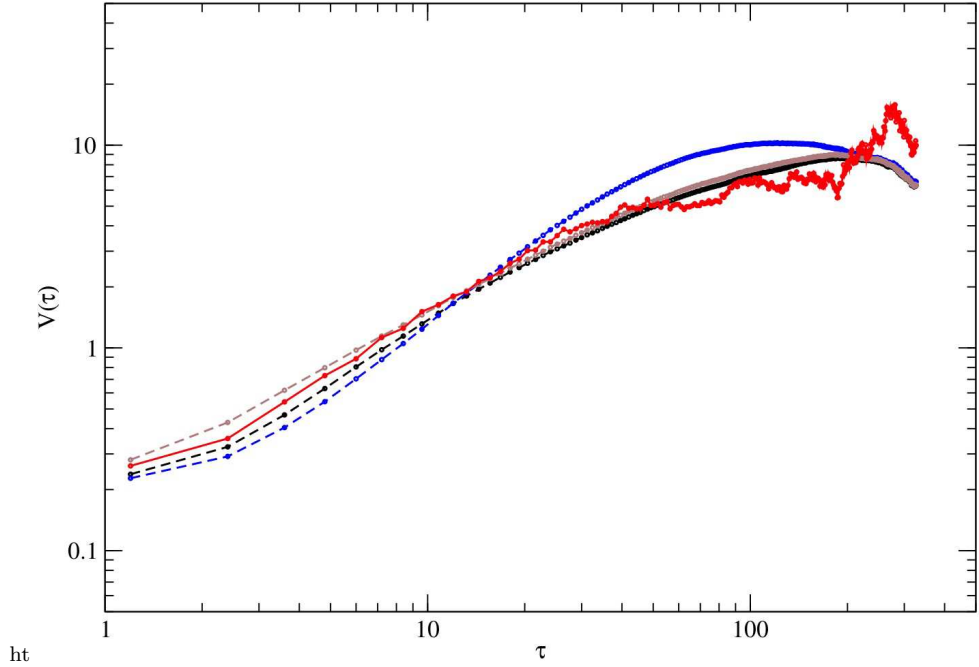}
      \caption[Observed and best fitting structure functions.]{Comparison of the observed structure function (red) and the structure functions of the best fitting single broken PSD (brown, $ 92 \% $ of  acceptance), the best fitting double broken PSD (black, $ 94 \% $), and a single broken PSD with $ 68 \% $ of acceptance (break at 210 min, blue). For details see text.}
         \label{fit}
\end{figure*}

The steps are as follows:

\begin{itemize}

\item Starting with a double broken power-law PSD of the form of Equation~(\ref{brokenpsd}) we generate 5000 light curves (as described above) for a number of combinations of the parameters ($\alpha_{1}$, $\alpha_{2}$,  $\alpha_{3} $, $ f_{a} $, $ f_{b} $). Each light curve has a length of $ 4 \cdot 10^{6} $ min and a sampling of 1 min and is re-sampled to the cadence of the observed data. 

\item We calculate the structure function for each surrogate light curve in the same way as for the observed data sample.

\item We define a goodness parameter for the comparison of the individual structure function with the ``average" structure function for each set of parameters. The probability of acceptance of a parameter set is defined as the percentage of the 5000 surrogate light curves that have a worse goodness-value with respect to the ``average" structure function than the observed sample. 

\end{itemize} 

\cite{2009ApJ...694L..87M} used standard $ \chi^{2} $-values and an arithmetic average of the structure functions for the estimation of the acceptance. To account for a possible non-Gaussian distribution of the structure function values for a given separation $ \tau $, we prefer a modified $ \chi^{2} $ estimation (\citealt{2010MNRAS.404..931E}):

\begin{equation}
\chi^{2}_{\rm{ps}} = \sum_{k}\left(\frac{\langle\log[V(\tau_{k})]\rangle - \log[V(\tau_{k})]}{\sigma_{k}} \right)^{2}
\end{equation}
with $ \sigma_{k} $ the standard variation of $ \log[V(\tau_{k})] $.
We use the logarithm to make the skewed distribution of each structure function value more ``symmetric". In particular the difference of the mean and the most probable value, which is a consequence of the skewness, is reduced making the modified $ \chi^{2} $ a measure of the distance to the most probable rather than to the average structure function, as it is necessary for a maximum likelihood approach.   

Since $ f_{a} $ corresponds to much shorter timescales than the overall length of the observed light curve ($3.6 \cdot 10^{6}$ min), and the power-law slope of the PSD at small frequencies is very flat ($ \sim -0.3 $, \citealt{2009ApJ...694L..87M}), it is not necessary to produce much longer light curves to avoid red-noise leakage. Also a higher sampling rate and a subsequent smoothing in order to simulate the effect of the detector integration does not change the results of our simulations. 

We first explored the parameter space by manual fitting and then defined the range of parameters, for which we set up the Monte Carlo simulation. For the structure function we used time separation bins of 1.2 min and concentrated on the first 287 points (up to a time separation of $ \sim  340$ min) for the estimation of the acceptance values, using the constant slope of $ \alpha_{1} = 0.3 $ found by \cite{2009ApJ...694L..87M} for the long timescales. The well fitting combinations are constrained by the fact that the normalization step makes the choice of e.g. $ \alpha_{3} $ dependent on the choice of $ f_{b} $. Additionally, the differences between $ f_{a} $ and $ f_{b} $ should correspond to a timescale that still can be measured within a typical observation length ($ \sim 130 $ min). Finally, the difference between $ \alpha_{2} $ and $ \alpha_{3} $ should still be big enough to differentiate the double broken PSD model from a single broken model.
For a double broken power-law PSD we tested all combinations of the following parameter set:

%Since the main power-law break is at much shorter timescales than the overall length of the observed light curve ($3.6 \cdot 10^{6}$ min), and the power-law slope of the PSD at small frequencies is very flat ($ \sim -0.3 $, \citealt{2009ApJ...694L..87M}), it is not necessary to produce much longer light curves to avoid red-noise leakage. Also a higher sampling rate and a subsequent smoothing in order to simulate the effect of the detector integration does not change the results of our simulation. 
%
%We first explored the parameter space by manual fitting and then defined the range of parameters, for which we set up the Monte Carlo simulation. For the structure function we used time separation bins of 1.2 min and concentrated on the first 400 points (up to a time separation of $ \sim  340$ min) for the estimation of the acceptance values, using the constant slope of $ a_{1} = 0.3 $ found by \cite{2009ApJ...694L..87M} for the long timescales. The well fitting combinations are constrained by the fact that the normalization step makes the choice of e.g. $ \alpha_{3} $ dependent on the choice of $ f_{b} $. Additionally, the differences between $ f_{a} $ and $ f_{b} $ should correspond to a timescale that still can be measured within a typical observation length ($ \sim 130 $ min). Finally, the difference between $ \alpha_{2} $ and $ \alpha_{3} $ should still be big enough to differentiate the double broken PSD model from a single broken model.
%For a double broken power-law PSD we tested all combinations of the following parameter set:

\begin{eqnarray}
\alpha_{1} & = & 0.3 \nonumber \\
\alpha_{2} &  = &  1.8/1.9/2.0 \nonumber \\
\alpha_{3} &  = &  2.5/2.8/3.3 \\
f_{a} &  =  & 0.001/0.0017/0.0025   \nonumber \\
f_{b} &  = &  0.0133/0.0333/0.0533~~,   \nonumber 
\end{eqnarray}

with $ f_{a} $ and $ f_{b} $ given in $ \rm{min}^{-1} $ , and to obtain a finer grid, additionally all combinations of

\begin{eqnarray}
\alpha_{1} & = & 0.3 \nonumber \\
\alpha_{2} &  = &  1.85/1.9/1.95 \nonumber \\
\alpha_{3} &  = &  2.65/2.8/3.05 \\
f_{a} &  =  & 0.0014/0.0017/0.002   \nonumber \\
f_{b} &  = &  0.0233/0.0333/0.0433~~.   \nonumber 
\end{eqnarray}

For a single broken power-law PSD we tested the conbinations of:

\begin{eqnarray}
\alpha_{1} & = & 0.3 \nonumber \\
\alpha_{2} &  = &  2.0/2.3/2.6/2.9 \nonumber \\
f_{a} &  =  & 0.002/0.004/0.008/0.01~~.
\end{eqnarray}

We find the highest probability of acceptance of $ 96 \% $ for a double broken power-law with slopes of $ \alpha_{2} = 1.9$ and $ \alpha_{3} = 3.3$ and break timescales at $ f_{b}= 20$~min and $ f_{a}= 590$~min, respectively. Several combinations in the ranges of $\alpha_{2} = 1.8-2.0$, $\alpha_{3} = 2.5-3.3$ for the slopes, and 20-40 min and 500-700 min for the break timescales reach acceptance values higher than $ 90 \% $. Considering the typical statistical fluctuations of the acceptance values of about 2 percentage points they can be considered as equivalent. In comparison the single broken PSD models with a break in the range of the $90 \%$ confidence interval of \cite{2009ApJ...694L..87M} ($ 154^{+124}_{-87} $ min) only reach $ 75\% $ at maximum. On the other hand, the high acceptance values are typically reached for values of $ f_{b}=0.0017/0.002 $ and $ \alpha_{2}=1.8-2.0 $, independent of the values for $ f_{a} $ and $ \alpha_{3}$. Indeed, a single broken power-law PSD with $ \alpha_{2} = 2.0 $ and $ f_{a} = 0.002 $ (500 min) also reaches an acceptance level of $ 92 \% $, making the question if the true PSD has more structure than a single break, undecidable on the base of the presented dataset. The behavior of all here discussed parameter sets for timescales longer than 500 min (not depicted here) match the measured structure function.

In Figure~\ref{fit} we show the most probable structure function estimated from 5000 surrogate datasets for a high probability double broken PSD model with break timescales at 30 min and 590 min ($ 94 \% $ of acceptance) and the highly probable single broken PSD model ($ 92 \% $ of acceptance, break at 500 min). Both are almost identical. Additionally, we present the most probable structure function for a single broken PSD model with a lower break timescale at 210 min and a probability of acceptance of $ 68 \% $. We can clearly see that all three structure functions show a flattening towards longer timescales beginning at about 25 min, independent of their dominant timescale or the shape of the PSD. This shows that the features in the observed structure function starting at $ \sim 25 $ min are dominated by the influence of the window function of the dataset and cannot be interpreted as intrinsic.

Light curves generated with the double broken PSD with $ 92 \% $ acceptance and with the single broken PSD with $ 68 \% $ acceptance are shown in Figure~\ref{simlc94} and Figure~\ref{simlc68}. Figure~\ref{simlclong} shows the comparison of long light curves for the high probability single and double broken cases, demonstrating their similarity on long timescales. Also on shorter timescales the high probability single broken models generate light curves which are not obviously different from the high probability double broken cases. It is plausible that a difference that is difficult to find in time continuous data can not be significant in data with a sparse time coverage. Note that due to the dominant timescale in both light curves of about 500-600 min there are time intervals (sometimes even more than a day) where the brighter flares start from higher flux density levels than the normal minimum level. In the case of a sparse data coverage this can lead to a misinterpretation of these minimum level differences as variability on long timescales, and from this point of view the interpretation of these differences as variations on timescales of weeks and month given in \cite{2011ApJ...728...37D} is not the only possible explanation.

We have to make two comments on error bars and confidence levels: First, there is no good method to provide error bars for the observed structure function, because it would require knowledge about the true PSD and its interaction with the window function of the observation. Additionally, the distribution of the single point in the structure function can be very skewed and it is questionable whether e.g. a standard deviation over the MC light curves can be established as a good error estimate. Secondly, the break timescale of 500 min found is statistically equal to the value found by  \cite{2009ApJ...694L..87M} ( which lies in our $ 90 \% $ confidence interval). However, confidence levels deduced from acceptance values take into account the probability that the observed sample is not representative for the true variability but a statistical ``outlier", a possibility that, considering the coverage fraction of $ 0.4 \% $, makes any conclusion unreliable. For example, on a $ 90 \% $ confidence level, the light curves generated from PSDs with acceptance values of $ 68 \% $ and $ 94 \% $, respectively, (Fig.~\ref{simlclong}) are indistinguishable. On the other hand, under the assumption that we actually are looking at a typical sample that represents the variability of a continuous 15000 min data piece quite well, we find an argument against the $ 68\% $-PSD: The generated light curve seems to show too many flares on the 20-30 mJy level. This demonstrates that the insignificance of a break at lower timescales due to identical acceptance values is more fundamental than the criticism that we might look at an exotic realization of the underlying process, or that the analysis suffers from an accidental selection effect: With the cadence of our observations, especially with the day-night gap, we cannot decide on question whether a PSD-break at timescales between 25 and 100 minutes is characteristic for the variability, even if we assume that our data sample is representative.

\begin{figure}[h]
   \centering
   \includegraphics[width=7.5cm]{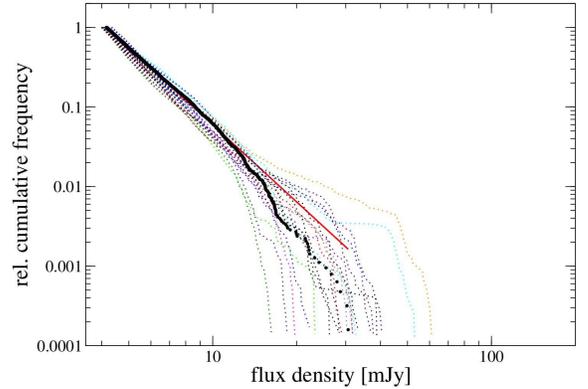}
   \caption[The observed CDF and 20 randomly selected CDFs of time correlated power-law surrogate data.]{The observed CDF (black) of flux densities and 20 randomly selected CDFs of time correlated power-law surrogate data (colored dashed lines). The best fit CDF is shown as a continuous red line.}
   \label{CDFcorr}
\end{figure}

As a last step we can use the 5000 re-sampled light curves with the best fitting structure function to test the plausibility of the power-law assumption as described in section~\ref{pl} (now taking into account the time correlation). We find a goodness parameter of $ q = 0.79 $ (as defined in section~\ref{pl}), much higher than the value of $ q = 0.2 $ for independent data, firmly establishing the plausibility for the power-law description of the probability density. The observed CDF and the CDFs of 20 randomly selected surrogate light curves are shown in Figure~\ref{CDFcorr}. The values of the parameters in Equation~(\ref{powerre}) can be confirmed also for the case of correlated data, the uncertainties are slightly bigger (0.15 mJy for $ x_{\rm{0}} $ and $ x_{\rm{min}} $, and 0.3 for $ \alpha $).

\section{Extreme flux density excursions} \label{echo}
In this section we investigate the consequences of an extrapolation of the measured time correlated power-law to high flux density levels and long timescales. It is obvious that infinitely bright outbursts are unphysical, raising the question of a physical constraint for the maximum flux densities that can be expected.
\subsection{Maximum expected NIR flux density}
\label{section:inversecompton}
Recent measurements of the VLBI source Sgr~A* at 230~GHz resulted in a lower limit to the brightness temperature
of $T_b = 2 \times 10^{10}$K (\citealt{2008Natur.455...78D}).
For the case of a synchrotron source the relations between observed and intrinsic flux densities and frequencies
\begin{eqnarray}
S_{obs} & = & \delta^{3-\alpha} S_{int} \\
\nu_{obs} & = & \delta \nu_{int} \\
\delta & = & (1-\beta^{2})^{\frac{1}{2}}\cdot [1-\beta \cdot \cos(\phi)]^{-1} \\
\beta & = & \frac{v}{c}~~,
\end{eqnarray}
with $ v $ the bulk motion of the synchrotron region, $ \phi $ the viewing angle toward the emitting component, $ \alpha $ the spectral slope, and $\nu_{int}$ the self-absorption peak frequency in GHz,
result in a brightness temperature
(including the possibility of beaming):
\begin{equation}
T_{b}= \frac{1.22~S_{obs}}{\theta^2 \nu^2_{obs}}
= \frac{1.22~S_{int}}{\theta^2 \nu^2_{int}} \delta^{1-\alpha}~~~,
\end{equation}
$\theta$ is the source diameter in milliarcseconds and $T_{b}$ is the equipartition brightness temperature
in units of $10^{12}$K (see \citealt{2012A&A...537A..52E}). Thus, for constant $ \theta $ and $ \delta $ the brightness temperature is proportional to the flux density.
From 345~GHz and 690~GHz SMA measurements, there is evidence that
the self-absorption peak frequency of most synchrotron source
components in Sgr~A* peak around 345~GHz
(\citealt{2006PhDT........32M,2006ApJ...640..308M,2006JPhCS..54..354M}, see also \citealt{2012A&A...537A..52E}). With the lower limit for the brightness temperature mentioned before, which has been obtained close to this peak frequency, we can calculate a flux density limit which extreme outbursts in the NIR must reach, assuming
that the size of the luminous region stays approximately constant and that
the brightness temperature is linked to optically thin
infrared emission. The maximum brightness of the synchrotron source Sgr~A* is certainly given by the
inverse Compton limit of $10^{12}$K (or a few times $10^{12}$K in the case of boosting).
With both constraints the maximum brightness of Sgr~A* can be expected to be
about 100 times brighter than regular flare amplitudes.
Since these amplitudes reach K-band flux densities of around 20 to 30 mJy, extreme values about 3 Jy can be expected.
Only for smaller source sizes and lower self-absorption frequencies,
or higher bulk motions during the flare
the expected extreme K-band flux density may be lower.

\subsection{Extreme X-ray flares in the past}
\label{section:echo}
The present X-ray luminosity of Sgr~A* is more than 10 orders of magnitude less than its Eddington luminosity.
The observation of hard X-ray emission and iron fluorescence from some of the massive molecular
clouds surrounding the Galactic Center has been interpreted as a light echo of a
luminous past flare,
that may have happened up to 400 years ago (\citealt{2004A&A...425L..49R}, \citealt{1998MNRAS.297.1279S}, \citealt{2010ApJ...719..143T}).
\cite{2010ApJ...719..143T} report the observation of a clear decay of the hard X-ray emission from the
molecular cloud Sgr B2 during the past seven years. They argue that this decay strengthens the case for
such a bright flare in the past and significantly weakens the
alternative explanations involving low-energy cosmic rays.

The authors also argue that the luminosity of the event 
which may have been responsible for the observed light echo
was 1.5-5$\times$10$^{39}$ ~erg~s$^{-1}$.
Capelli et al. (2012) present evidence for the fact that the luminosity of SgrA* decreased during the past 150 years
(in good agreement with Muno et al. 2007) with upwards and downwards trends lasting from a few to perhaps 10 years.

If we assume that the infrared flares are linked to X-ray flares by the underlying radiation mechanism, it is interesting to compare the flux densities of the brightest X-ray events in the past with our infrared flare statistics that we 
obtained over the recent $\sim$7 year period, even if the timescales of those events may have been significantly different.
In this picture the short and long term variability would be due to the same underlying physical process and the
present time is close to an 'off-state' while the flare events lasting for a few to 10 years during the
past 150-400 years corresponds to 'on-states' in the SgrA* light curve. 

Lets us assume that such a bright flare in the X-ray domain occurred between 150 and 400 years ago. 
This means that its occurrence is between a factor $ 5 \times 10^{-2} $ and $7 \times 10^{-5}$ less frequent
 than the brightest infrared flares observed until today (one 30 mJy event every 15000 minutes - as observed -
 to one 30 mJy event every 7 years in the case that the observed flare was the brightest over 
the total period of seven years). 
The bulk of the X-ray flares observed in recent years lies between $10^{34}$ to 3$\times$ $10^{35}$ ergs/s 
(\citealt{2001Natur.413...45B,2003ApJ...591..891B},
\citealt{2003A&A...407L..17P,2008A&A...488..549P}).
Assuming an optically thin spectrum between the NIR and X-ray domain,
and assuming that the brightest K-band flares (of 10 to 30~mJy) sample the brightest X-ray flares,
we can expect that the K-band flux density associated with the flare 150-400 years ago has been 
$0.05 \times10^5$ to $5 \times10^5$ times higher (corresponding to 50-15000 Jy)
than what we have observed until today.
This is 2 to 4 orders of magnitude above the K-band flux density limit we estimated
for the case of the inverse Compton limit. Hence an optically thin NIR/X-ray flare
can be excluded as the source of the light echo.
\begin{figure*}[t]
\centering
\includegraphics[width=10.5cm]{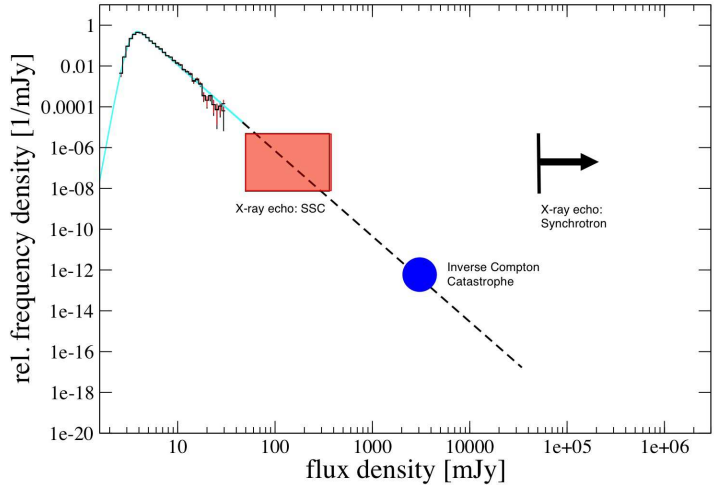}
\caption[The likelihood of extreme flux density excursions.]{Illustration of the likelihood of extreme flux density excursions extrapolated from the statistics of the observed variability. The expected maximum flux density given by the inverse Compton catastrophe and an estimate of its uncertainty are shown as the blue circle, the SSC infrared flux density for a bright X-ray outburst as expected from the observed X-ray echo is depicted as the red rectangular.The lower limit for the NIR flux density in the case of a pure synchrotron model is indicated with the black arrow.}
\label{extended}
\end{figure*}
However, the requirements on the K-band flare brightness are considerably reduced if
we assume a synchrotron self-Compton (SSC) mechanism as a source of the
bright X-ray flare.
We assume a synchrotron radio spectrum
with a turnover between the optically thick and the thin part at
a frequency of $\nu_m$ (in GHz) and a flux density S$_m$ (in Jy)
with an optically thin spectral index $\alpha$
\begin{equation}
S_{\nu}\propto \nu^{-\alpha}.
\end{equation}
Adapting the formulae given by \cite{1983ApJ...264..296M,2009arXiv0909.2576M} for the case
of the Galactic Center, we then find that the SSC X-ray flux density
$S_{X,SSC}$ (in $\mu$Jy),
is given by
\begin{eqnarray}
\label{eq1}
S_{X,SSC} & = & d(\alpha)\ln\left(\frac{\nu_2}{\nu_m}\right)\theta^{-2(2\alpha+3)}\nu_m^{-(3\alpha+5)} \nonumber \\
&& \times S_m^{2(\alpha+2)}E_X^{-\alpha}\delta^{-2(\alpha+2)}~~,
\end{eqnarray}
\noindent
where $d(\alpha)$ is a dimensionless parameter and $E_X$ the X-ray photon energy in $keV$.
Relativistic boosting, denoted by $\delta$, may occur for
anisotropic motion along a jet, or due to orbital motion close to
the black hole, or due to a rapid isotropic expansion
of a source component
(see discussion by \citealt{2012A&A...537A..52E}).
With the $S_{X,SSC} \propto S_m^{2(\alpha+2)}$ dependency
the 10$^4$ to 10$^5$ increase in X-ray luminosity can be
achieved by an increase of the K-band flux density by a factor of 5 to 12.
Here we used a synchrotron spectral index $\alpha$ of 0.7$\pm$0.3
(\citealt{2007ApJ...667..900H}, see also \citealt{2011A&A...532A..26B}) and
assumed that source sizes and turnover frequencies
are similar to what had been obtained for the recently
observed flares.
This indicates that the light echo could be produced by a
50 to 360~mJy flare that is a factor of $5 \times 10^{-2}$ to $7 \times 10^{-5}$
times less frequent than the recently observed brightest Ks-band flares.
These are flux densities that are well below those derived for
the inverse Compton limit. Assuming that the frequency of occurrence (measured in $ \rm{min}^{-1} $) of peak flux densities of rare events is proportional to the frequency of flux densities in the flux density histogram (measured in $ \rm{mJy}^{-1} $), and with a frequency of about $ 10^{-4} \;\rm{mJy}^{-1}$ for flux densities around 30 mJy we obtain a range of $ 5 \times 10^{-6}$ - $ 7 \times 10^{-9} \;\rm{mJy}^{-1}$ and 50-360 mJy in the frequency-flux density diagram.
Fig.~\ref{extended} shows that (within the uncertainties) these values are in agreement with the
extrapolated probability density of the K-band flux densities as obtained over
the past 7 years. Also the long light curves in Fig.~\ref{simlclong} show that flares with amplitudes up to 100 mJy can be expected already for continuous light curves of $ 2\cdot 10^{6} $ min, and for 5000 light curves with the observational cadence we find typically 5-7 datasets with maximum flux densities over 250 mJy.

While the presence of bright NIR flares may be inferred from the SgrA* flux distribution histogram,
the occurrence of these flares and the flare length must be discussed in the framework of the
PSD derived for SgrA*:
Gonzales-Martin and S. Vaughan (2012) have investigated the PSDs of 104 nearby AGNs
of which 72\% show strong variability in at least one XMM band and a high percentage 
of low-luminosity sources does not show a significant variability over the monitoring 
time span within the limited XMM lifetime.
Only for 17 sources (15\%) PSD breaks could be found which appear to follow a 
linear scaling relations of break timescales with the black hole mass. 
For 85\% of the total number of sources no break time scale at all was found and especially for 
the sources with significant variability the vast majority (77\%) have not reported break 
timescale (Gonzales-Martin and S. Vaughan (2012) although their black hole masses certainly will 
compare well with the mass range covered by those few for which a break frequency was detected.
This implies that the majority of sources shows a significant power at variability 
frequencies lower than the break frequencies expected for their specific black hole masses
(if 17 sources are regarded as representative).
From our present knowledge of variability of galactic nuclei it is unclear if 
a source for which a PSD break frequency has been detected over a typical observing 
timescale of only $le$10~years, does always have to show such a break frequency. 
Given the current statistics (Gonzales-Martin \& Vaughan (2012) we may in general assume that,
if sampled at other times, these sources are more likely 
to show variability characteristics closer to that of the majority of the sources.
In that case 150 to 400 years ago SgrA* may very well have shown significant power 
at variability frequencies lower than the break frequencies. The required flare lengths of 
a few to about 10 years (Capelli et al. 2012) are short compared to this time span.

In addition, extraordinary events may occur:

\cite{2003AIPC..686..109F} found that for Milky Way type
galaxies the probability of an interaction between a super massive
black hole and a main-sequence star (MSS), a white dwarf (WD)
and a neutron-star or stellar black hole (NS/SBH) is of the
order of 10$^{-5}$, 10$^{-7}$ and 10$^{-8}$ events per year, respectively.
The orbital decay and a possibly significant mass transfer may last for
several months or even years before the actual merger
(e.g. Capozziello et al. 2009), such that
as a consequence significantly enhanced accretion rates and flares
are likely to be observed.
It is also speculated that these merger events actually contribute
significantly to the mass content of SMBH accretion disks.
The flux densities associated with these merger events will be
appreciable (two cases for which this phenomenon has recently been
discussed are Swift J2058.4+0516 at z=1.2, \citealt{2011arXiv1107.5307C,2011ApJ...743..134K}; and Swift J1644+57 at z=0.3, Metzger et al. 2012).
It can also not be excluded that an increase in the accretion rate
and therefore an increase in luminosity can be caused by
encounters with members of the central stellar cluster
as proposed by \cite{2012Natur.481...51G}.
It needs to be investigated how fast a stationary
process can recover from such events.
\section{Conclusion} \label{concl}

We summarize our results:

\begin{itemize}
\item The Ks-band flux density distribution of Sgr~A* as obtained from the last seven years of observations is in convincing agreement with a pure power-law distribution, giving no indication for a break or two-state behavior.

\item We could find an upper limit of the intrinsic mean flux density of about $ 1.7 \pm 0.15 $ mJy, and with a power-law extrapolation to flux densities below the detection limit a mean of $ 0.9 \pm 0.15 $ mJy (including extinction correction). 

\item We found an algorithm to statistically simulate light curves that show the same flux density distribution and time correlation as the observed sample. It is based on the algorithm by \cite{1995A&A...300..707T} to generate linearly time-correlated surrogate samples, but includes a transformation to account for the non-linear appearance of the NIR flux densities of Sgr~A*. In a first approximation broken power-law PSDs are invariant under this transformation. This statistical model (best fitting PSD, flux density distribution and the algorithm) does not provide immediate information on the physical system, but serves as a statistical ``summary" of the observed variability, defining constrains for every physical model. Furthermore, it allows a straight forward power-law extrapolation to higher flux density levels, flux density levels below the detection limit, and long timescales. 

\item This extrapolation demonstrates that high flux density excursions,
as required to explain the supposed X-ray light echo in the molecular
clouds surrounding Sgr~A*, are well within the expected statistical extreme
values of the flux density distribution that we observed at much lower flux densities. This implies that even if we include the flare responsible for the X-ray echo as the most extreme event that is suggested by observations there is no evidence for a two-state variability behavior. 

\item Within our statistical model the concept of a ``quiescent state" and the differentiation between continuous variability and ``off" states (as investigated by \citealt{2011ApJ...728...37D}) turn out to be problematic: In this description, Sgr~A* is always variable and the probability to find it at a flux density level of zero is actually zero. But any flux density interval starting with zero is represented with a higher probability than any other interval of the same length, allowing for arbitrarily faint flux density states. So from an observational point of view, this model predicts ``off" states due to the instrument dependent limited resolution and sensitivity. In the case of Ks-band observations with NACO the detection limit is very close to the mean flux density of the intrinsic distribution.  This means that the distribution of about half of the intrinsic flux density states can not be accessed with our current resolutions and sensitivity. Next generation instruments like LINC-NIRVANA at the Large Binocular Telescope and GRAVITY at the VLTI will allow to probe the lower flux density range.

\item The question whether timescales comparable to the orbital timescale at the innermost stable orbit play a role for the variability is principally undecidable on the base of this data. In order to access this problem a significant amount of continuous light curves with a length of more than 1500 min would be needed. This, however, does not exclude the presence of orbital signatures in polarimetric data as reported in \cite{2010A&A...510A...3Z}. Here we are investigating all light curves in total under the assumption of stationarity. Thus, we cannot make statements on a possible time development of the dominant timescale or the role of shorter timescales as a transient phenomenon.  

\end{itemize}

\acknowledgments
{\small
We thank A. Witzel, S. Britzen, E. Angelakis, A. Ghez, W. Huchtmeier, J. Girard, A. Stolte for fruitful discussions. For the time series analysis we used IDL-routines by S. Vaughan. We thank the anonymous referees for their constructive comments and support. 
Part of this work was supported by the German Deutsche 
Forschungsgemeinschaft, DFG, via grant SFB 956
and fruitful discussions with members of the 
European Union funded COST Action MP0905: 
Black Holes in a violent Universe and PECS project No. 98040. B. Shahzamanian and M. Valencia-S. are members of the International Max Planck Research School (IMPRS) for Astronomy and Astrophysics at the MPIfR Bonn and the Universities of Bonn and Cologne. Rainer Sch\"odel acknowledges the support by Ram\'on y Cajal programme, by grants AYA2010-17631 and AYA2009-13036 of the Spanish Ministry of Science and Innovation, and by grant P08-TIC-4075 of the Junta de Andaluc\'ia. Vladimir Karas acknowledges MSMT project ME09036. N. Sabha is member of the Bonn Cologne Graduate School (BCGS) for Physics and Astronomy supported by
the Deutsche Forschungsgemeinschaft.
{\it Facilities:} \facility{VLT:Yepun (NACO)}. This paper is based on observations conducted with the European Southern Observatory telescopes obtained from the ESO/STECF Science Archive Facility.
Work on LINC/NIRVANA at the LBT and GRAVITY at the VLTI is in parts supported
by the German Federal Department for Education and Research (BMBF) under the
grant Verbundforschung BMBF~05A08PK1 and  BMBF~05A08PKA.
}
\bibliographystyle{aa}
\bibliography{mybib}{}

\clearpage
\begin{appendix}

\section{Two state probability model}
\label{domo}
As a description for the observed distribution of the flux density $ x $ (measured in mJy) \cite{2011ApJ...728...37D} suggest the following probability density:    

\begin{eqnarray}
\label{dodds1}
P_{\rm{0+err}}(x) & = & \int P_{\rm{0}}(x')\frac{1}{\sqrt{2 \pi}\sigma_{\rm{obs}}(x)}\exp\left[ \frac{-(x-x')^{2}}{2\sigma_{\rm{obs}}(x)^{2}}\right]dx' 
\end{eqnarray}

i.e., a convolution of an Gaussian error with flux density-dependent width 

\begin{equation}
\frac{\sigma_{\rm{obs}}}{\rm{mJy}}~=~0.174~\cdot~\left( \frac{x}{\rm{mJy}}\right)^{0.5} 
\end{equation}

and an intrinsic two state probability density:

\begin{equation}
P_{\rm{0}}(x) = \left \{ \begin{array}{r@{\quad:\quad}l}
k P_{\rm{logn}}(x) & x \leq x_{b}+x_{t} \\
k P_{\rm{logn}}(x_{t}+x_{b})\left( \frac{x-x_{b}}{x_{t}}\right)^{-s}  & x > x_{t}+x_{b}
\end{array} \right.
\end{equation}

with $ k $ a  dimensionless normalizing factor, and the log-normal part defined as

\begin{eqnarray}
\label{dodds2}
P_{\rm{logn}}(x) & = & \frac{1}{\sqrt{2 \pi}\sigma_{\ast}\left(x-x_{b}\right)}\exp\left\lbrace -\frac{\left[\ln\left(\frac{x-x_{b}}{\rm{mJy}}\right)-\mu_{\ast}\right]^{2}}{2\sigma_{\ast}^{2}}\right\rbrace
\end{eqnarray}

The parameters of their best fit model ($\chi^{2}/dof=2.99$) are $ \sigma_{\ast} = 0.75 \pm 0.05 $, $ \mu_{\ast}=0.05 \pm 0.06$, $ s = 2.7 \pm 0.14$, $ x_{t} = 4.6 \pm 0.5 $~mJy, and $ x_{b} = 3.59 \pm 0.06$~mJy (the unit of the probability density is $ \rm{mJy}^{-1} $).\\
\newpage
\section{Data quality}
\label{apqual}
Here we give an overview about the observation conditions. All plots show the data after quality cut.
\begin{figure}[h!]
   \centering
   \includegraphics[width=15.5cm]{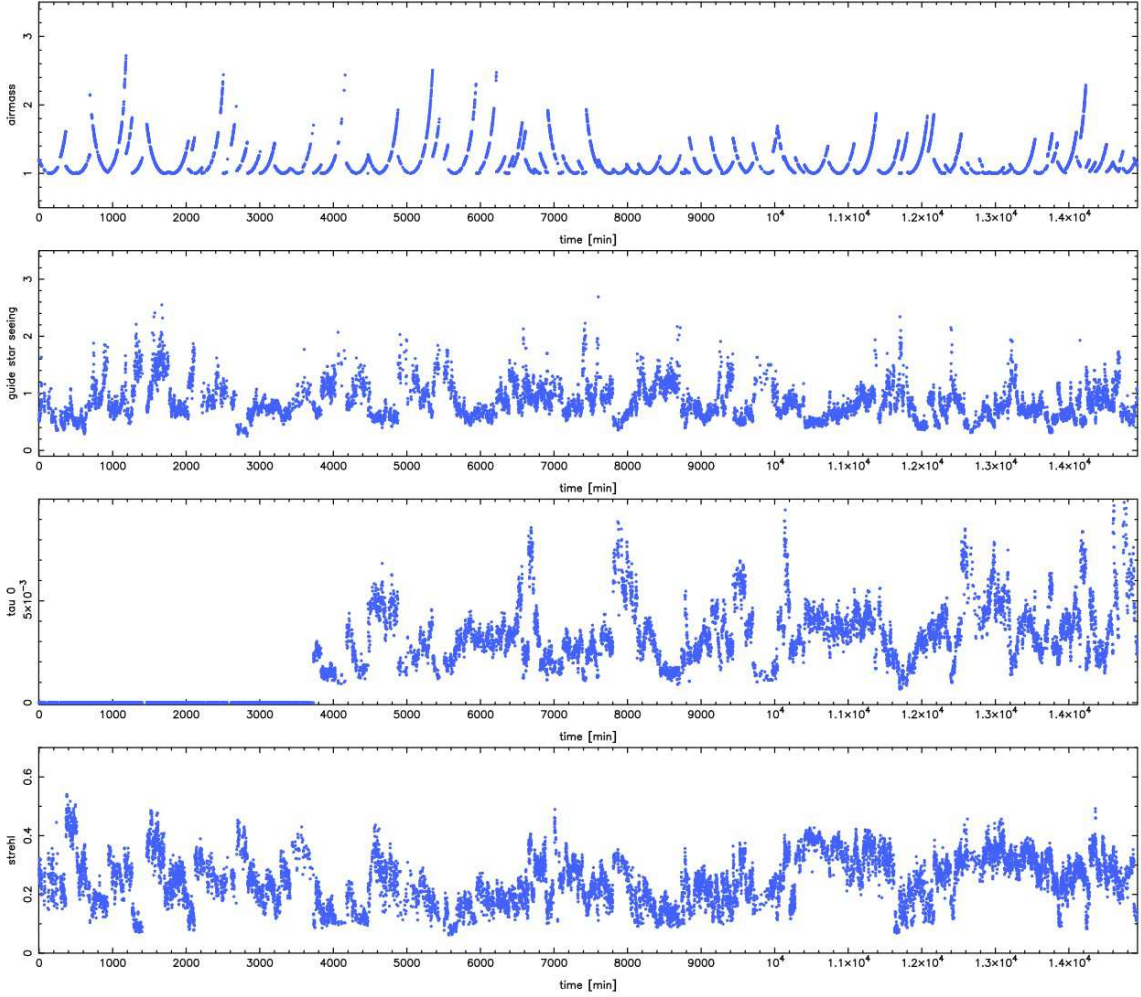}
      \caption[Time series of quality parameters.]{Time series of quality parameters. We show the airmass during observation, the seeing values obtained by a measurement on the active optics guide star, the atmosphere coherence $ \tau_{\rm{0}} $ (not available for all frames), and the Strehl ratio obtained from the extracted PSF. For a histogram representation see Figure~\ref{qualhis}.}
         \label{quality}
\end{figure}

\begin{figure}
 
   \includegraphics[width=6.5cm]{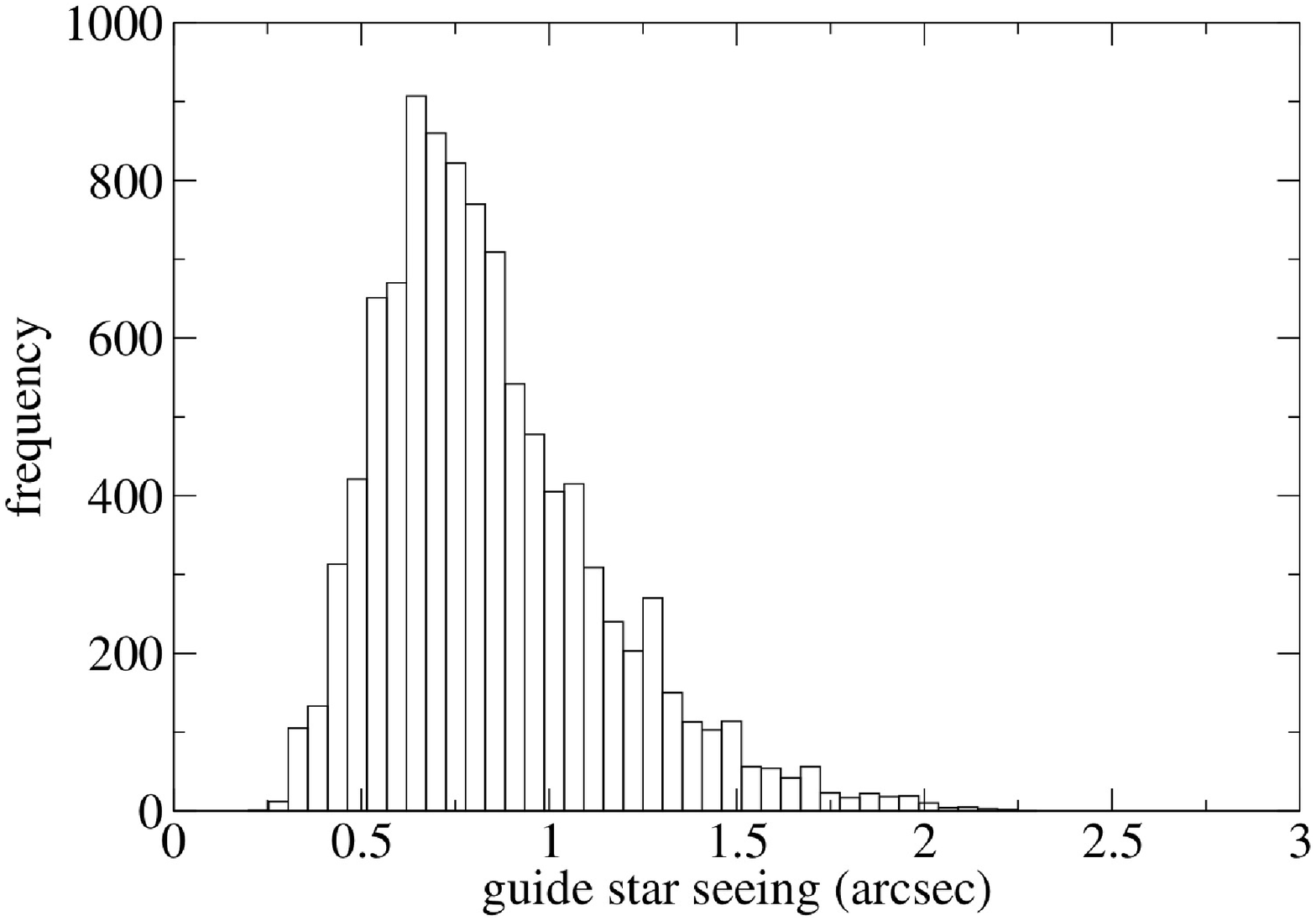}
   \includegraphics[width=6.5cm]{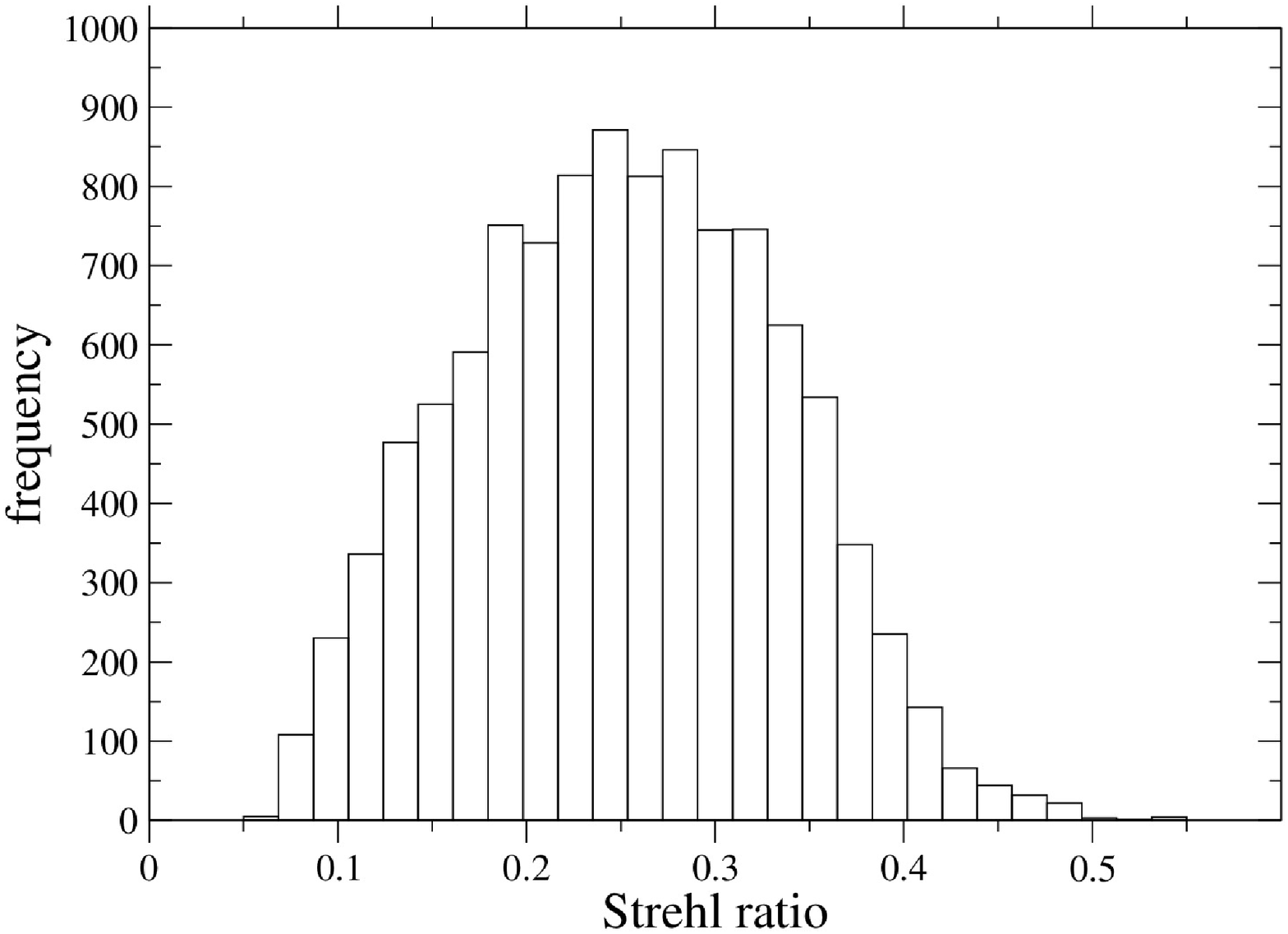}
   \centering
   \includegraphics[width=6.5cm]{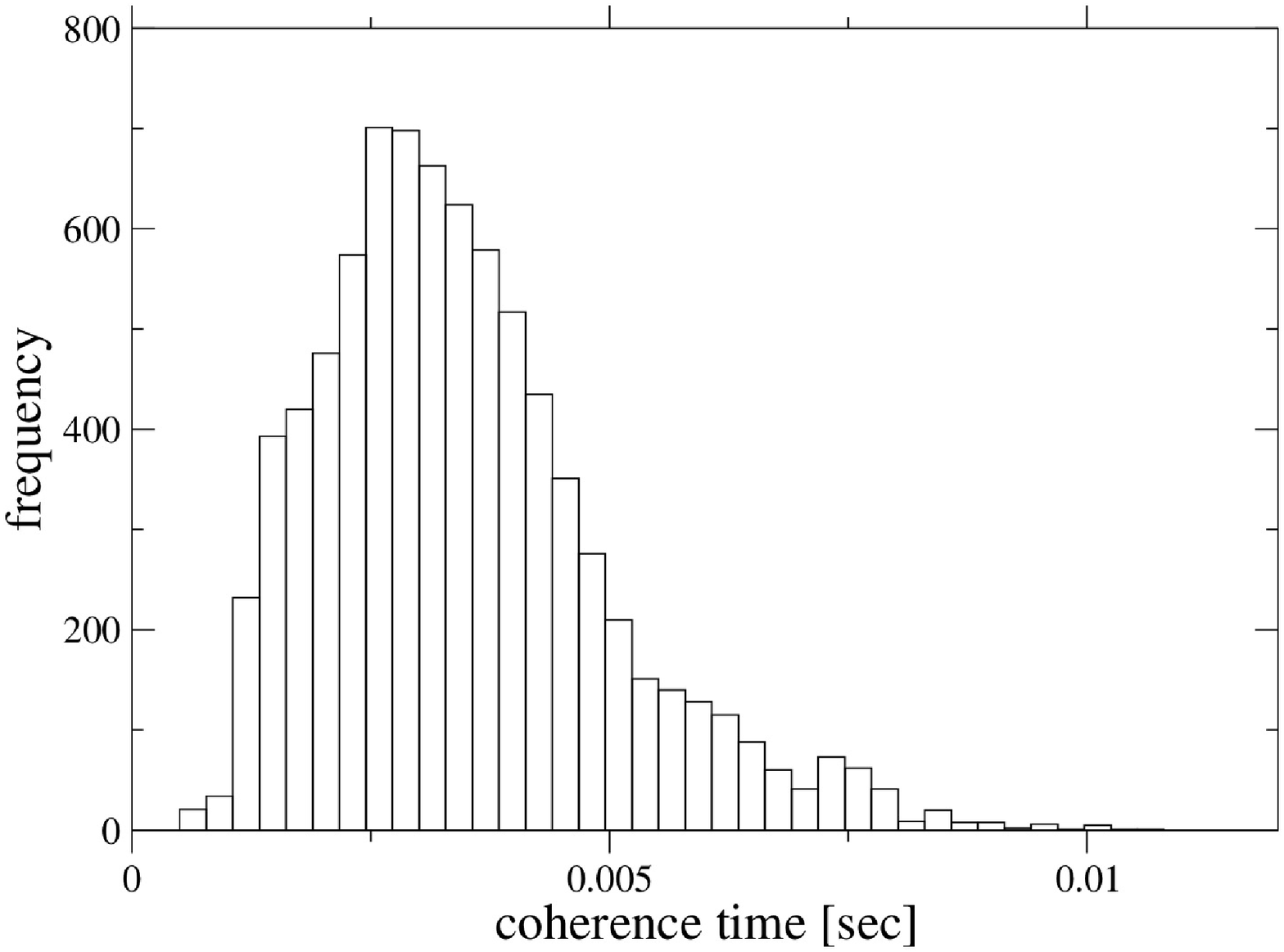}
      \caption[Histograms of the guide star seeing, the Strehl ratio, and the atmospheric coherence time.]{Histograms of the guide star seeing, the Strehl ratio, and the atmospheric coherence time.}
         \label{qualhis}
\end{figure}

\begin{figure}
   \includegraphics[width=7.5cm]{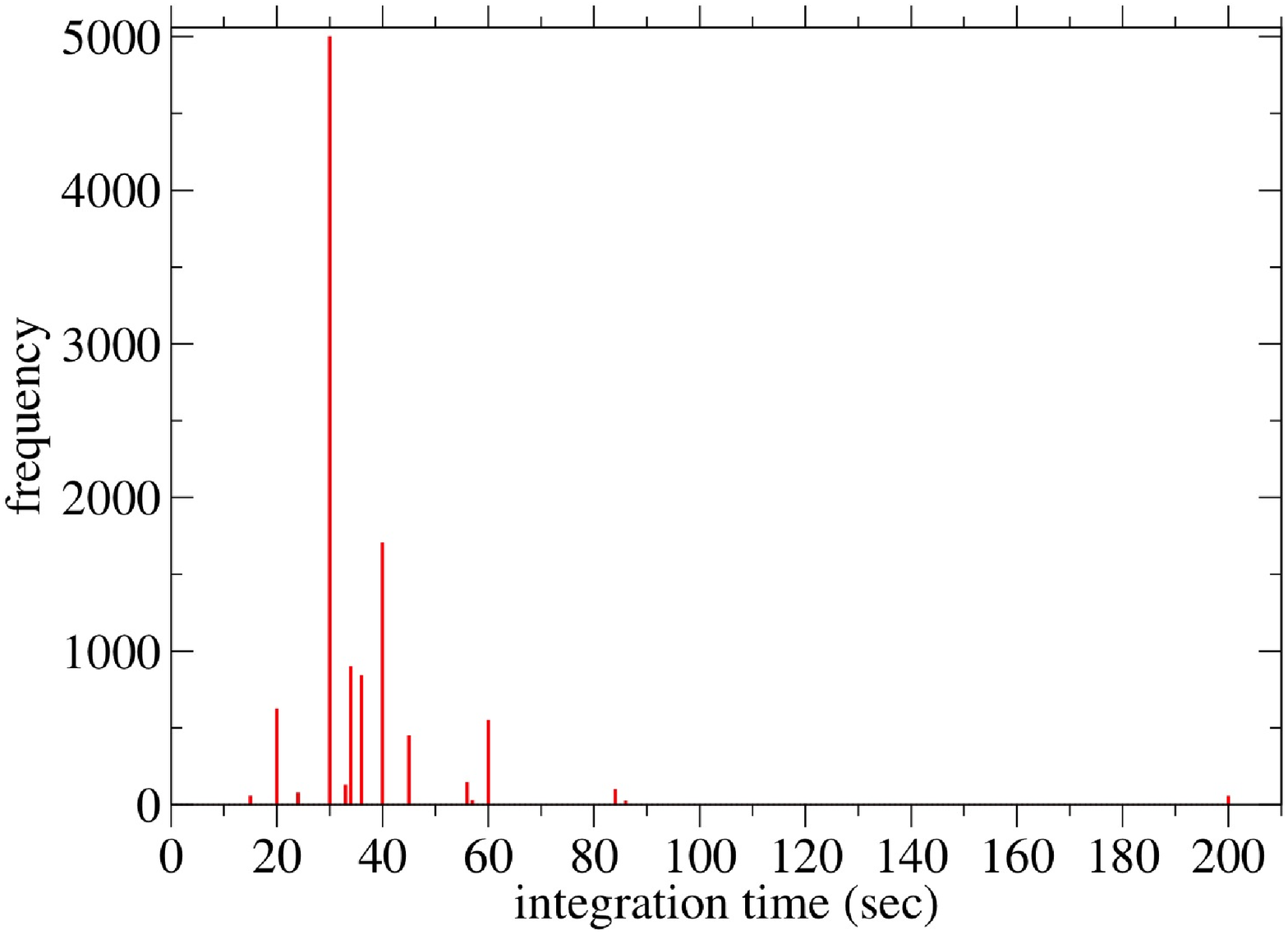}
   \includegraphics[width=7.5cm]{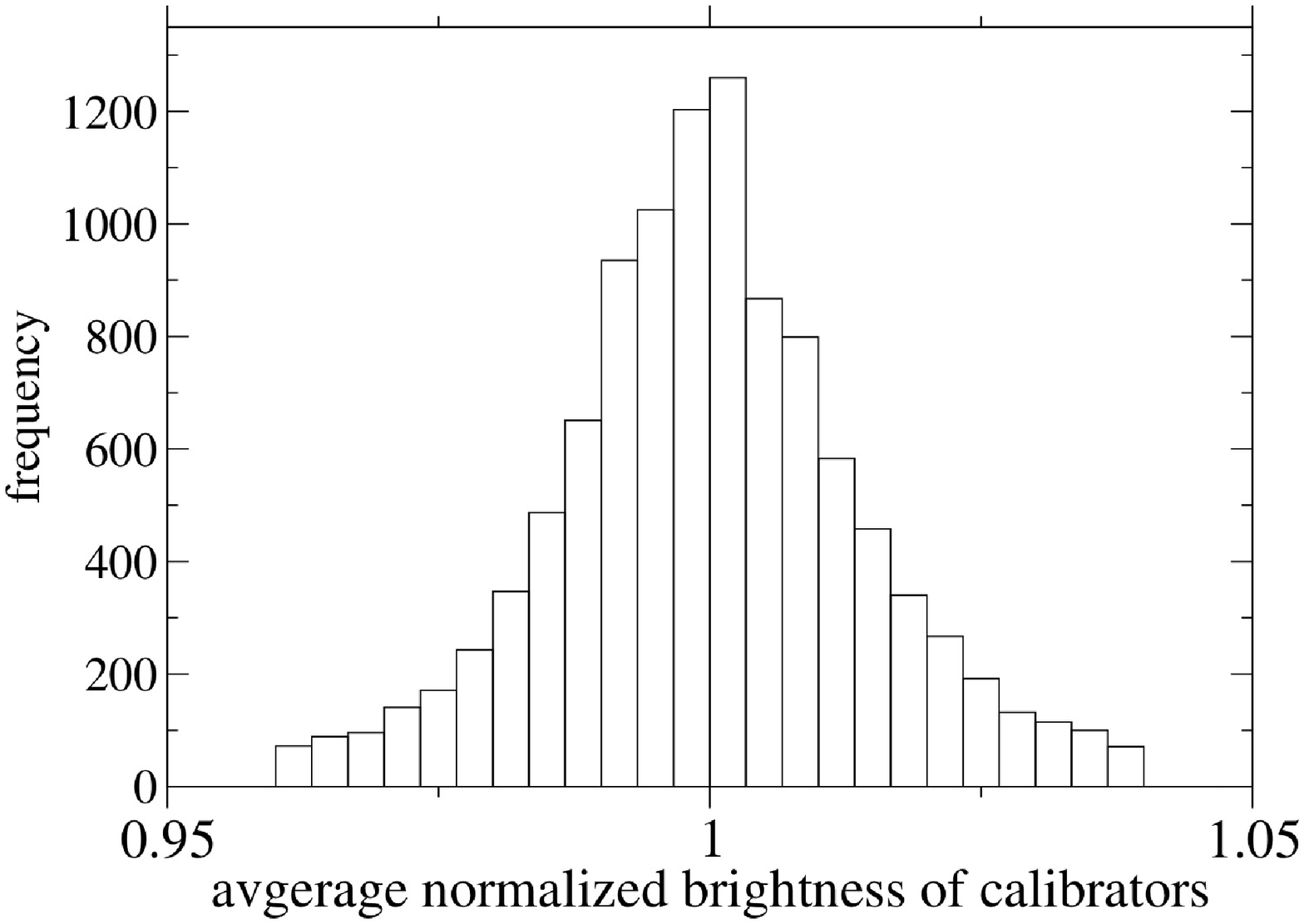}
      \caption[Integration times and average calibrator flux densities.]{The left panel shows the integration times used for our data sample,the right a histogram of the average calibrator flux densities (see description in section~\ref{red}).}
         \label{int}
\end{figure}

\newpage
\clearpage

\section{Light curves}
\label{detlc}

Here we show a more detailed presentation of the data of Figure~\ref{conclc} (blue points). Each box represents a continuous data piece without interruption longer than 30 min. Additionally to the data of Sgr~A* we show the light curves of the calibrator S7 (green), of the average calibrator flux density (red), and the average background apertures (b-apertures in Figure~\ref{calib}, grey). Please note that the $x$-axis is scaled differently, and in the case of the very bright flare of August 5th 2008 the $y$-axis as well. The missing points of S7 are caused by the rejection algorithm described in section~\ref{red}. Since it is very difficult to estimate a reliable error for the individual point due to the changing correction conditions of the AO system and its interplay with the extended background, the confusion and the deconvolution algorithm, we did not include error bars. The point to point scatter of the comparison star and the calibration can serve as an estimate for the individual dataset. The overall error statistics are described in section~\ref{pl}, and in average we find a Gaussian error of about $ \sigma =0.3$ mJy. Furthermore, we present a Table (Tab.~\ref{datasets}) of all datasets included in this analysis with all important information, including average sampling, length of dataset, and maximum flux density.  

\vspace{2cm}

\begin{figure}[h]
   \centering
   \includegraphics[angle=-90, width=15cm]{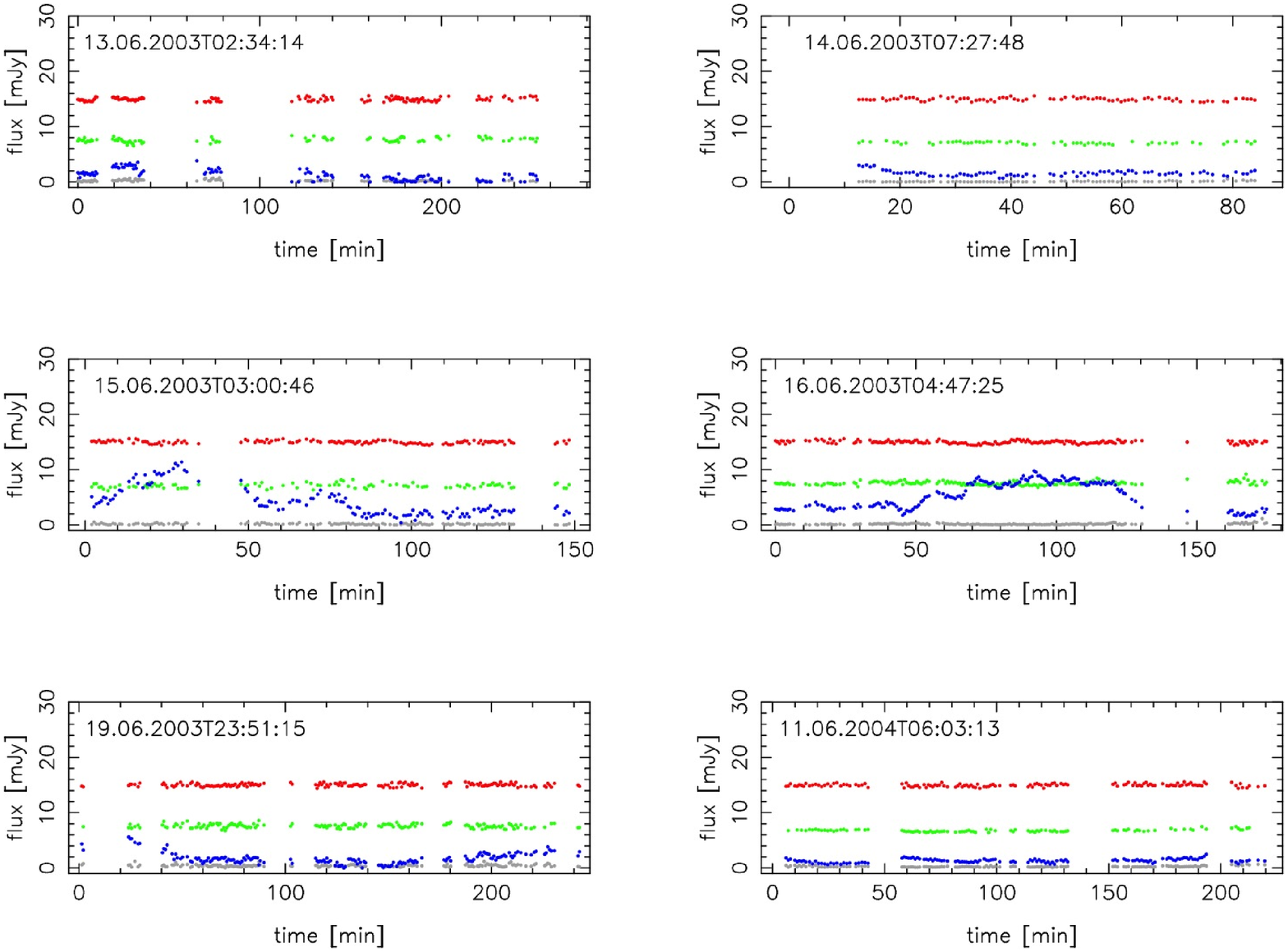}  
\end{figure}

\newpage
\clearpage 

\begin{figure}
   \centering
   \includegraphics[angle=-90, width=15cm]{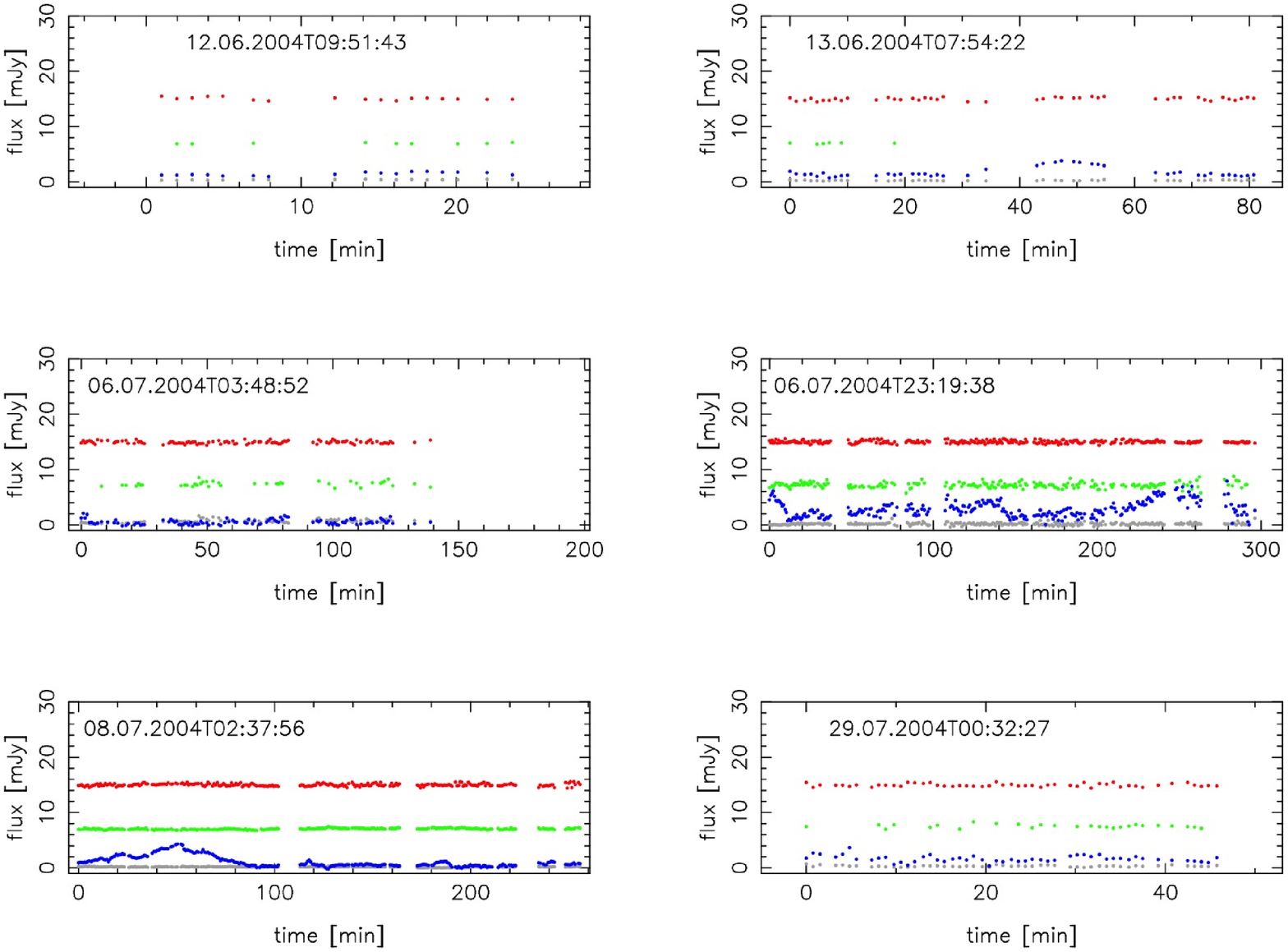}

\end{figure}

\begin{figure}

\vspace{0.5cm}
   \centering
   \includegraphics[angle=-90, width=15cm]{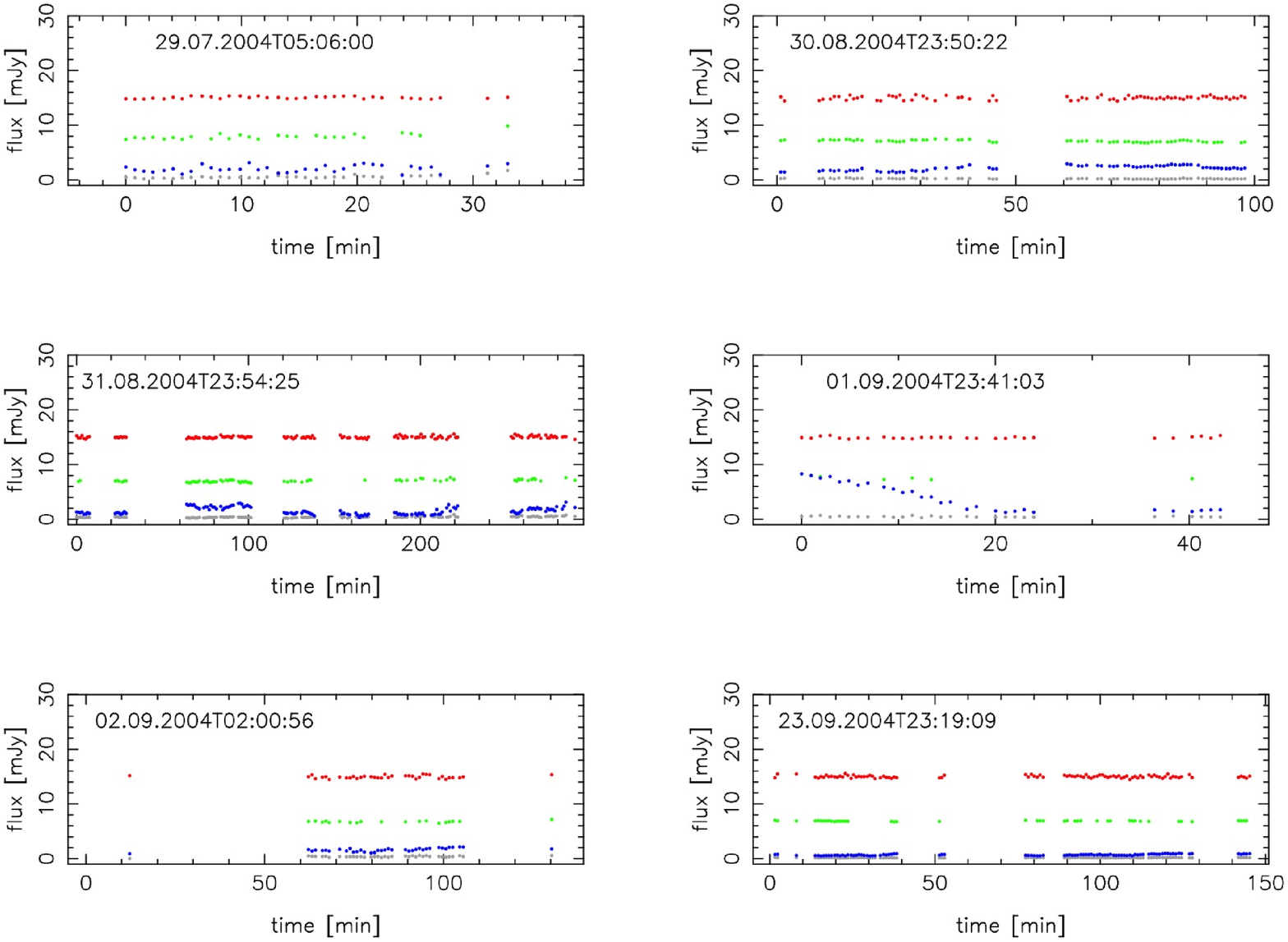}

\end{figure}

\newpage
\clearpage

\begin{figure}
   \centering
   \includegraphics[angle=-90, width=15cm]{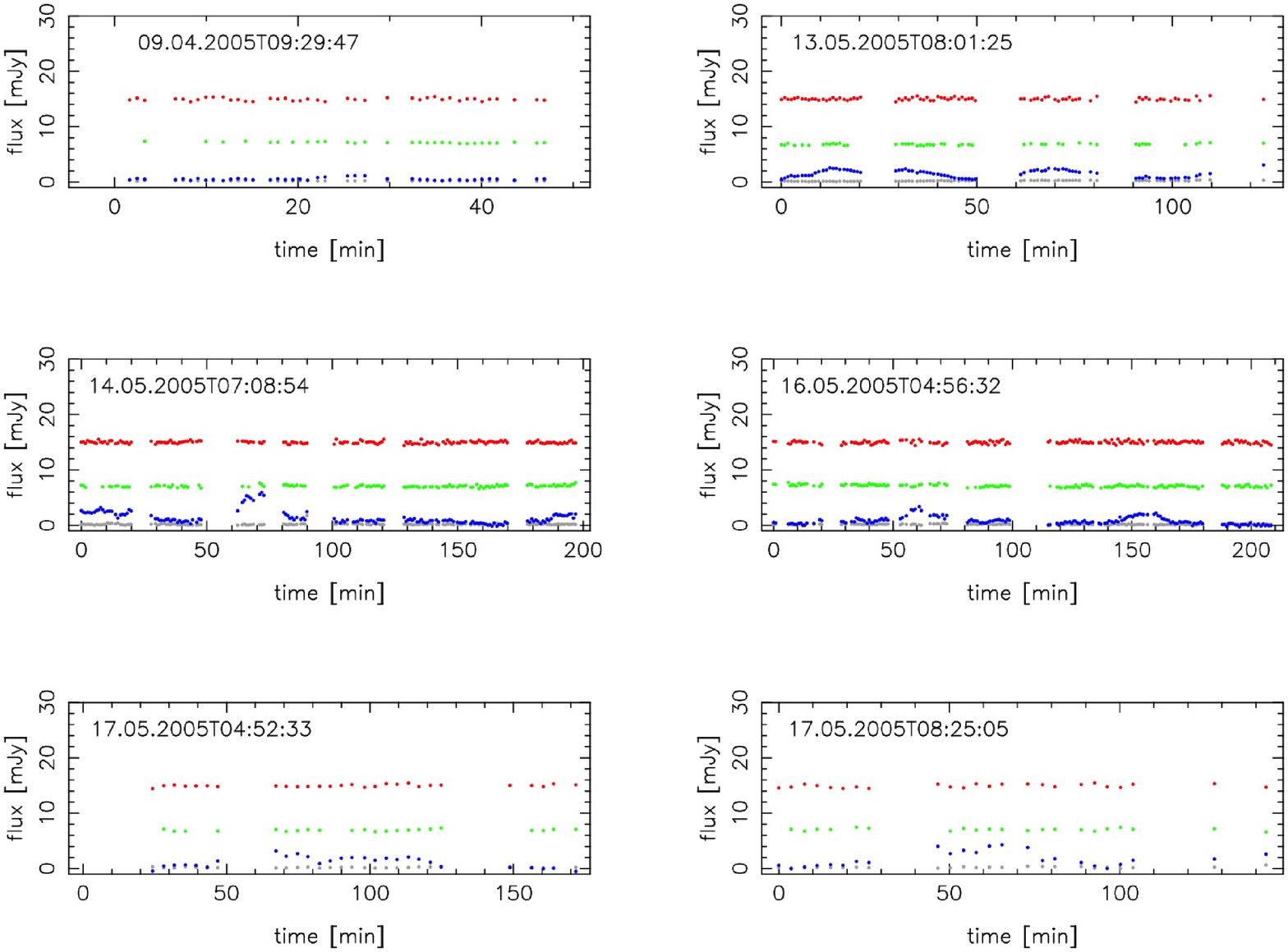}

\end{figure}

\begin{figure}

\vspace{0.5cm}
   \centering
   \includegraphics[angle=-90, width=15cm]{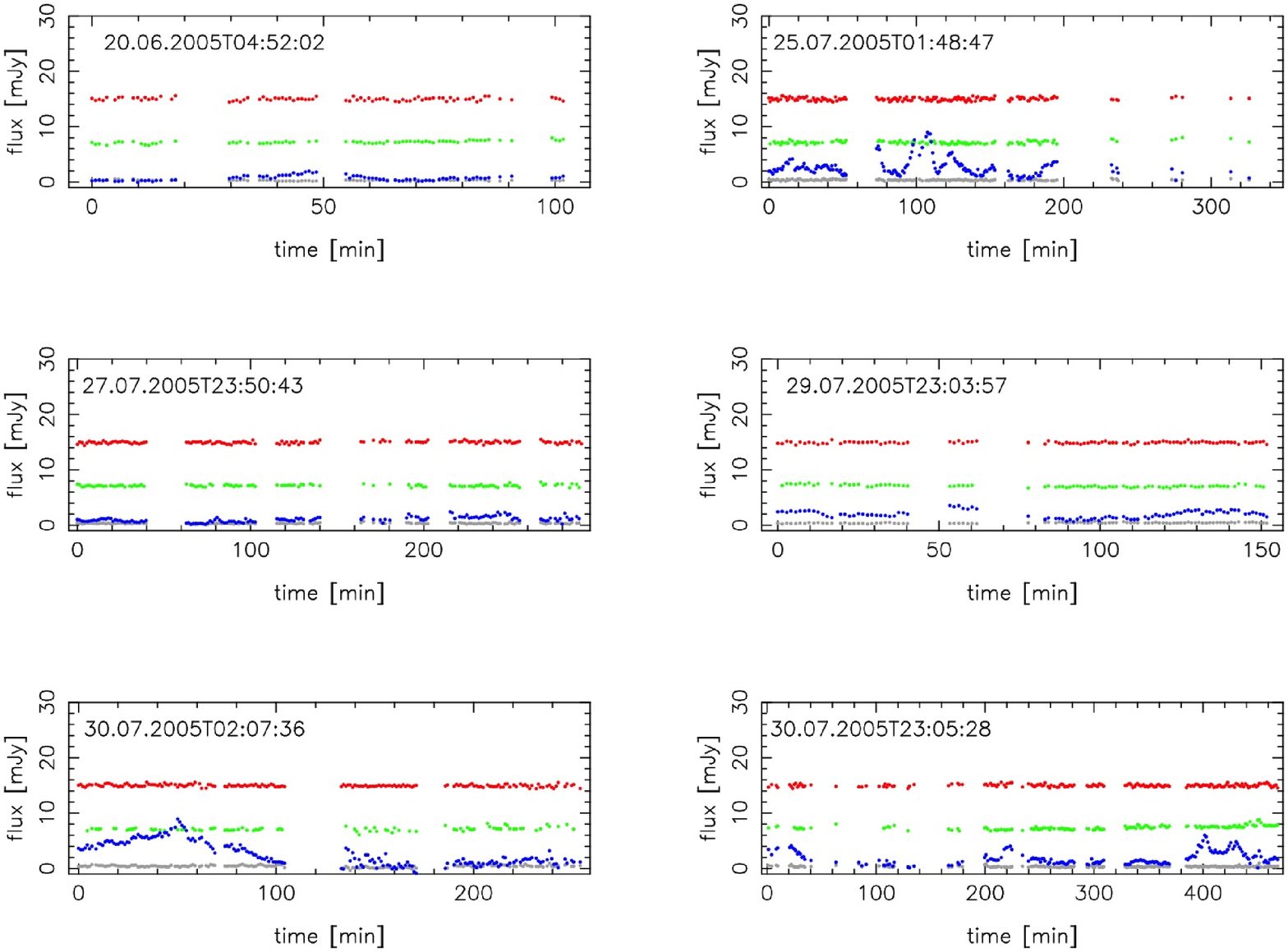}

\end{figure}
\newpage
\clearpage

\begin{figure}
   \centering
   \includegraphics[angle=-90, width=15cm]{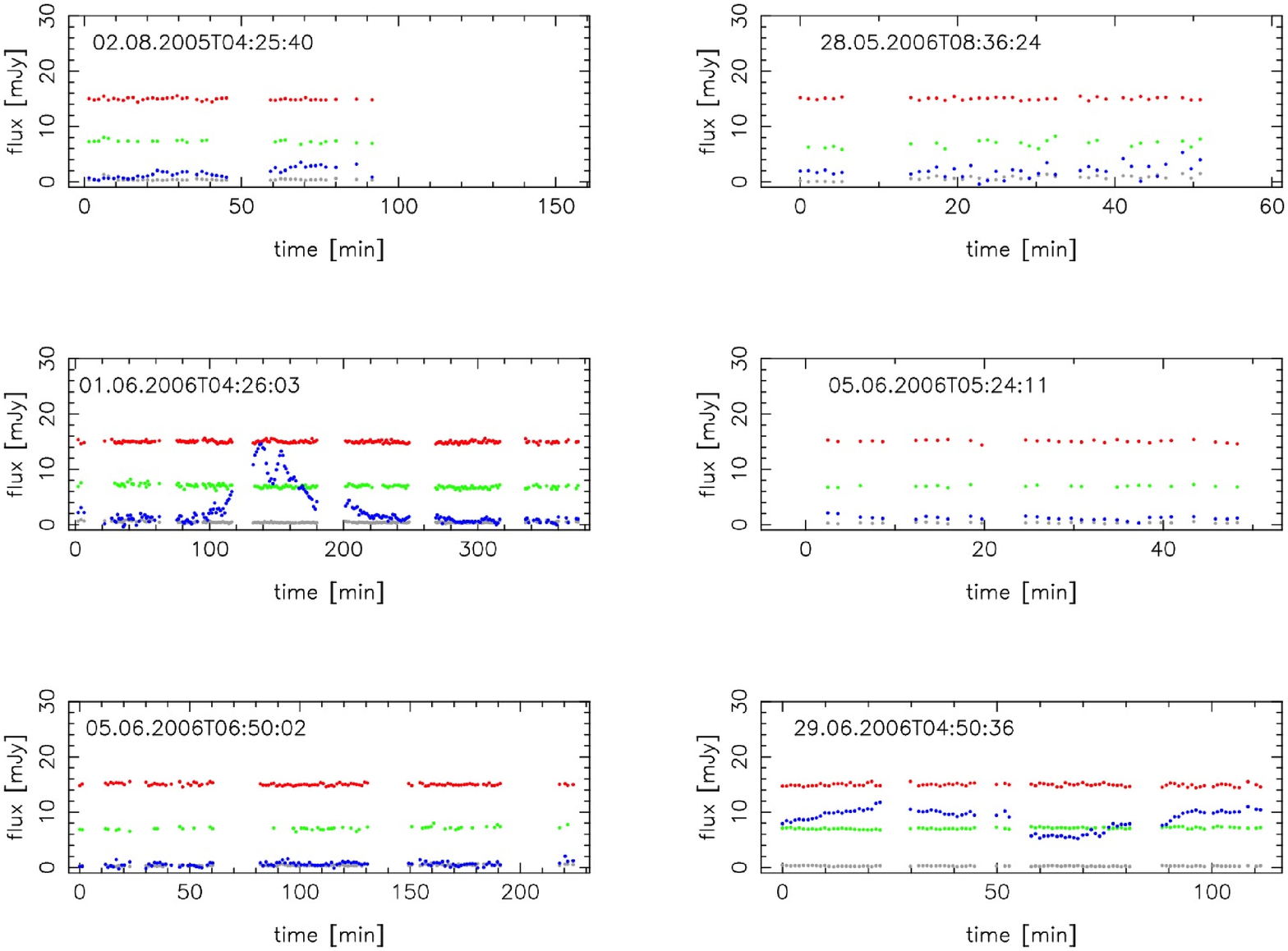}

\end{figure}

\begin{figure}

\vspace{0.5cm}
   \centering
   \includegraphics[angle=-90, width=15cm]{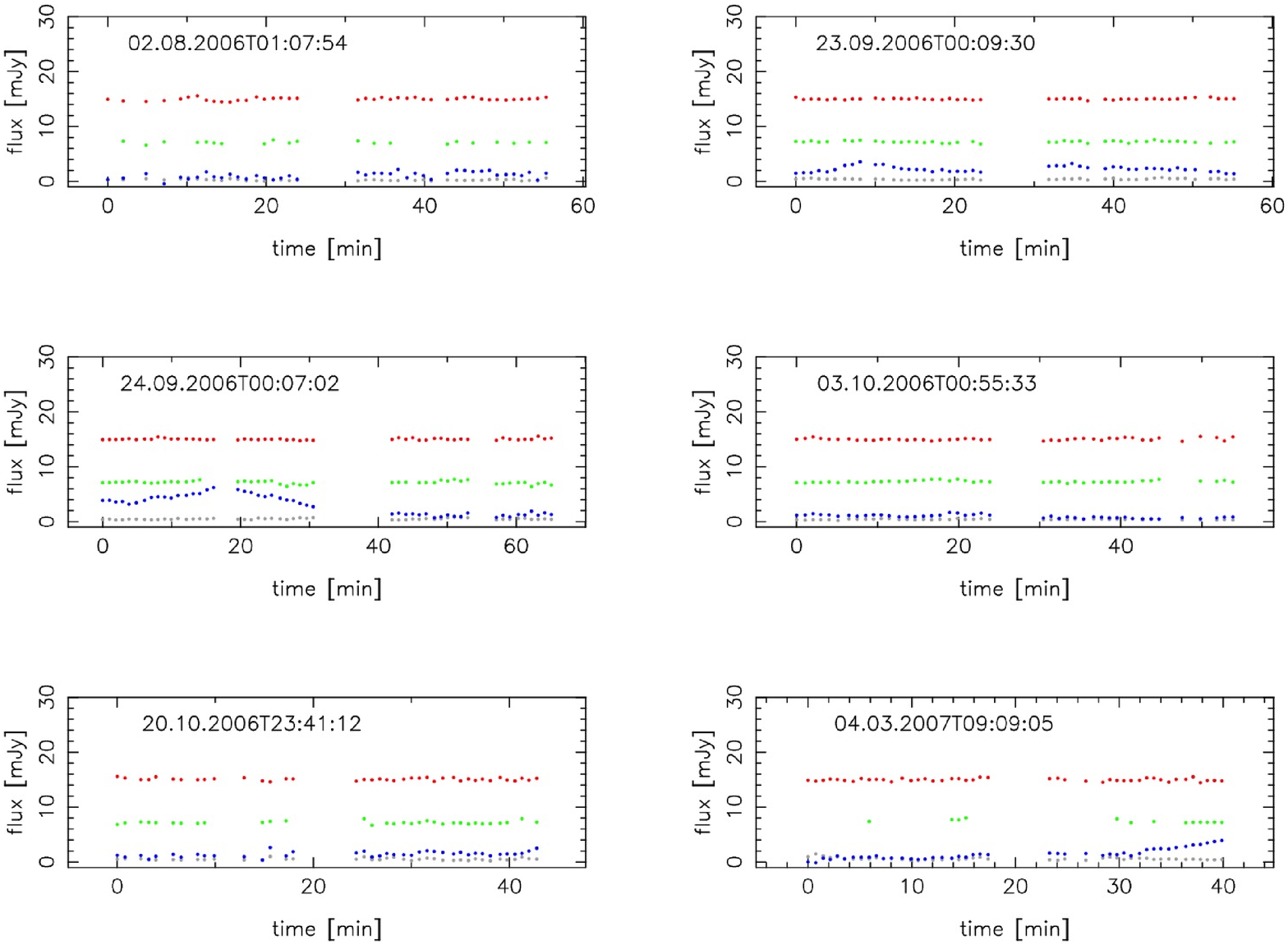}

\end{figure}

\newpage
\clearpage

\begin{figure}
   \centering
   \includegraphics[angle=-90, width=15cm]{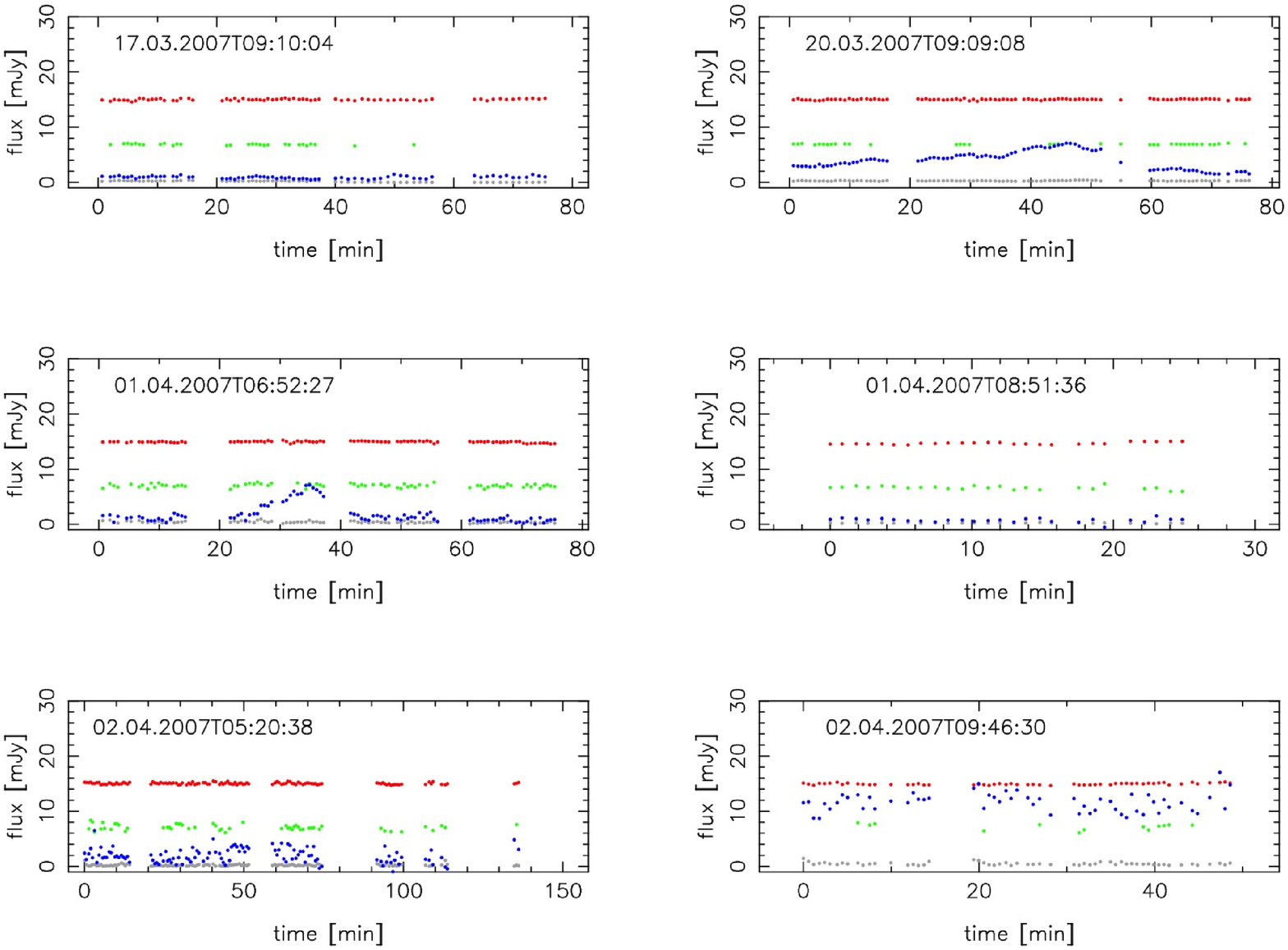}

\end{figure}

\begin{figure}

\vspace{0.5cm}
   \centering
   \includegraphics[angle=-90, width=15cm]{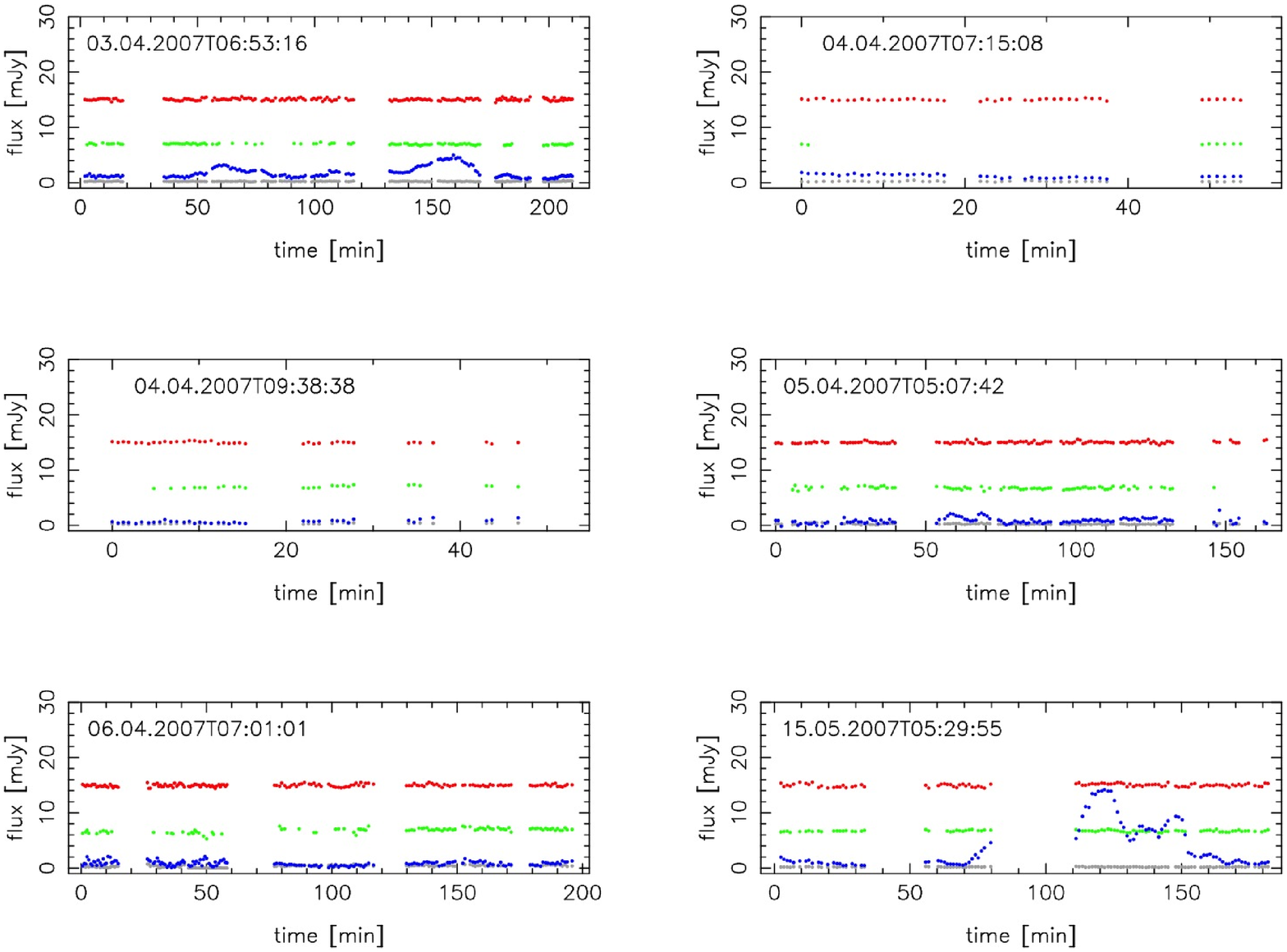}

\end{figure}

\newpage
\clearpage

\begin{figure}
   \centering
   \includegraphics[angle=-90, width=15cm]{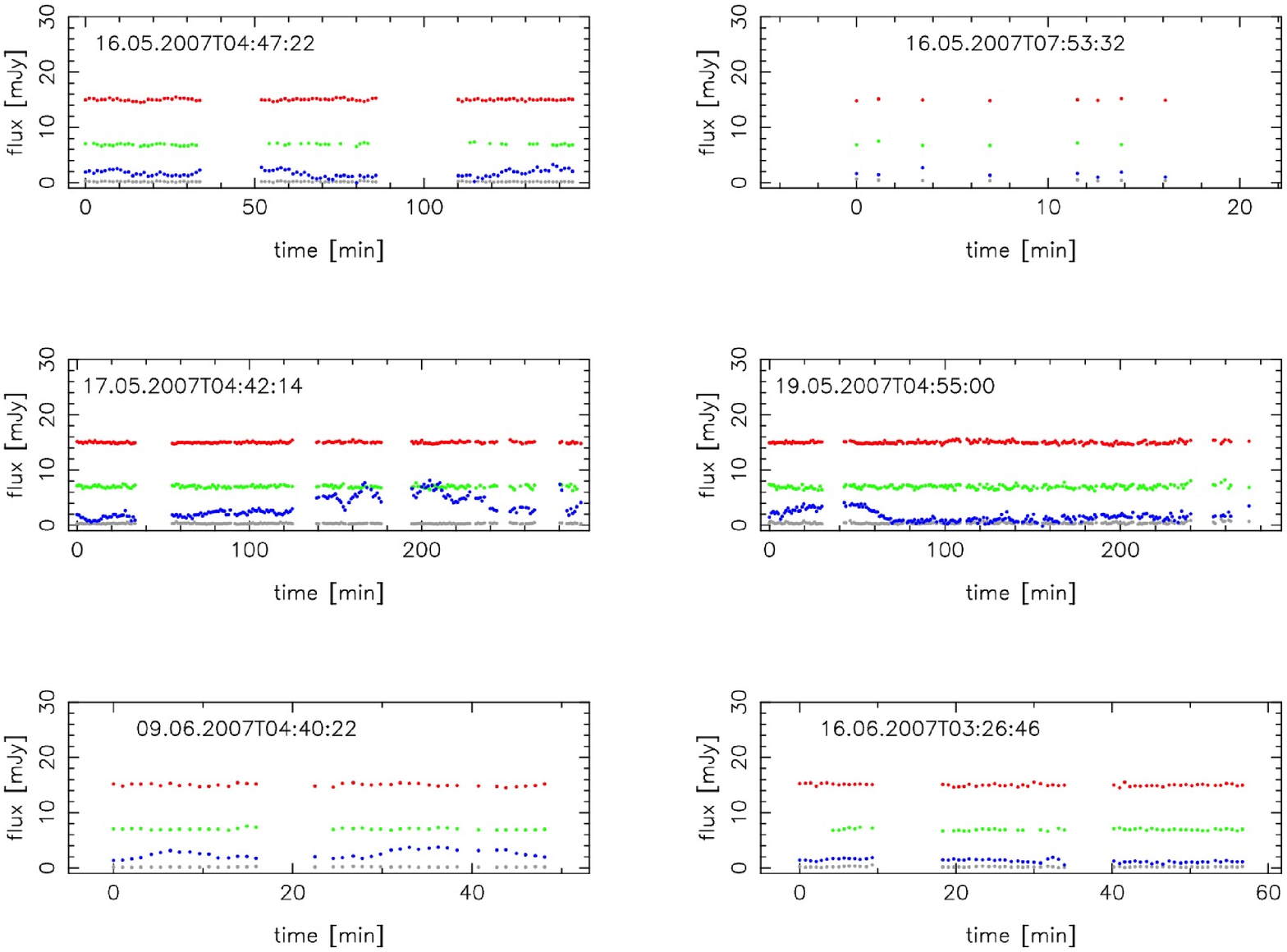}

\end{figure}

\begin{figure}

\vspace{0.5cm}
   \centering
   \includegraphics[angle=-90, width=15cm]{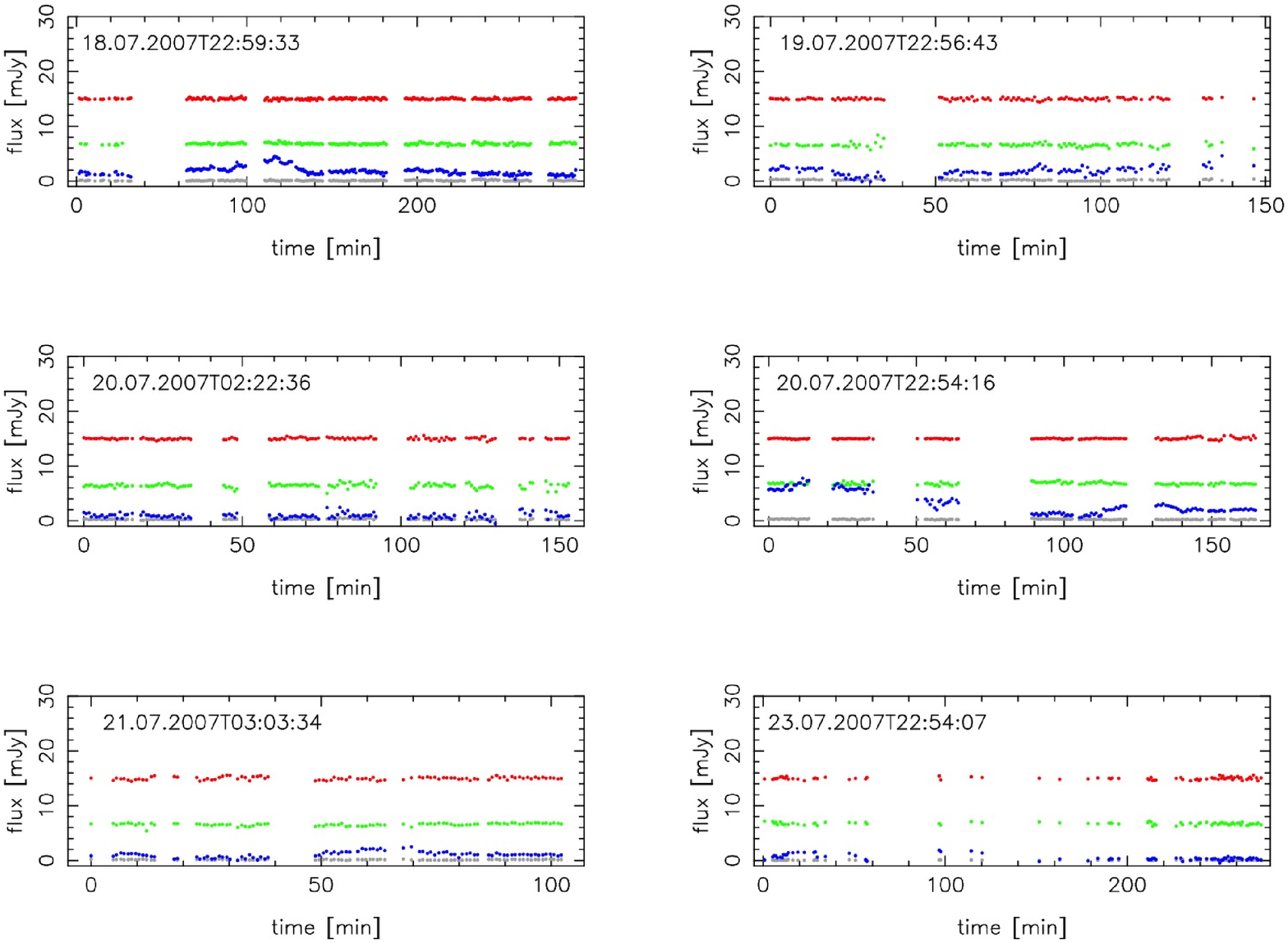}

\end{figure}

\newpage
\clearpage

\begin{figure}
   \centering
   \includegraphics[angle=-90, width=15cm]{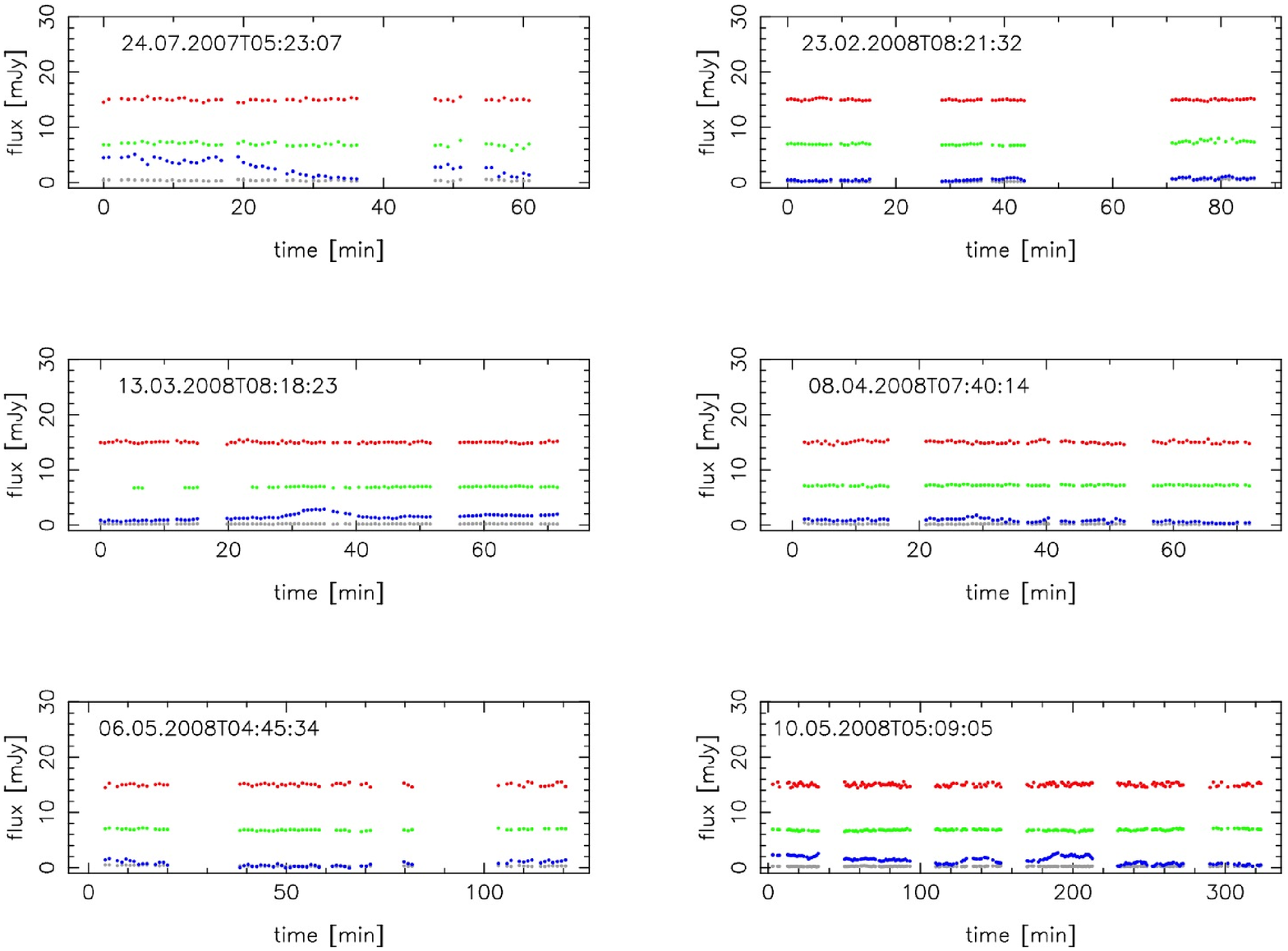}

\end{figure}

\begin{figure}

\vspace{0.5cm}
   \centering
   \includegraphics[angle=-90, width=15cm]{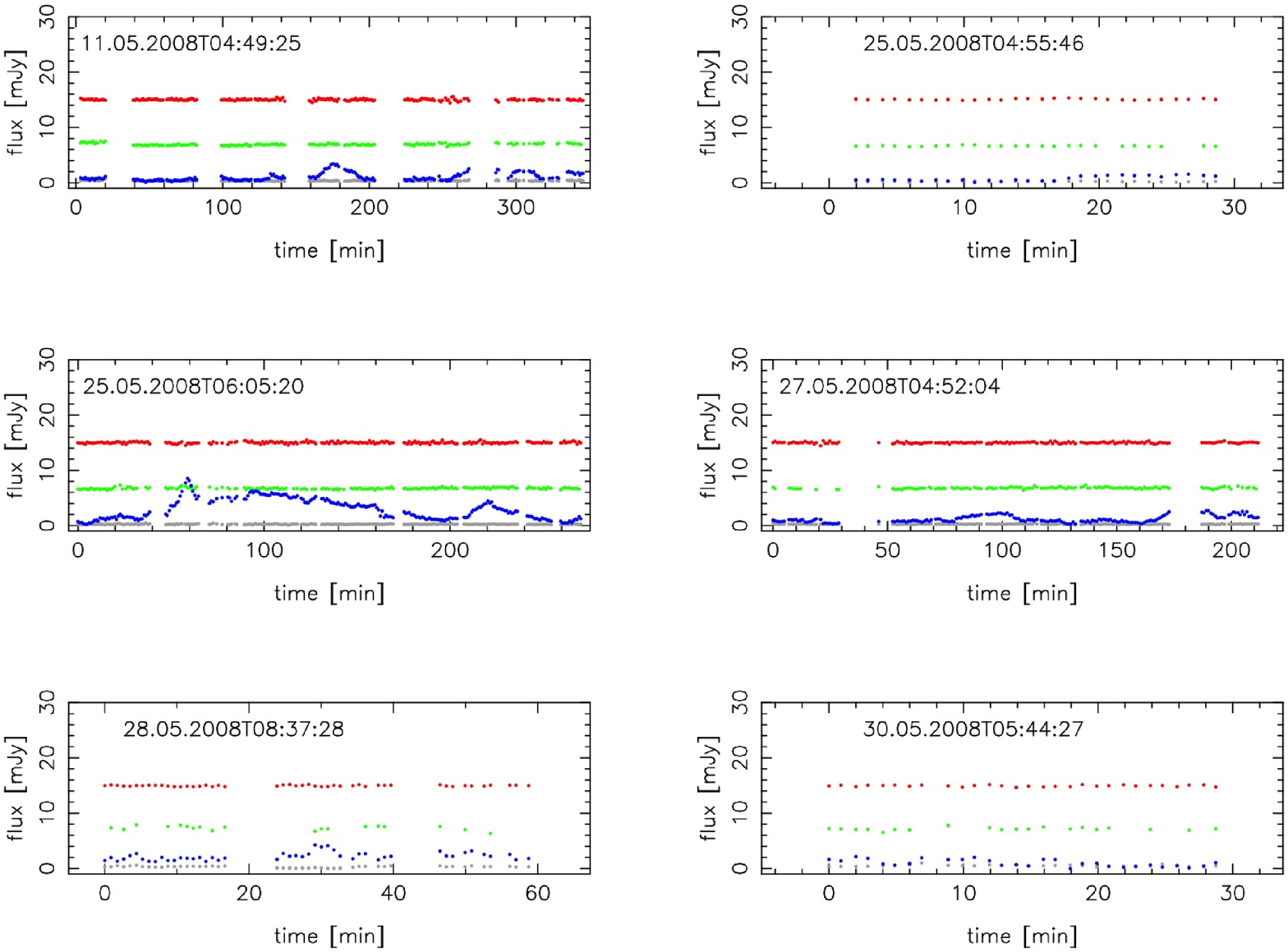}

\end{figure}

\newpage
\clearpage

\begin{figure}
   \centering
   \includegraphics[angle=-90, width=15cm]{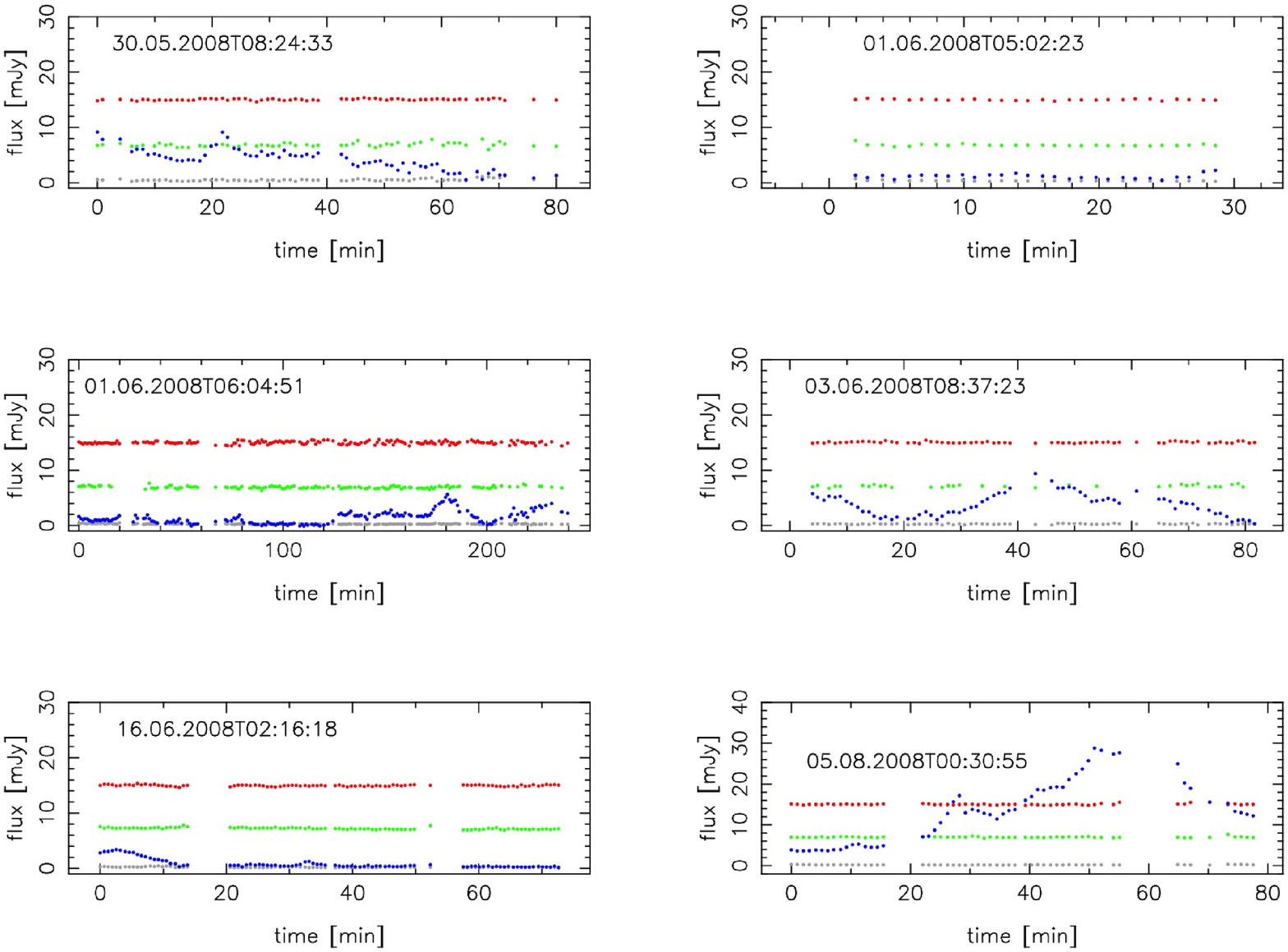}

\end{figure}

\begin{figure}

\vspace{0.5cm}
   \centering
   \includegraphics[angle=-90, width=15cm]{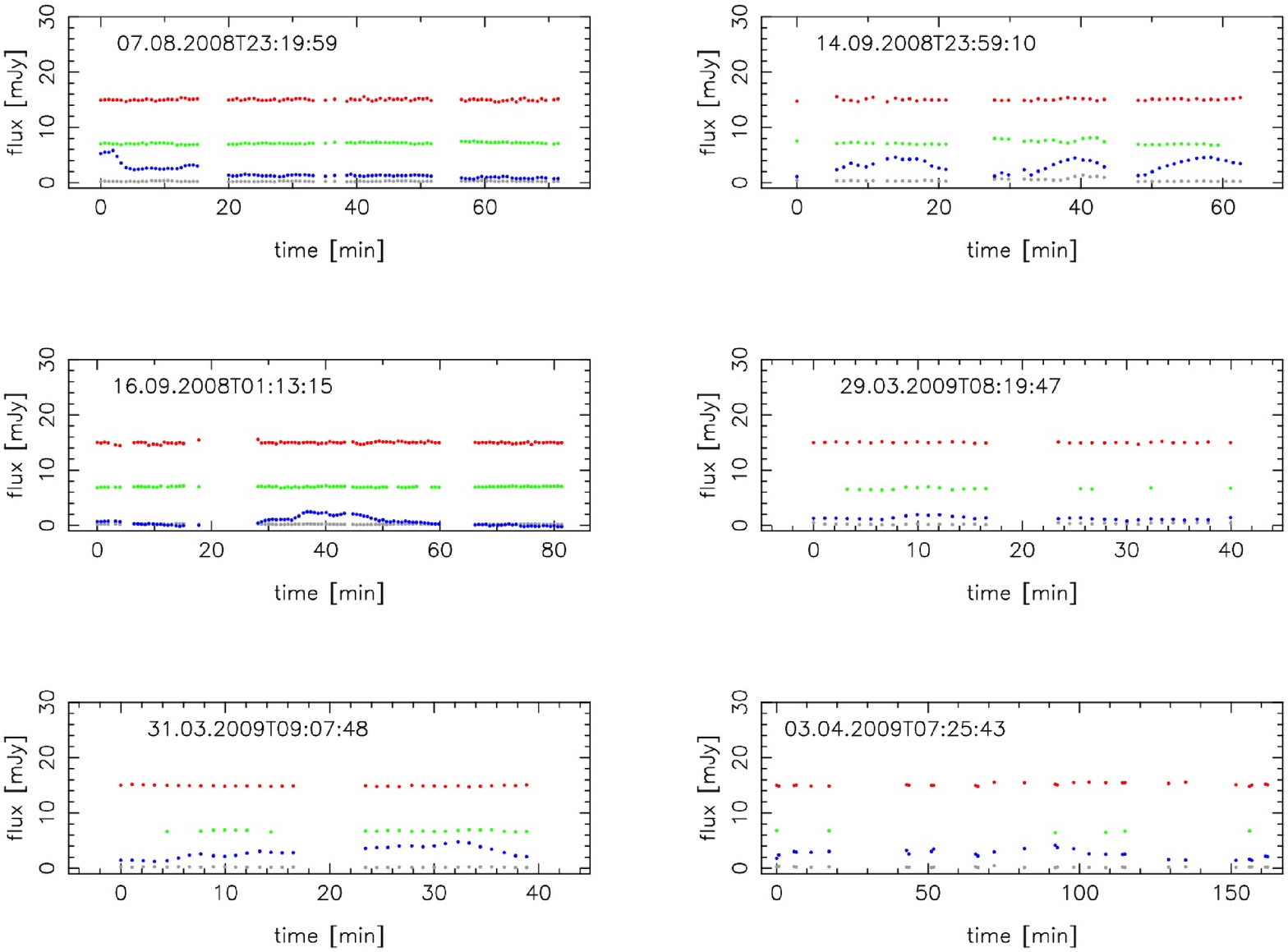}

\end{figure}

\newpage
\clearpage

\begin{figure}
   \centering
   \includegraphics[angle=-90, width=15cm]{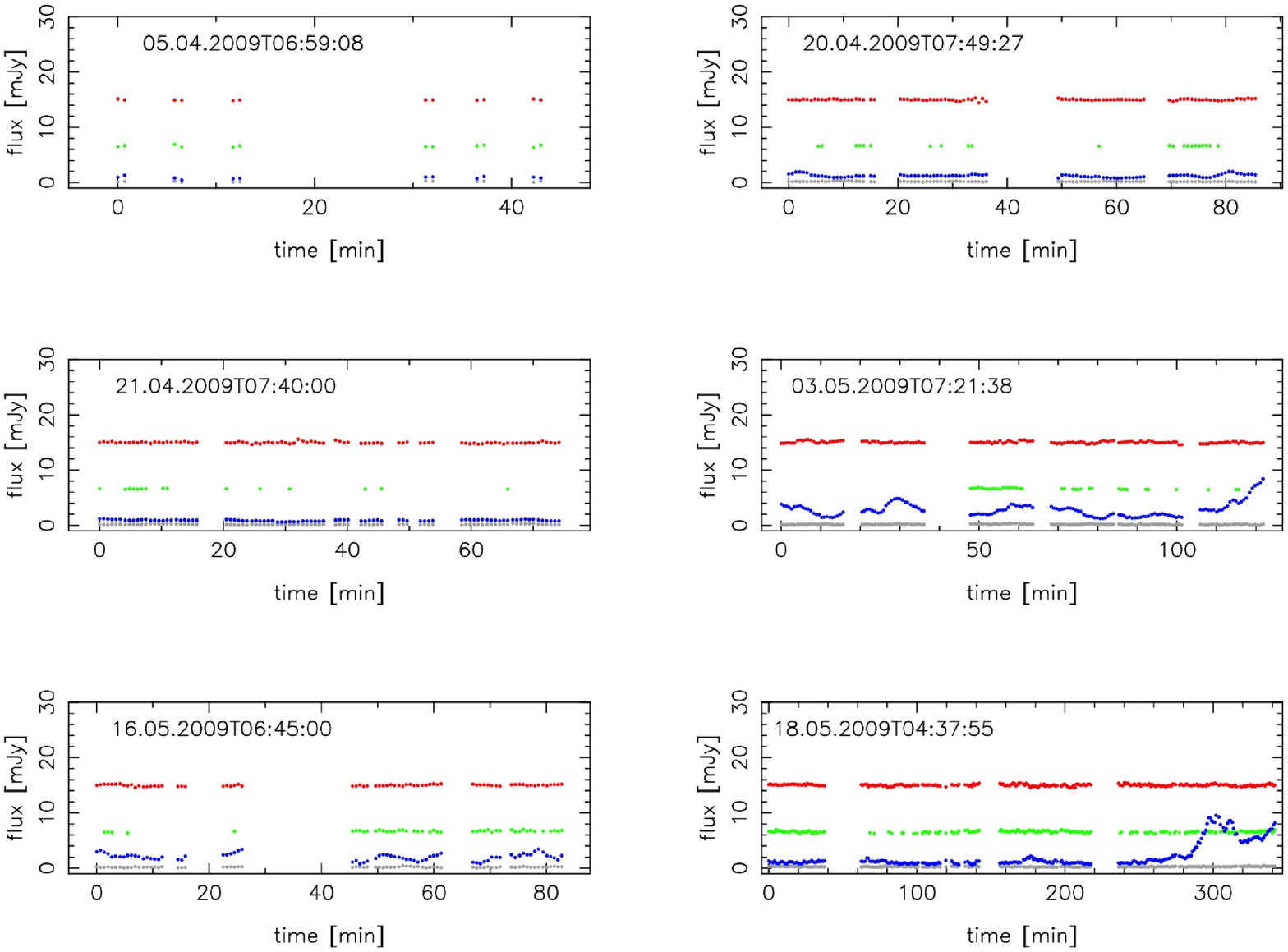}

\end{figure}

\begin{figure}

\vspace{0.5cm}
   \centering
   \includegraphics[angle=-90, width=15cm]{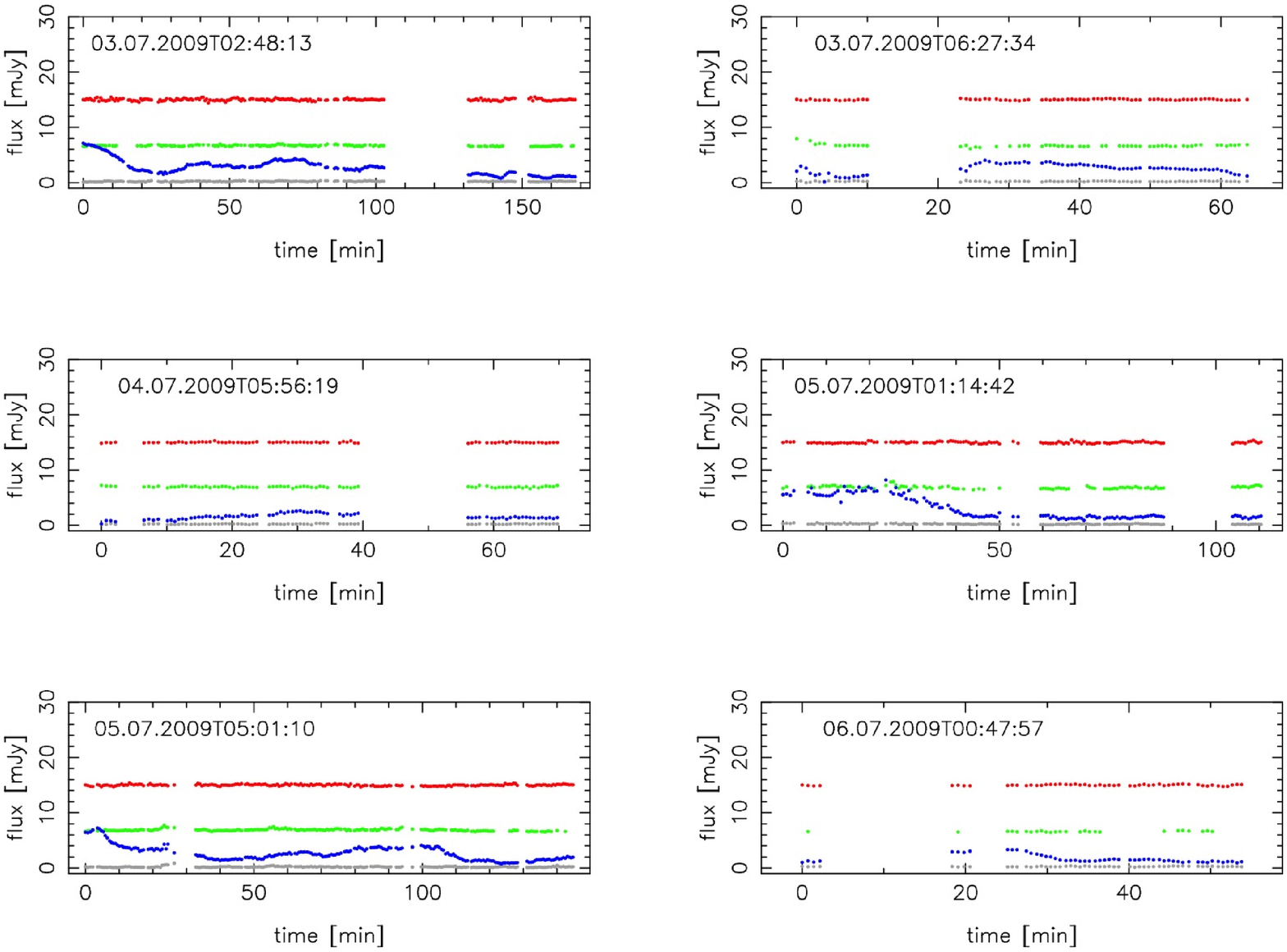}
  
\end{figure}

\newpage
\clearpage

\begin{figure}
   \centering
   \includegraphics[angle=-90, width=15cm]{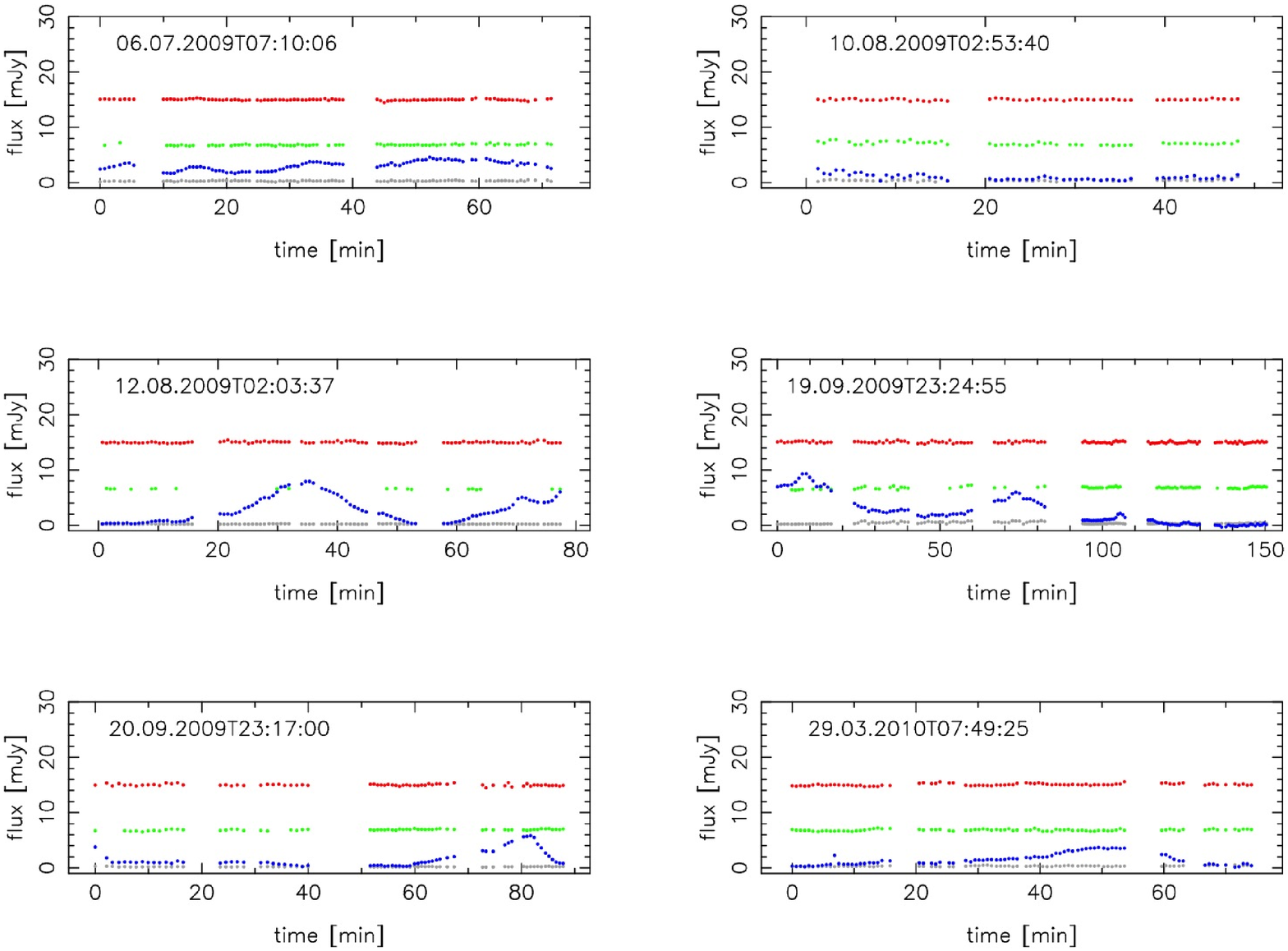}
   
\end{figure}

\begin{figure}

\vspace{0.5cm}
   \centering
   \includegraphics[angle=-90, width=15cm]{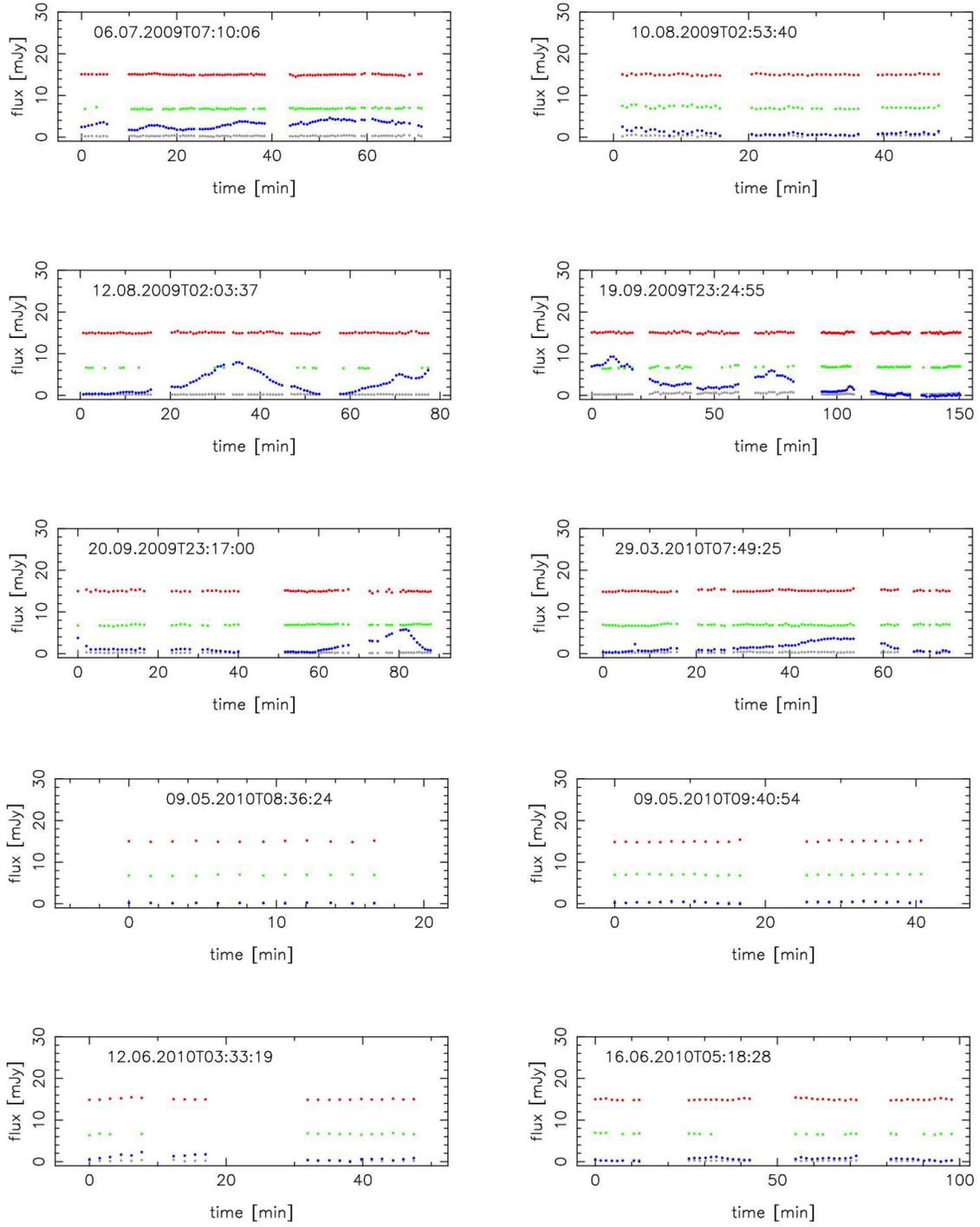}
         \label{lcs}
        \caption[Light curves of Sgr~A*.]{Light curves of Sgr~A* (blue), S7 (green), average calibrator flux density (red), and background.}
\end{figure}

\clearpage
%\LongTables
\begin{landscape}
%\begin{table}%[h!]
\begin{center}
\scriptsize
%\caption{Observations of Sgr~A* in the K-band.}
\label{datasets}
\begin{tabular}{ccllllllll}
%\rotate
\tableline
\tableline
start & stop & length & \# frames & max. flux density & avg. sampl. & integr. time & mode & camera & project ID \\
&&[min]&&[mJy]&[min]&[sec]&&&\\
\tableline
13.06.2003 02:34:14.23 & 13.06.2003 07:10:52.51 & 276.64 & 253 & 3.84 & 1.097 & 15/20 & I/P & S13 & 713-0078(A) \\
14.06.2003 07:27:48.75 & 14.06.2003 08:51:47.09 & 83.97 & 120 & 3.07 & 0.705 & 20 & I & S13 & 713-0078(A) \\
15.06.2003 03:00:46.50 & 15.06.2003 05:30:23.90 & 149.62 & 181 & 11.38 & 0.831 & 20 & I & S13 & 713-0078(A) \\
16.06.2003 04:47:25.19 & 16.06.2003 07:42:52.46 & 175.45 & 208 & 9.80 & 0.847 & 20 & I & S13 & 713-0078(A) \\
19.06.2003 23:51:15.31 & 20.06.2003 03:53:58.14 & 242.71 & 219 & 5.60 & 1.113 & 20 & I & S13 & 271.B-5019(A) \\
11.06.2004 06:03:13.30 & 11.06.2004 09:45:46.88 & 222.56 & 154 & 2.50 & 1.454 & 20 & P & S13 & 073.B-0084(A) \\
12.06.2004 09:51:43.42 & 12.06.2004 10:15:20.04 & 23.61 & 22 & 1.89 & 1.124 & 20 & P & S13 & 073.B-0084(A) \\
13.06.2004 07:54:22.95 & 13.06.2004 09:15:08.79 & 80.76 & 70 & 3.84 & 1.170 & 20 & P & S13 & 073.B-0084(A) \\
06.07.2004 03:48:52.53 & 06.07.2004 07:05:59.31 & 197.11 & 217 & 2.11 & 0.912 & 30 & I & S27 & 073.B-0775(A) \\
06.07.2004 23:19:38.98 & 07.07.2004 04:27:41.56 & 308.04 & 344 & 7.92 & 0.898 & 30 & I & S13 & 073.B-0775(A) \\
08.07.2004 02:37:56.76 & 08.07.2004 06:53:45.70 & 255.82 & 285 & 4.33 & 0.900 & 30 & I & S13 & 073.B-0775(A) \\
29.07.2004 00:32:27.25 & 29.07.2004 01:20:29.53 & 48.04 & 60 & 3.68 & 0.814 & 30 & I & S13 & 273.B.5023(C) \\
29.07.2004 05:06:00.25 & 29.07.2004 05:40:33.66 & 34.56 & 43 & 3.16 & 0.822 & 30 & I & S13 & 273.B.5023(C) \\
30.08.2004 23:50:22.82 & 31.08.2004 01:28:19.90 & 97.95 & 100 & 2.93 & 0.989 & 30 & I & S13 & 073.B-0775(A) \\
31.08.2004 23:54:25.97 & 01.09.2004 04:44:21.31 & 289.92 & 150 & 3.13 & 1.945 & 30 & I(P) & S13 & 073.B-0085(C) \\
01.09.2004 23:41:03.92 & 02.09.2004 00:24:17.26 & 43.22 & 32 & 8.32 & 1.394 & 30 & P & S13 & 073.B-0775(B) \\
02.09.2004 02:00:56.18 & 02.09.2004 04:15:06.65 & 134.17 & 102 & 2.15 & 1.328 & 30 & I/P & S13 & 073.B-0775(B) \\
23.09.2004 23:19:09.64 & 24.09.2004 01:45:11.86 & 146.04 & 115 & 1.00 & 1.281 & 30 & I & S13 & 073.B-0085(C) \\
09.04.2005 09:29:47.38 & 09.04.2005 10:16:38.57 & 46.85 & 53 & 1.14 & 0.901 & 30 & I & S13 & 073.B-0085(I) \\
13.05.2005 08:01:25.07 & 13.05.2005 10:04:44.69 & 123.33 & 108 & 3.09 & 1.152 & 30 & I & S27 & 073.B-0085(E) \\
14.05.2005 07:08:54.21 & 14.05.2005 10:26:43.81 & 197.83 & 176 & 5.97 & 1.130 & 30 & I & S13 & 073.B-0085(D) \\
16.05.2005 04:56:32.18 & 16.05.2005 08:24:54.94 & 208.38 & 191 & 3.36 & 1.096 & 30 & I & S13 & 073.B-0085(D) \\
17.05.2005 04:52:33.23 & 17.05.2005 07:44:17.20 & 171.73 & 34 & 3.19 & 5.204 & 200 & I & S13 & 073.B-0085(D) \\
17.05.2005 08:25:05.68 & 17.05.2005 10:48:01.76 & 142.93 & 30 & 4.31 & 4.928 & 200 & I & S13 & 073.B-0085(D) \\
20.06.2005 04:52:02.95 & 20.06.2005 06:34:37.16 & 102.57 & 100 & 1.94 & 1.036 & 30 & I & S27 & 073.B-0085(F) \\
25.07.2005 01:48:47.44 & 25.07.2005 07:32:33.24 & 343.76 & 330 & 8.98 & 1.044 & 30 & I & S13 & 271.B-5019(A) \\
27.07.2005 23:50:43.95 & 28.07.2005 04:41:49.31 & 291.09 & 158 & 2.41 & 1.854 & 60 & I & S13 & 075.B-0093(C) \\
29.07.2005 23:03:57.80 & 30.07.2005 01:35:41.92 & 151.74 & 101 & 3.55 & 1.517 & 30/60 & I/P & S13 & 075.B-0093(C) \\
30.07.2005 02:07:36.13 & 30.07.2005 06:21:40.41 & 254.07 & 187 & 8.94 & 1.365 & 30 & P & S13 & 075.B-0093(C) \\
30.07.2005 23:05:28.75 & 31.07.2005 06:53:58.74 & 468.50 & 266 & 6.01 & 1.767 & 60 & I & S13 & 075.B-0093(C) \\
02.08.2005 04:25:40.21 & 02.08.2005 07:01:26.26 & 155.77 & 80 & 3.49 & 1.971 & 60 & I & S13 & 075.B-0093(C) \\
28.05.2006 08:36:24.43 & 28.05.2006 09:32:40.35 & 56.27 & 46 & 5.32 & 1.250 & 30 & I & S27 & 077.B-0552(A) \\
01.06.2006 04:26:03.16 & 01.06.2006 10:44:27.63 & 378.41 & 244 & 14.60 & 1.557 & 30 & I/P & S13/S27 & 077.B-0552(A) \\
05.06.2006 05:24:11.62 & 05.06.2006 06:12:28.06 & 48.27 & 40 & 2.12 & 1.237 & 30 & P & S13 & 077.B-0552(A) \\
05.06.2006 06:50:02.57 & 05.06.2006 10:36:32.66 & 226.50 & 126 & 2.05 & 1.812 & 30 & P & S13 & 077.B-0552(A) \\
29.06.2006 04:50:36.43 & 29.06.2006 06:41:55.29 & 111.31 & 96 & 11.79 & 1.171 & 33.4 & I & S13 & 077.B-0014(C) \\
02.08.2006 01:07:54.32 & 02.08.2006 02:03:15.71 & 55.36 & 48 & 2.17 & 1.177 & 33.6 & I & S13 & 077.B-0014(D) \\
23.09.2006 00:09:30.22 & 23.09.2006 01:04:39.36 & 55.15 & 48 & 3.56 & 1.173 & 33.4 & I & S13 & 077.B-0014(F) \\
24.09.2006 00:07:02.00 & 24.09.2006 01:12:07.83 & 65.10 & 53 & 6.20 & 1.251 & 33.4 & I & S13 & 077.B-0014(F) \\
03.10.2006 00:55:33.17 & 03.10.2006 01:49:23.80 & 53.84 & 48 & 1.67 & 1.145 & 33.4 & I & S13 & 077.B-0014(F) \\
20.10.2006 23:41:12.70 & 21.10.2006 00:24:00.13 & 42.79 & 47 & 2.64 & 0.930 & 33.6 & I & S13 & 078.B-0136(A) \\
04.03.2007 09:09:05.79 & 04.03.2007 09:48:57.48 & 39.86 & 48 & 3.91 & 0.848 & 36 & I & S13 & 078.B-0136(B) \\
17.03.2007 09:10:04.97 & 17.03.2007 10:27:46.76 & 77.70 & 78 & 1.45 & 1.009 & 33.6/56 & I & S27 & 078.B-0136(B) \\
20.03.2007 09:09:08.59 & 20.03.2007 10:25:19.74 & 76.19 & 96 & 7.10 & 0.801 & 34.4 & I & S13 & 078.B-0136(B) \\
01.04.2007 06:52:27.86 & 01.04.2007 08:08:27.40 & 75.99 & 96 & 7.16 & 0.799 & 30 & I & S27 & 179.B-0261(A) \\
01.04.2007 08:51:36.27 & 01.04.2007 09:18:19.08 & 26.71 & 30 & 1.55 & 0.921 & 30 & P & S13 & 179.B-0261(A) \\
02.04.2007 05:20:38.21 & 02.04.2007 07:53:36.91 & 152.98 & 150 & 6.52 & 1.026 & 30 & I & S13/S27 & 179.B-0261(A) \\
02.04.2007 09:46:30.23 & 02.04.2007 10:35:40.93 & 49.18 & 72 & 17.03 & 0.692 & 30 & I & S27 & 179.B-0261(A) \\
03.04.2007 06:53:16.77 & 03.04.2007 10:25:34.11 & 212.29 & 188 & 4.98 & 1.135 & 30 & I/P & S13 & 179.B-0261(A) \\
04.04.2007 07:15:08.41 & 04.04.2007 08:08:55.54 & 53.79 & 46 & 1.84 & 1.195 & 30 & P & S13 & 179.B-0261(A) \\
04.04.2007 09:38:38.23 & 04.04.2007 10:28:30.48 & 49.87 & 63 & 1.41 & 0.804 & 30 & I & S13 & 179.B-0261(A) \\
05.04.2007 05:07:42.08 & 05.04.2007 07:51:16.52 & 163.57 & 140 & 2.74 & 1.176 & 30 & P & S13 & 179.B-0261(A) \\
06.04.2007 07:01:01.93 & 06.04.2007 10:18:49.13 & 197.79 & 175 & 2.10 & 1.136 & 30 & I/P & S13/S27 & 179.B-0261(A) \\
15.05.2007 05:29:55.42 & 15.05.2007 08:31:48.45 & 181.88 & 116 & 14.20 & 1.581 & 40 & I & S13 & 079.B-0018(A) \\
16.05.2007 04:47:22.50 & 16.05.2007 07:11:11.68 & 143.82 & 90 & 3.27 & 1.615 & 40 & P & S13 & 079.B-0018(A) \\
16.05.2007 07:53:32.58 & 16.05.2007 08:10:43.02 & 17.17 & 16 & 2.69 & 1.144 & 40 & P & S13 & 079.B-0018(A) \\
17.05.2007 04:42:14.84 & 17.05.2007 09:34:40.15 & 292.42 & 192 & 8.11 & 1.531 & 40 & P & S13 & 079.B-0018(A) \\
\tableline
\end{tabular}
\end{center}
\begin{center}
\scriptsize
\begin{tabular}{ccllllllll}
\tableline
\tableline
start & stop & length & \# frames & max. flux density & avg. sampl. & integr. time & mode & camera & project ID \\
&&[min]&&[mJy]&[min]&[sec]&&&\\
\tableline
19.05.2007 04:55:00.53 & 19.05.2007 09:41:46.46 & 286.77 & 244 & 4.10 & 1.180 & 40 & P & S13 & 079.B-0018(A) \\
09.06.2007 04:40:22.39 & 09.06.2007 05:28:28.78 & 48.11 & 40 & 3.75 & 1.233 & 56 & I & S27 & 179.B-0261(H) \\
16.06.2007 03:26:46.45 & 16.06.2007 04:23:27.76 & 56.69 & 62 & 1.93 & 0.929 & 34.4 & I & S13 & 179.B-0261(H) \\
18.07.2007 22:59:33.75 & 19.07.2007 03:52:43.66 & 293.17 & 260 & 4.52 & 1.131 & 30 & P & S13 & 179.B-0261(D) \\
19.07.2007 22:56:43.85 & 20.07.2007 01:23:08.82 & 146.42 & 138 & 4.61 & 1.068 & 30 & P & S13 & 179.B-0261(D) \\
20.07.2007 02:22:36.82 & 20.07.2007 04:55:26.07 & 152.82 & 140 & 2.44 & 1.099 & 30 & P & S13 & 179.B-0261(D) \\
20.07.2007 22:54:16.15 & 21.07.2007 01:39:04.00 & 164.80 & 140 & 7.75 & 1.185 & 24/30 & P & S13 & 179.B-0261(D) \\
21.07.2007 03:03:34.27 & 21.07.2007 04:45:44.95 & 102.18 & 102 & 2.50 & 1.011 & 30 & P & S13 & 179.B-0261(D) \\
23.07.2007 22:54:07.39 & 24.07.2007 03:27:34.93 & 273.46 & 248 & 1.90 & 1.107 & 30 & P & S13 & 179.B-0261(D) \\
24.07.2007 05:23:07.91 & 24.07.2007 06:27:33.88 & 64.43 & 60 & 5.12 & 1.092 & 30 & P & S13 & 179.B-0261(D) \\
23.02.2008 08:21:32.28 & 23.02.2008 09:47:38.65 & 86.11 & 72 & 1.19 & 1.212 & 34.4 & I & S13 & 179.B-0261(L) \\
13.03.2008 08:18:23.01 & 13.03.2008 09:29:52.25 & 71.49 & 96 & 2.89 & 0.752 & 34.4 & I & S13 & 179.B-0261(L) \\
08.04.2008 07:40:14.17 & 08.04.2008 08:52:12.72 & 71.98 & 96 & 1.79 & 0.757 & 34 & I & S13 & 179.B-0261(M) \\
06.05.2008 04:45:34.40 & 06.05.2008 06:47:08.09 & 121.56 & 80 & 1.64 & 1.538 & 45 & P & S13 & 179.B-0261(P) \\
10.05.2008 05:09:05.13 & 10.05.2008 10:40:50.62 & 331.76 & 224 & 2.71 & 1.487 & 45 & P & S13 & 179.B-0261(P) \\
11.05.2008 04:49:25.76 & 11.05.2008 10:35:01.18 & 345.59 & 232 & 3.39 & 1.496 & 45 & P & S13 & 179.B-0261(P) \\
25.05.2008 04:55:46.22 & 25.05.2008 05:24:22.93 & 28.61 & 30 & 1.54 & 0.986 & 40 & P & S13 & 081.B-0648(A) \\
25.05.2008 06:05:20.32 & 25.05.2008 10:35:38.65 & 270.31 & 250 & 8.51 & 1.085 & 40 & P & S13 & 081.B-0648(A) \\
26.05.2008 04:53:25.66 & 26.05.2008 05:21:58.26 & 28.54 & 30 & 0.00 & 0.984 & 40 & P & S13 & 081.B-0648(A) \\
27.05.2008 04:52:04.92 & 27.05.2008 08:29:38.07 & 217.55 & 184 & 2.74 & 1.188 & 40 & P & S13 & 081.B-0648(A) \\
28.05.2008 08:37:28.23 & 28.05.2008 09:39:43.95 & 62.26 & 58 & 4.26 & 1.092 & 40 & I & S27 & 081.B-0648(A) \\
30.05.2008 05:44:27.73 & 30.05.2008 06:13:11.96 & 28.74 & 30 & 2.12 & 0.990 & 40 & P & S13 & 081.B-0648(A) \\
30.05.2008 08:24:33.51 & 30.05.2008 09:45:25.69 & 80.87 & 80 & 9.18 & 1.023 & 40 & P & S13 & 081.B-0648(A) \\
01.06.2008 05:02:23.63 & 01.06.2008 05:31:01.00 & 28.62 & 30 & 2.24 & 0.986 & 40 & P & S13 & 081.B-0648(A) \\
01.06.2008 06:04:51.56 & 01.06.2008 10:10:26.78 & 245.59 & 240 & 5.62 & 1.027 & 40 & P & S13 & 081.B-0648(A) \\
03.06.2008 08:37:23.56 & 03.06.2008 09:58:58.85 & 81.59 & 80 & 9.36 & 1.032 & 40 & P & S13 & 081.B-0648(A) \\
16.06.2008 02:16:18.38 & 16.06.2008 03:28:57.79 & 72.66 & 96 & 3.39 & 0.764 & 34.4 & I & S13 & 179.B-0261(T) \\
05.08.2008 00:30:55.22 & 05.08.2008 01:48:26.93 & 77.53 & 64 & 28.82 & 1.230 & 56/57 & I & S27 & 179.B-0261(N) \\
07.08.2008 23:19:59.48 & 08.08.2008 00:31:25.59 & 71.44 & 96 & 5.81 & 0.751 & 34.4 & I & S13 & 179.B-0261(N) \\
14.09.2008 23:59:10.24 & 15.09.2008 01:02:36.99 & 63.45 & 49 & 4.59 & 1.321 & 56 & I & S27 & 179.B-0261(U) \\
16.09.2008 01:13:15.06 & 16.09.2008 02:34:32.07 & 81.28 & 103 & 2.47 & 0.796 & 57/34 & I & S13 & 179.B-0261(U) \\
29.03.2009 08:19:47.54 & 29.03.2009 08:59:45.50 & 39.97 & 32 & 1.96 & 1.289 & 60 & I & S27 & 179.B-0261(X) \\
31.03.2009 09:07:48.95 & 31.03.2009 09:47:44.71 & 39.93 & 32 & 4.77 & 1.288 & 60 & I & S27 & 179.B-0261(X) \\
03.04.2009 07:25:43.81 & 03.04.2009 10:07:48.85 & 162.08 & 42 & 4.16 & 3.953 & 30 & I & S27 & 082.B-0952(A) \\
05.04.2009 06:59:08.13 & 05.04.2009 07:42:07.38 & 42.99 & 12 & 1.33 & 3.907 & 30 & I & S27 & 082.B-0952(A) \\
20.04.2009 07:49:27.08 & 20.04.2009 09:14:51.76 & 85.41 & 96 & 1.99 & 0.899 & 36 & I & S13 & 178.B-0261(W) \\
21.04.2009 07:40:00.63 & 21.04.2009 08:54:11.99 & 74.19 & 96 & 1.25 & 0.780 & 36 & I & S13 & 178.B-0261(W) \\
03.05.2009 07:21:38.94 & 03.05.2009 09:23:23.02 & 121.73 & 144 & 8.41 & 0.851 & 36 & I & S13 & 183.B-0100(G) \\
16.05.2009 06:45:00.27 & 16.05.2009 08:07:48.12 & 82.80 & 78 & 3.46 & 1.075 & 36 & I & S13 & 183.B-0100(G) \\
18.05.2009 04:37:55.08 & 18.05.2009 10:19:54.10 & 341.98 & 286 & 9.54 & 1.199 & 40 & P & S13 & 083.B-0331(A) \\
03.07.2009 02:48:13.57 & 03.07.2009 05:36:24.08 & 168.18 & 216 & 7.13 & 0.782 & 30/36 & I & S13 & 183.B-0100(D) \\
03.07.2009 06:27:34.75 & 03.07.2009 07:31:17.42 & 63.71 & 80 & 4.04 & 0.806 & 30 & I & S13 & 183.B-0100(D) \\
04.07.2009 05:56:19.24 & 04.07.2009 07:06:02.72 & 69.72 & 80 & 2.66 & 0.882 & 30 & I & S13 & 183.B-0100(D) \\
05.07.2009 01:14:42.14 & 05.07.2009 03:05:06.21 & 110.40 & 139 & 8.20 & 0.800 & 30 & I & S13 & 183.B-0100(D) \\
05.07.2009 05:01:10.12 & 05.07.2009 07:25:56.07 & 144.77 & 224 & 7.22 & 0.649 & 30 & I & S13 & 183.B-0100(D) \\
06.07.2009 00:47:57.05 & 06.07.2009 01:41:45.51 & 53.81 & 56 & 3.35 & 0.978 & 30 & I & S13 & 183.B-0100(D) \\
06.07.2009 07:10:06.50 & 06.07.2009 08:22:39.60 & 72.55 & 104 & 4.59 & 0.704 & 30 & I & S13 & 183.B-0100(D) \\
10.08.2009 02:53:40.41 & 10.08.2009 03:41:47.05 & 48.11 & 62 & 2.51 & 0.788 & 36 & I & S13 & 183.B-0100(I) \\
12.08.2009 02:03:37.66 & 12.08.2009 03:20:56.96 & 77.32 & 101 & 7.92 & 0.773 & 36 & I & S13 & 183.B-0100(I) \\
19.09.2009 23:24:55.71 & 20.09.2009 01:55:16.81 & 150.35 & 132 & 9.30 & 1.147 & 60 & I & S13/S27 & 183.B-0100(J) \\
20.09.2009 23:17:00.08 & 21.09.2009 00:44:50.73 & 87.84 & 80 & 5.83 & 1.111 & 60 & I & S13/S27 & 183.B-0100(J) \\
29.03.2010 07:49:25.57 & 29.03.2010 09:03:33.59 & 74.13 & 96 & 3.70 & 0.780 & 36 & I & S13 & 183.B-0100(L) \\
09.05.2010 08:36:24.16 & 09.05.2010 08:53:01.83 & 16.63 & 12 & 0.23 & 1.511 & 63 & I & S13 & 183.B-0100(T) \\
09.05.2010 09:40:54.93 & 09.05.2010 10:23:02.79 & 42.13 & 24 & 0.68 & 1.831 & 84 & I & S13 & 183.B-0100(T) \\
12.06.2010 03:33:19.65 & 12.06.2010 04:20:46.86 & 47.45 & 24 & 2.30 & 2.063 & 86 & I & S13 & 183.B-0100(T) \\
16.06.2010 05:18:28.69 & 16.06.2010 06:56:15.44 & 97.78 & 48 & 1.38 & 2.080 & 84 & I & S13 & 183.B-0100(U) \\
\tableline
\end{tabular}

\tablecomments{This table shows basic information for each data set shown in the Figures before: start-stop times, length of the data set, number of frames, maximum flux density occurring in the set, average time sampling over the set, the used integration time, the observational mode (imaging or polarimetric), the camera (S13 with 13'' FOV or S27 with 27''), and the project ID of the observations. Note that start-stop times, frame numbers and length average are given for the datasets before applying the quality cut, average sampling and maximum flux density after the cut.}
\end{center}
%\end{table}
\clearpage
\end{landscape}

\clearpage

\section{Flux density statistics}
\label{apstat}
In this Appendix we present supplement information for the statistical analysis described in section~\ref{rep}. We present histograms of the Ks-Band flux densities of Sgr~A* with different than the optimal binning, showing that the linear behavior of the histogram is not binning dependent, and the CDFs of independently drawn power-law distributed surrogate data in comparison to the observed CDF (see section~\ref{pl}).

\begin{figure}[h!]
   \includegraphics[width=7.0cm]{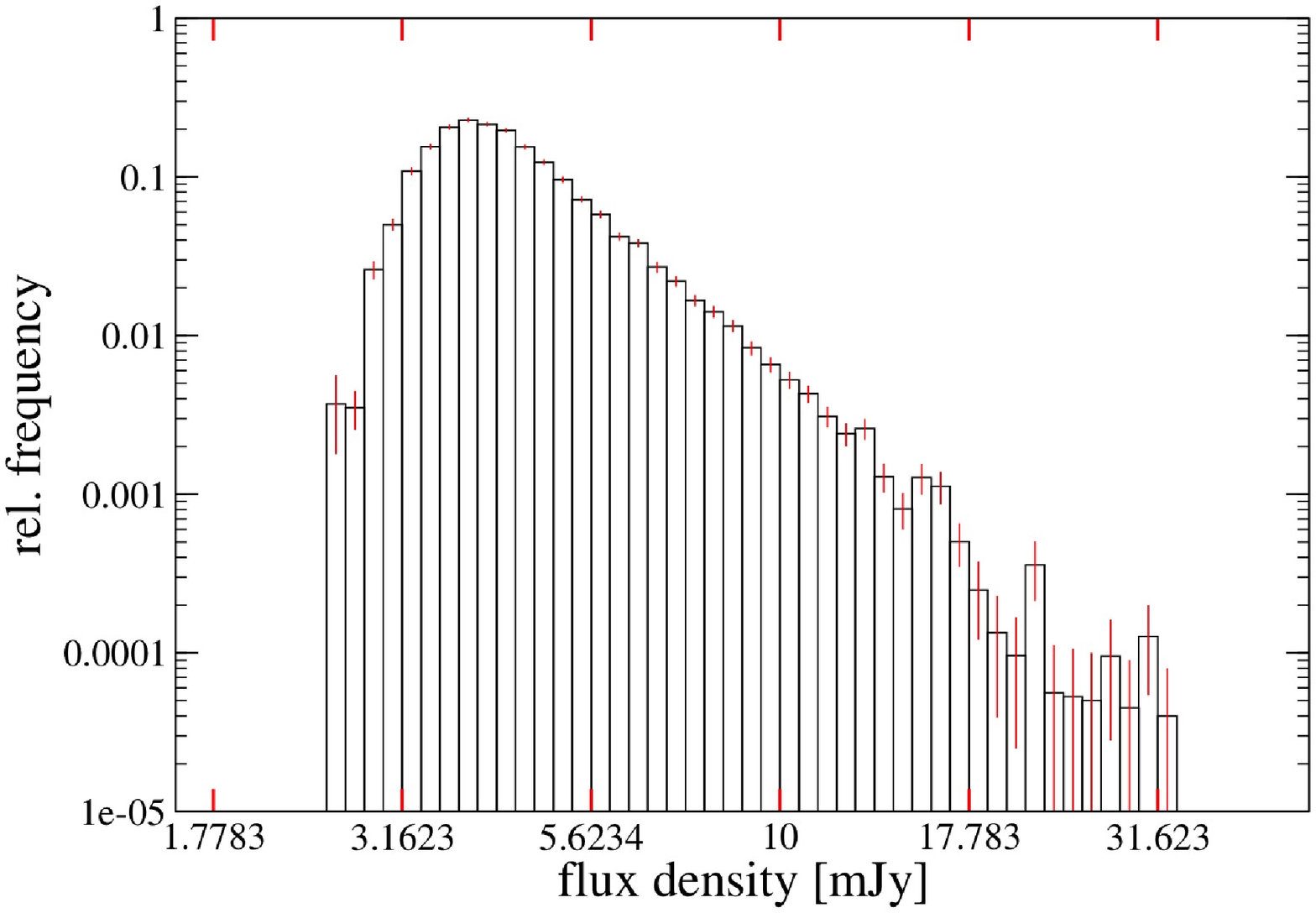}
   \includegraphics[width=7.0cm]{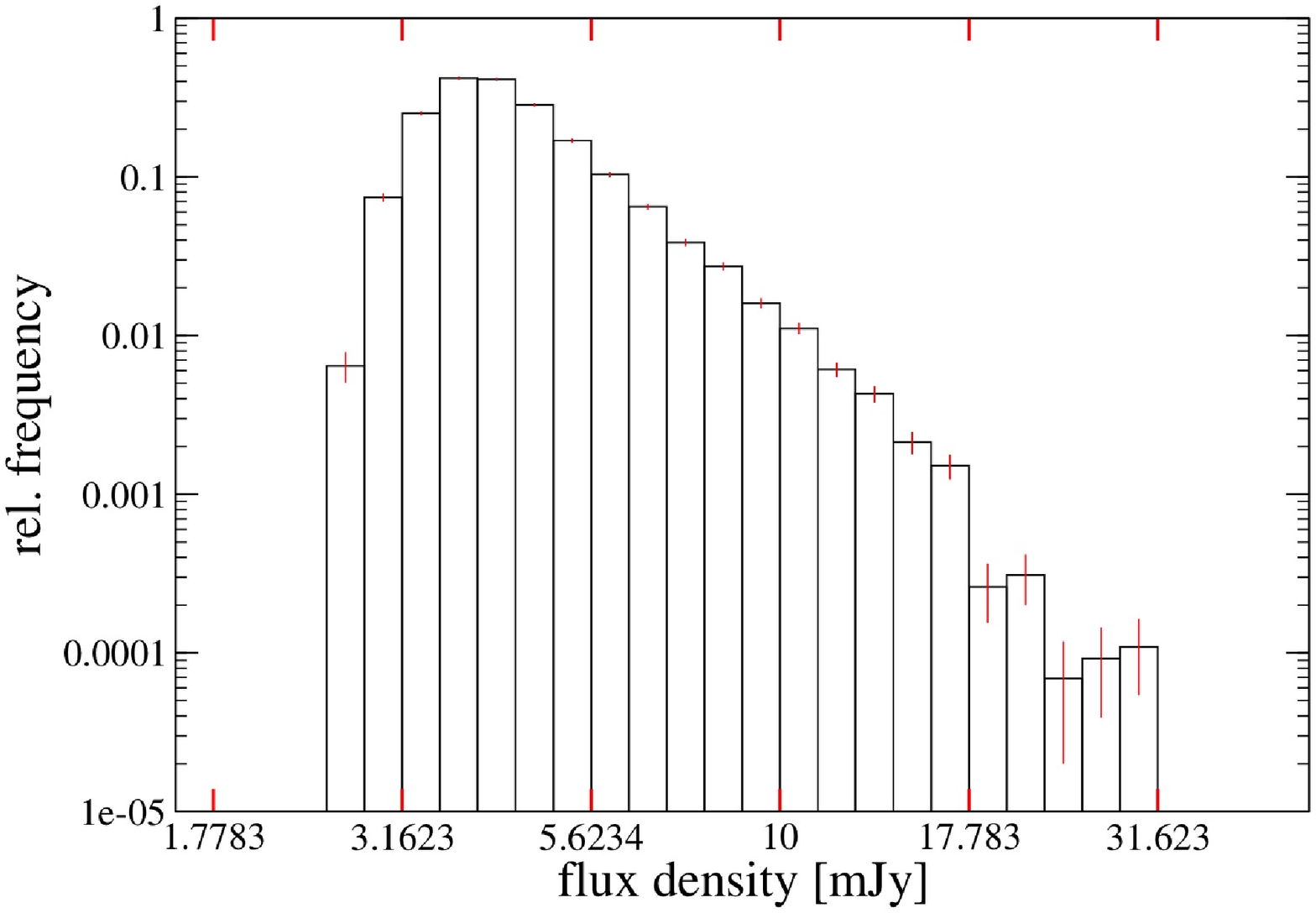}
    \centering
    \includegraphics[width=7.0cm]{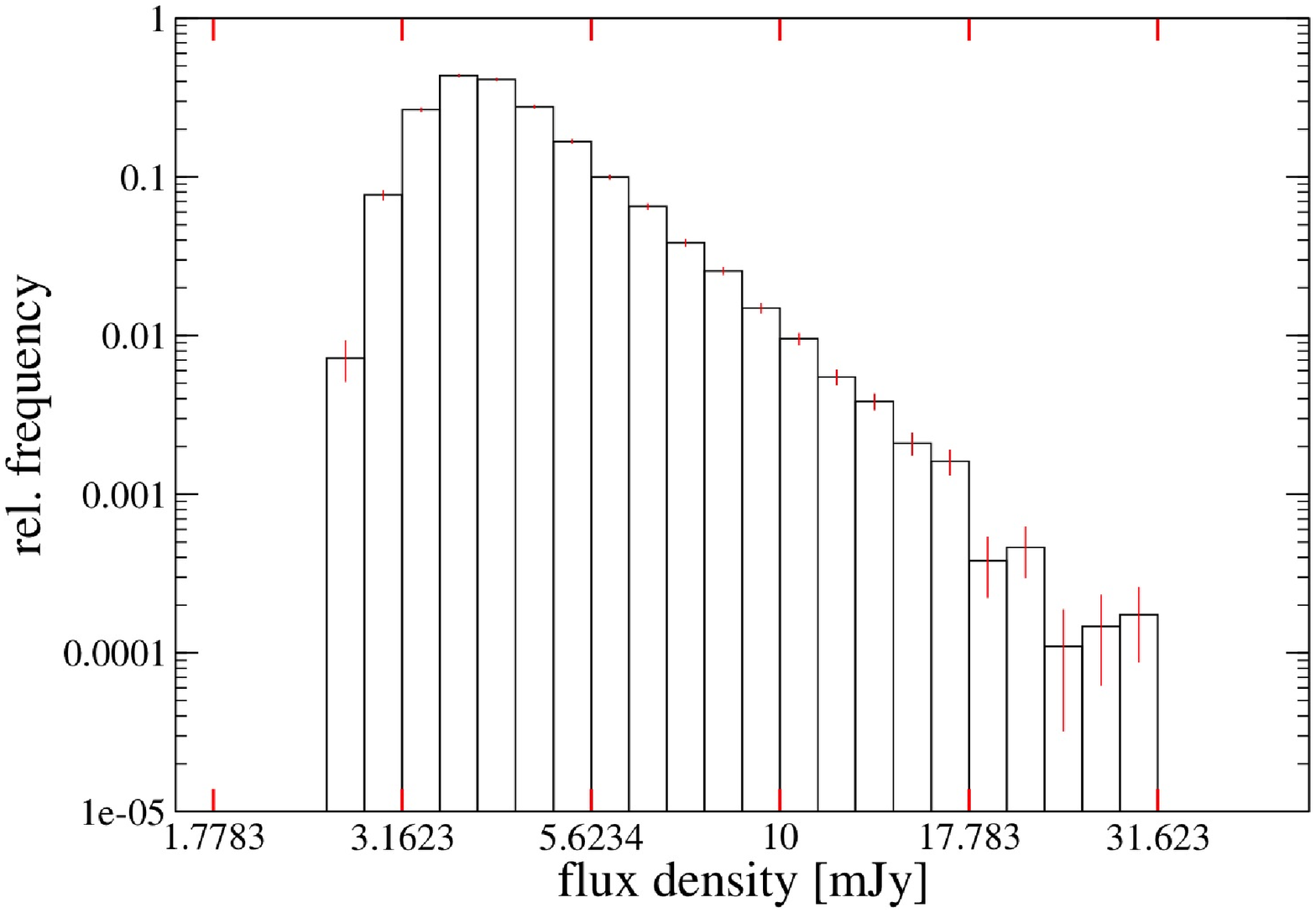}
      \caption[Flux density histograms with different binning.]{The observed flux densities in histograms with bin numbers higher (upper left, 45 bins) and lower (upper right, 22, compare \citealt{2011ApJ...728...37D}) than optimal (32). The latter we also show in a integration time weighted version (as conducted by \citealt{2011ApJ...728...37D}), finding now significant difference (lower plot).}
         \label{nonoptbin}
\end{figure}

\begin{figure}[h!]
   \includegraphics[width=7.5cm]{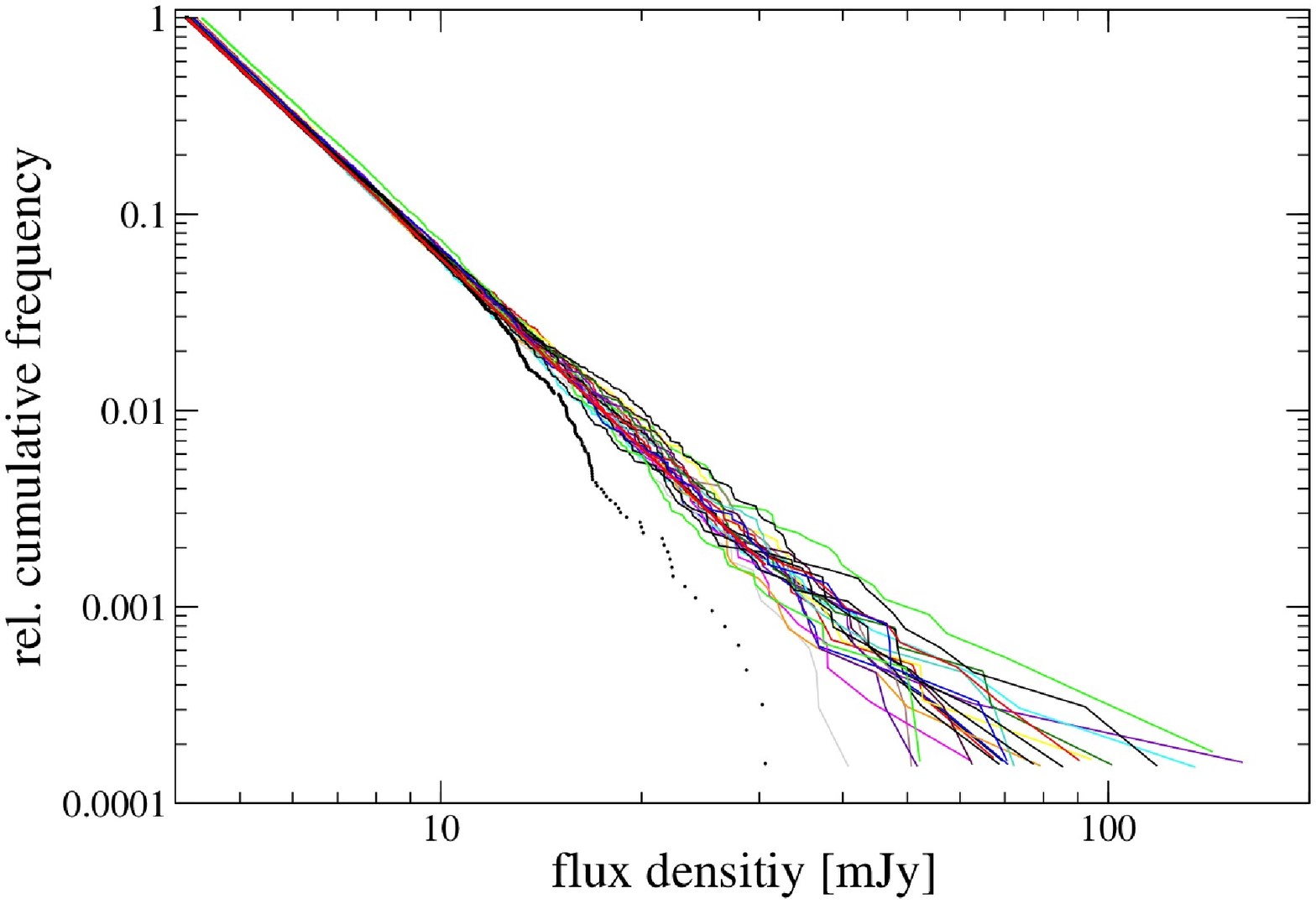}
      \includegraphics[width=7.5cm]{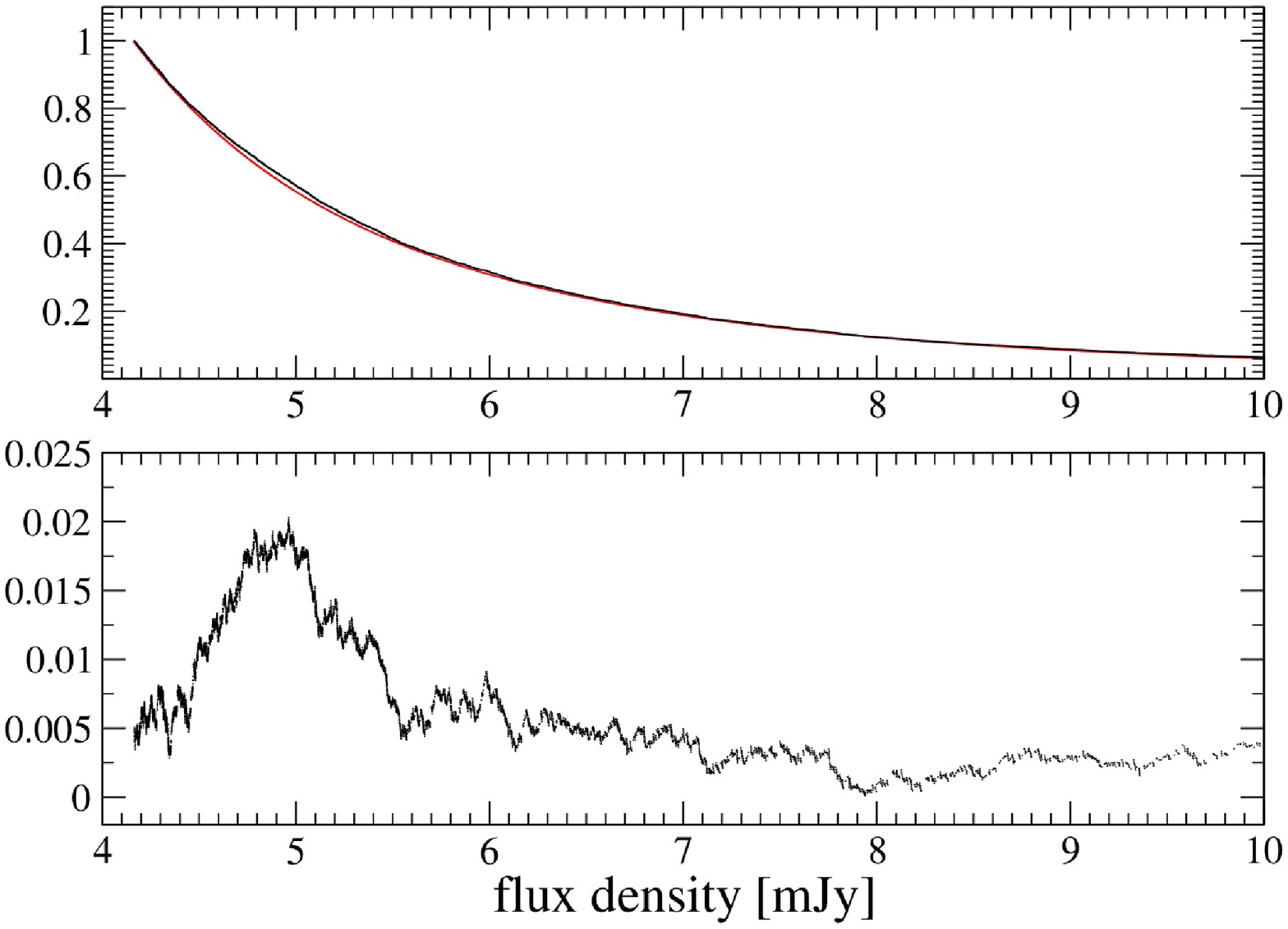}
      \caption[Comparison of the observed CDF with surrogate data.]{Comparison of the observed CDF with surrogate data drawn independently from a power-law according to Equation~(\ref{powerre}) (section~\ref{pl}). Although the most obvious differences seem to be at highest flux densities, this impression is caused by the logarithmic presentation (left). The KS statistics used for determining the best $ x_{\rm{min}} $ and the plausibility of a power-law model in section~\ref{pl} is dominated by differences at lower flux densities as demonstrated in the plots on the right, showing the comparison of the observed CDF and a surrogate CDF with a worse Kolmogorov-value than the observed data (upper right: CDFs in linear plotting, lower right: difference of the CDFs).}
         \label{CDFindepend}
\end{figure}

\newpage
\clearpage

\section{Generating Fourier Transform based surrogate data}
\label{surr}
Examples of red-noise light curves obtained by transformation according to Equation~(\ref{transf}). The transform was applied to a linear Gaussian light curve with single and double broken power-law PSDs as discussed in section~\ref{sim}.

\begin{figure*}[h!]
   \centering
   \includegraphics[width=14.5cm]{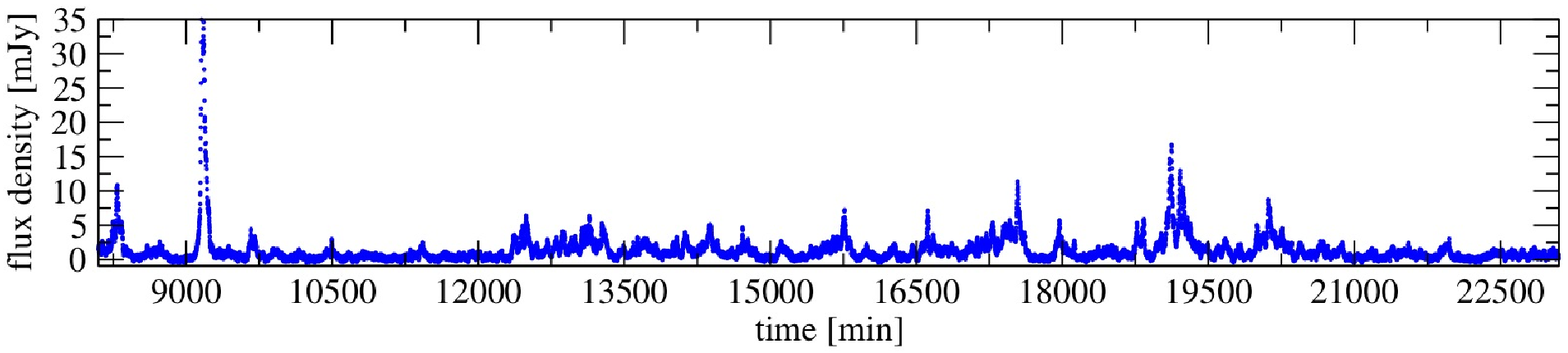}
\end{figure*}

\begin{figure*}[h!]
   \centering
   \includegraphics[width=14.5cm]{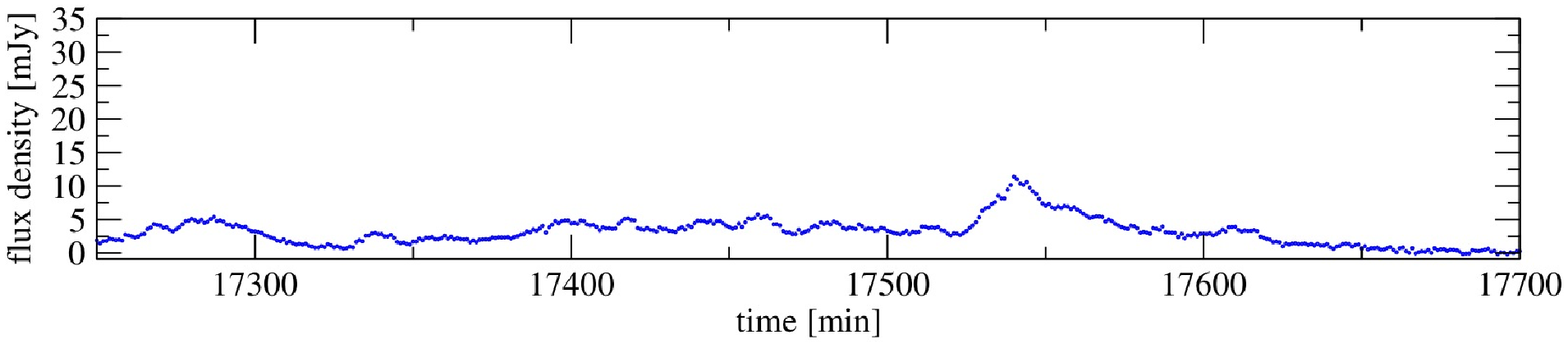}
\end{figure*}

\begin{figure*}[h!]
   \centering
   \includegraphics[width=14.5cm]{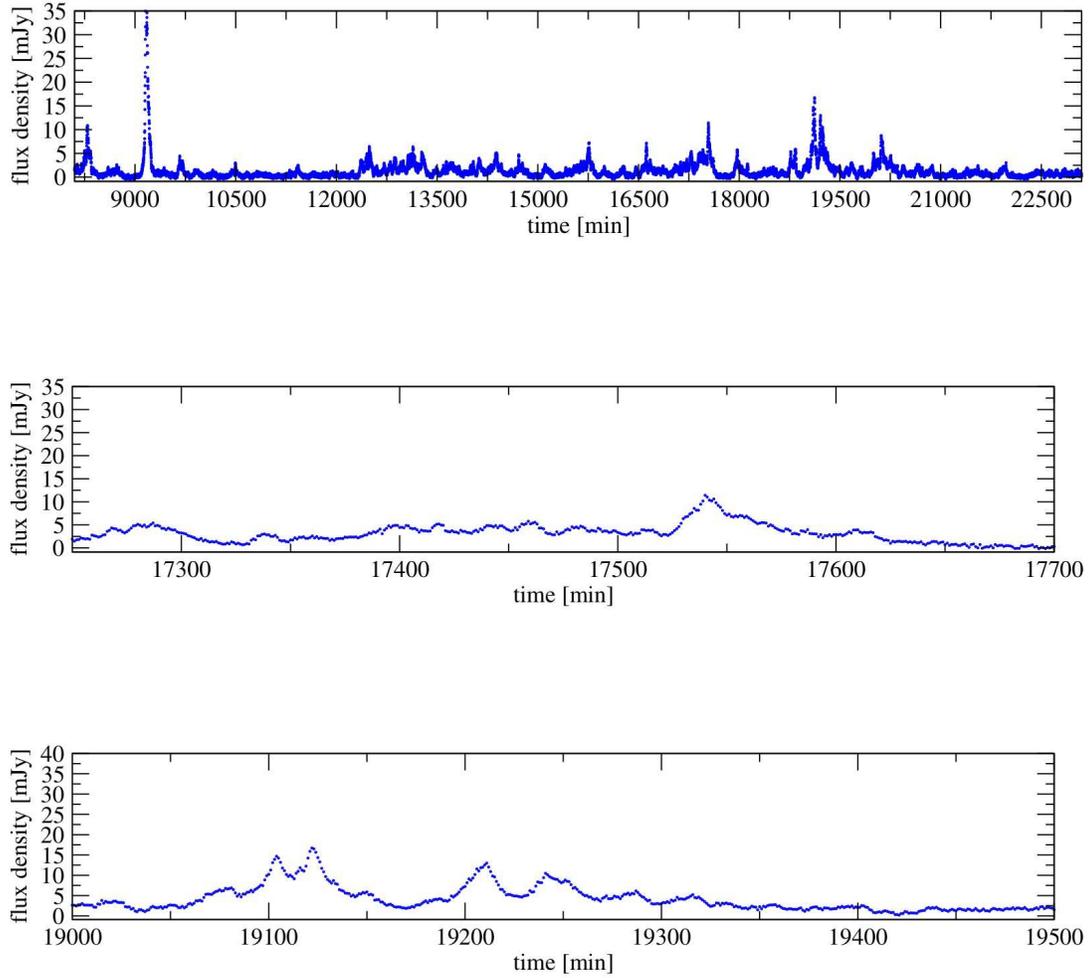}
      \caption[Simulated lightcurves I.]{Simulated lightcurves taken from a $ 4 \cdot 10^6 $ min time series created with the best double-broken power-law PSD. The upper panel shows typical 15\,000 min, the lower panels 500 min closeups with lower and higher flux density levels. Light curves created with the best single broken PSD show to a first order the same behavior as the presented case (compare Figure~\ref{simlclong}).}
         \label{simlc94}
\end{figure*}

\newpage
\clearpage

\begin{figure*}
   \centering
   \includegraphics[width=14.5cm]{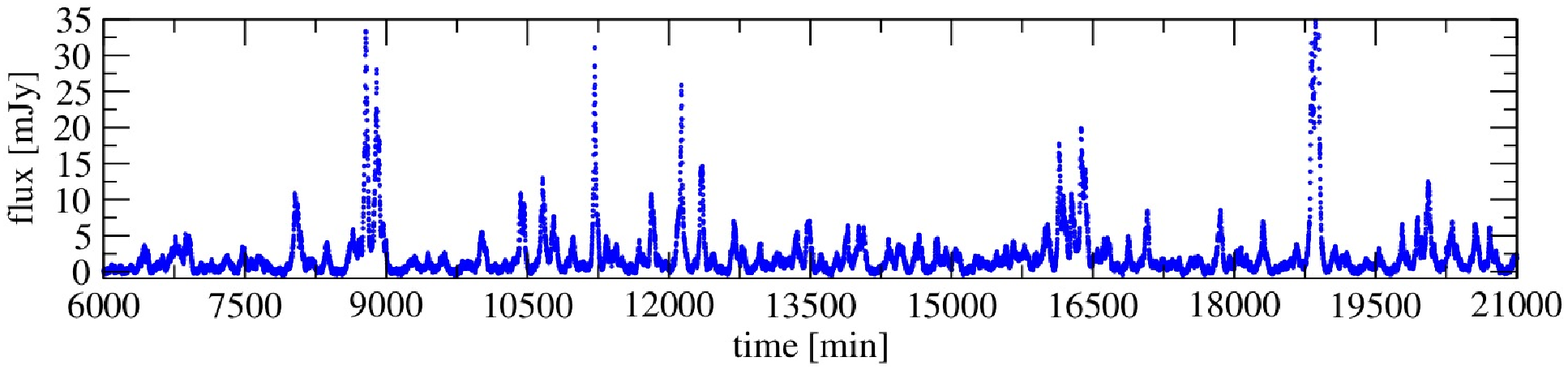}
\end{figure*}

\begin{figure*}
   \centering
   \includegraphics[width=14.5cm]{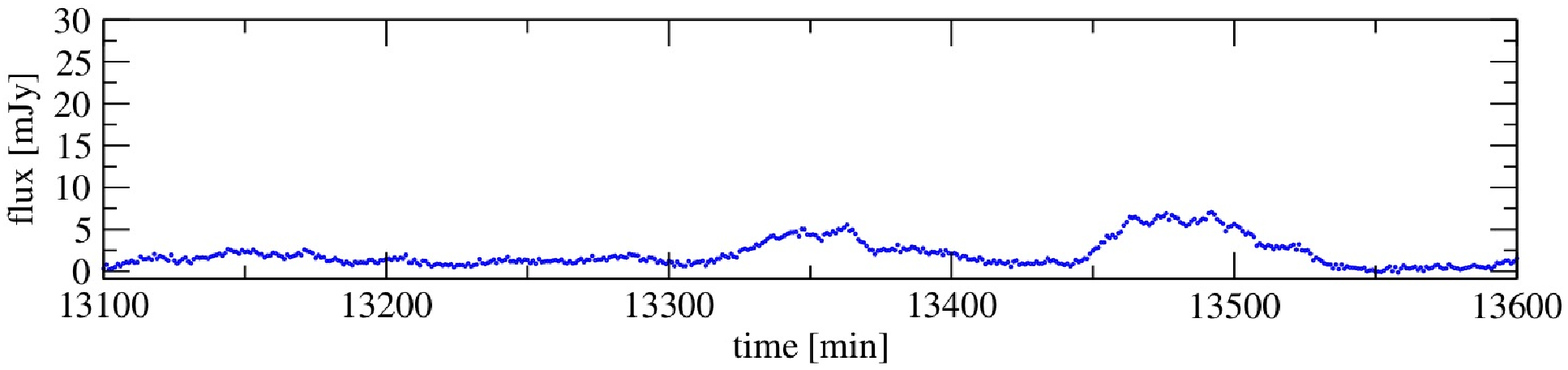}
\end{figure*}

\begin{figure*}
   \centering
   \includegraphics[width=14.5cm]{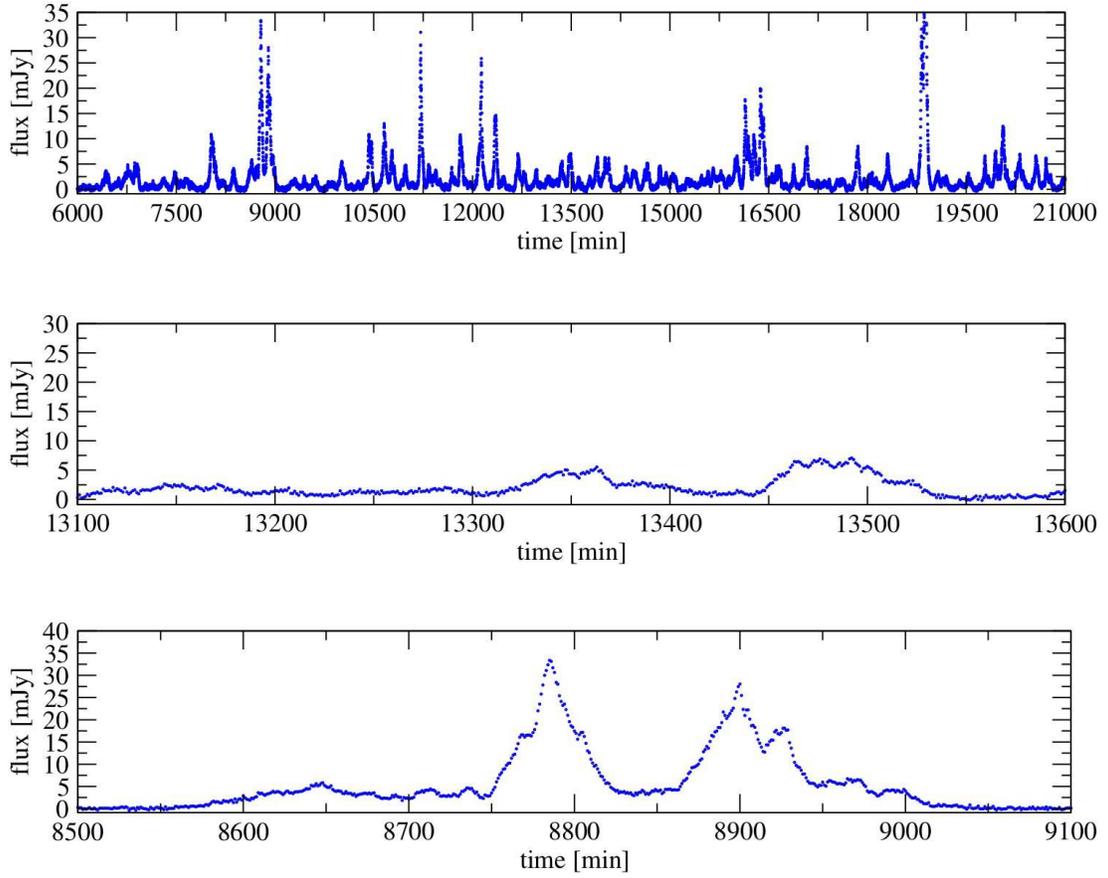}
      \caption[Simulated lightcurves II.]{Simulated lightcurves taken from a $ 4 \cdot 10^6 $ min time series created with a single-broken power-law PSD with an acceptance of $ 68\% $, corresponding to the blue structure function in Figure~\ref{fit}. The upper panel shows typical 15\,000 min, the lower panels 500 min closeups with lower and higher flux density levels.}
         \label{simlc68}
\end{figure*}

\begin{figure*}
  \centering
   \includegraphics[width=14.5cm]{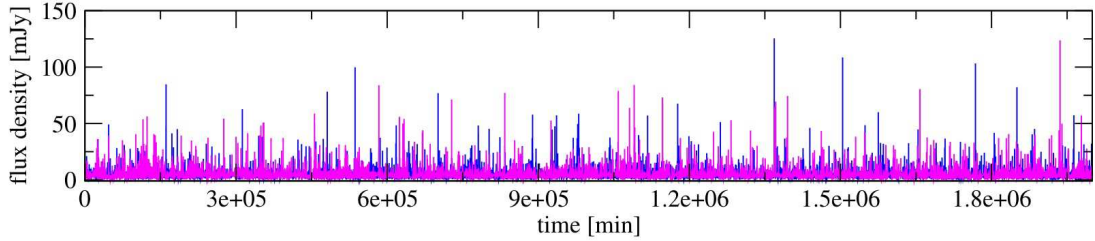}
      \caption[Simulated lightcurves III.]{Long term appearance of light curves generated with the best single (blue) and the best double (magenta) broken PSD. On a time period of $ 2 \cdot 10^{6} $ min no obvious difference is noticeable.}
         \label{simlclong}
\end{figure*}
\end{appendix}

\end{document}